\documentstyle[epsfig,rotate,12pt]{article}
\topmargin - 1.0 true cm
\def\sss{\scriptscriptstyle}
\def\barp{{\raise.35ex\hbox{${\sss (}$}}---{\raise.35ex\hbox{${\sss )}$}}}
\def\bdbarp{\hbox{$B_d$\kern-1.4em\raise1.4ex\hbox{\barp}}}
\def\bsbarp{\hbox{$B_s$\kern-1.4em\raise1.4ex\hbox{\barp}}}
\def\dbarp{\hbox{$D$\kern-1.1em\raise1.4ex\hbox{\barp}}}
\def\dcp{D^0_{\sss CP}}
\def\dbar{{\overline{D^0}}}
\def \zpc#1#2#3{{\it Z.~Phys.,} C#1 (19#2) #3}
\def \plb#1#2#3{{\it Phys.~Lett.,} B#1 (19#2) #3}

\def\rly#1{\mathrel{\raise.3ex\hbox{$#1$\kern-.75em\lower1ex\hbox{$\sim$}}}}

\newread\epsffilein 
\newif\ifepsffileok 
\newif\ifepsfbbfound 
\newif\ifepsfverbose 
\newdimen\epsfxsize 
\newdimen\epsfysize 
\newdimen\epsftsize 
\newdimen\epsfrsize 
\newdimen\epsftmp  
\newdimen\pspoints  
\def\epsfbox#1{\global\def\epsfllx{72}\global\def\epsflly{72}%
 \global\def\epsfurx{540}\global\def\epsfury{720}%
 \def\lbracket{[}\def\testit{#1}\ifx\testit\lbracket
 \let\next=\epsfgetlitbb\else\let\next=\epsfnormal\fi\next{#1}}%
\def\epsfgetlitbb#1#2 #3 #4 #5]#6{\epsfgrab #2 #3 #4 #5 .\\%
 \epsfsetgraph{#6}}%
\def\epsfnormal#1{\epsfgetbb{#1}\epsfsetgraph{#1}}%
\def\epsfgetbb#1{%
\openin\epsffilein=#1
\ifeof\epsffilein\errmessage{I couldn't open #1, will ignore it}\else
 {\epsffileoktrue \chardef\other=12
 \def\do##1{\catcode`##1=\other}\dospecials \catcode`\ =10
 \loop
  \read\epsffilein to \epsffileline
  \ifeof\epsffilein\epsffileokfalse\else
   \expandafter\epsfaux\epsffileline:. \\%
  \fi
 \ifepsffileok\repeat
 \ifepsfbbfound\else
 \ifepsfverbose\message{No bounding box comment in #1; using defaults}\fi\fi
 }\closein\epsffilein\fi}%
\def\epsfclipstring{}
\def\epsfsetgraph#1{%
 \epsfrsize=\epsfury\pspoints
 \advance\epsfrsize by-\epsflly\pspoints
 \epsftsize=\epsfurx\pspoints
 \advance\epsftsize by-\epsfllx\pspoints
 \epsfxsize\epsfsize\epsftsize\epsfrsize
 \ifnum\epsfxsize=0 \ifnum\epsfysize=0
  \epsfxsize=\epsftsize \epsfysize=\epsfrsize
  \epsfrsize=0pt
arithmetic! 
  \else\epsftmp=\epsftsize \divide\epsftmp\epsfrsize
  \epsfxsize=\epsfysize \multiply\epsfxsize\epsftmp
  \multiply\epsftmp\epsfrsize \advance\epsftsize-\epsftmp
  \epsftmp=\epsfysize
  \loop \advance\epsftsize\epsftsize \divide\epsftmp 2
  \ifnum\epsftmp>0
   \ifnum\epsftsize<\epsfrsize\else
    \advance\epsftsize-\epsfrsize \advance\epsfxsize\epsftmp \fi
  \repeat
  \epsfrsize=0pt
  \fi
 \else \ifnum\epsfysize=0
  \epsftmp=\epsfrsize \divide\epsftmp\epsftsize
  \epsfysize=\epsfxsize \multiply\epsfysize\epsftmp   
  \multiply\epsftmp\epsftsize \advance\epsfrsize-\epsftmp
  \epsftmp=\epsfxsize
  \loop \advance\epsfrsize\epsfrsize \divide\epsftmp 2
  \ifnum\epsftmp>0
  \ifnum\epsfrsize<\epsftsize\else
   \advance\epsfrsize-\epsftsize \advance\epsfysize\epsftmp \fi
  \repeat
  \epsfrsize=0pt
 \else
  \epsfrsize=\epsfysize
 \fi
 \fi
 \ifepsfverbose\message{#1: width=\the\epsfxsize, height=\the\epsfysize}\fi
 \epsftmp=10\epsfxsize \divide\epsftmp\pspoints
 \vbox to\epsfysize{\vfil\hbox to\epsfxsize{%
  \ifnum\epsfrsize=0\relax
  \includegraphics{#1}%
  \else
  \epsfrsize=10\epsfysize \divide\epsfrsize\pspoints  
  \includegraphics{#1}%
  \fi
  \hfil}}%
\global\epsfxsize=0pt\global\epsfysize=0pt}%
 {\catcode`\%=12
\global\let\epsfpercent=
\long\def\epsfaux#1#2:#3\\{\ifx#1\epsfpercent
 \def\testit{#2}\ifx\testit\epsfbblit
  \epsfgrab #3 . . . \\%
  \epsffileokfalse
  \global\epsfbbfoundtrue
 \fi\else\ifx#1\par\else\epsffileokfalse\fi\fi}%
\def\epsfempty{}%
\def\epsfgrab #1 #2 #3 #4 #5\\{%
\global\def\epsfllx{#1}\ifx\epsfllx\epsfempty
  \epsfgrab #2 #3 #4 #5 .\\\else
 \global\def\epsflly{#2}%
 \global\def\epsfurx{#3}\global\def\epsfury{#4}\fi}%
\def\epsfsize#1#2{\epsfxsize}
\let\epsffile=\epsfbox
\def\bsll{$b \rightarrow s \ell^+ \ell^- $ }
\def\bxsll{$B \rightarrow X_s \ell^+ \ell^- $ }
\def\bxsee{B \rightarrow X_s e^+ e^-  }
\def\bxsmm{B \rightarrow X_s \mu^+ \mu^-  }
\def\bxstt{B \rightarrow X_s \tau^+ \tau^- }

\def\bxsg{$B \rightarrow X_s \gamma $ }
\def\s{\hat{s}}

\newcommand{\ra}{\rightarrow}

\def\g{\gamma}

\def\mt{m_t}
\def\mb{m_b}

\def\mc{\hat{m}_c}
\def\lo{\hat{\lambda}_1}
\def\lt{\hat{\lambda}_2}

\def\q{\hat{q}}
\def\bra{\langle}
\def\ket{\rangle}
\def\bea{\begin{eqnarray}}
\def\eea{\end{eqnarray}}
\def\be{\begin{equation}}
\def\ee{\end{equation}}
\def\a{\alpha}

\def\g{\gamma}

\def\p{\pi}

\def\l{\lambda}

\def\G{\Gamma}

\def\mt{m_t}

\newcommand{\bgamaxs}{$B \to X _{s} + \gamma$}

\newcommand{\BGAMAXS}{B \ra X _{s} + \gamma}
\newcommand{\BGAMAXD}{B \ra X _{d} + \gamma}
\newcommand{\BBGAMAXS}{{\cal B}(B \ra  X _{s} + \gamma)}
\newcommand{\BBGAMAXD}{{\cal B}(B \ra  X _{d} + \gamma)}

\newcommand{\BGAMAKSTAR}{B \ra  K^{\star} + \gamma}
\newcommand{\GGAMAXD}{\Gamma(B \ra  X _{d} + \gamma)}

\newcommand{\GGAMAXS}{\Gamma (B \ra  X _{s} + \gamma)}

\def\beq{\begin{equation}}
\def\eeq{\end{equation}}
\def\Vcdabs{\vert V_{cd} \vert}
\def\Vus{V_{us}}
\def\Vusabs{\vert V_{us} \vert}

\def\Vcsabs{\vert V_{cs} \vert}
\def\Vud{V_{ud}}
\def\Vudabs{\vert V_{ud} \vert}
\def\Vbc{V_{cb}}

\def\Vcbabs{\vert V_{cb} \vert}
\def\Vbu{V_{ub}}

\def\Vubabs{\vert V_{ub}\vert}
\def\Vtd{V_{td}}
\def\Vtdabs{\vert V_{td} \vert}
\def\Vts{V_{ts}}
\def\Vtsabs{\vert V_{ts} \vert}

\def\Vtbabs{\vert V_{tb}\vert}

\newcommand{\abseps}{\vert\epsilon\vert}

\def\BS{B_s^0}
\def\BSB{\bar{B_s^0}}

\newcommand{\fbb}{f^2_{B_d}B_{B_d}}
\newcommand{\fbbs}{f^2_{B_s}B_{B_s}}
\newcommand{\fbd}{f_{B_d}}

\newcommand{\go}[1]{\gamma^{#1}}
\newcommand{\gu}[1]{\gamma_{#1}}

\newcommand{\delmd}{\Delta M_d}
\newcommand{\delms}{\Delta M_s}

\def\sw{\sin{^2}\theta _{W}}

\def\qbar{\overline q}

\def\dbar{\overline d}

\def\q5q{\qbar{{\lambda_a}\over 2} i\gamma_5 q}

\newcommand{\bgamaxd}{$B \to X _{d} + \gamma$}

\def\bxslll{$B \rightarrow X_s \ell^+ \ell^- $}
\def\to{\rightarrow}

\def\mb{m_b}

\def\xs{x_s}
\def\xd{x_d}
\newcommand{\kkbar}{$K^0$-${\overline{K^0}}$}
\newcommand{\bdbdbar}{$B_d^0$-${\overline{B_d^0}}$}
\newcommand{\bsbsbar}{$B_s^0$-${\overline{B_s^0}}$}

\def\as{\alpha _s}

\def\Vbc{V_{cb}}
\def\Vbu{V_{ub}}
\def\Vtd{V_{td}}
\def\Vts{V_{ts}}

\begin{document}
\begin{flushright}
DESY 96-248\\
December 1996\\
\end{flushright}
\begin{center}
{\large \bf
\centerline
{$B$ Decays in the Standard Model - Status and Perspectives
$\footnote{Lectures given at the XXXVI School of Theoretical Physics,
Zakopane, Poland, June 1 - 10, 1996;\\ to be published in Acta Physica
Polonica B, Vol. 27 (1996).}$}}
\vspace*{1.5cm}
{\large A.~Ali}
\vskip0.2cm
  Deutsches Elektronen Synchrotron DESY, Hamburg \\
\vspace*{1.5cm}
\centerline{\it Dedicated to the memory of Professor Abdus Salam}
\vskip1.0cm
{\Large Abstract\\}
\parbox[t]{\textwidth}{
\indent
These lectures review some of the progress made in the quantitative 
understanding of $B$ decays. The emphasis here is on applications of QCD
using perturbative and non-perturbative techniques. In some cases, 
however, phenomenological models must at present be invoked to make 
meaningful comparison with data. The resulting picture is 
consistent with the standard model (SM) and this agreement is
quantified in terms of the branching ratios, mixing probabilities, and
lifetimes which measure the charge current and effective flavour changing
neutral current transitions involving $B$ hadrons. This, in turn,
enables a determination of five of the nine elements of the quark mixing 
matrix. We discuss several proposals on improving the precision on the
parameters of this matrix in forthcoming experiments. Issues intimately 
related to the quark mixing matrix such as the profile of the unitarity
triangle and CP-violating asymmetries in $B$ decays are discussed. In 
particular, we emphasize the role of rare $B$ decays and
$B^0$ - $\overline{B^0}$ mixings in testing the SM quantitatively and in 
searching for physics beyond the SM.}
\end{center}
\thispagestyle{empty}
\newpage
\setcounter{page}{1}
\textheight 23.0 true cm

\noindent
\section{Introduction}

    The principal interest in the studies of $B$ decays
in the context of the standard model (SM) \cite{GSW} lies in that they 
provide valuable information on the weak rotation matrix - the 
Cabibbo-Kobayashi-Maskawa (CKM) matrix \cite{Cabibbo,KM}.
In fact, $B$ decays determine five of the nine CKM matrix elements: 
 $V_{cb}, V_{ub}, V_{td}, V_{ts}$, and $V_{tb}$.
 The dominant decays of $b$-quark 
 stem from the direct $bcW^-$ coupling; then there are
decay modes which stem from the CKM-suppressed $buW^-$ coupling. These two 
classes represent the direct charged current (CC) transitions.
 The electromagnetic penguins and particle-antiparticle mixing(s), 
representing
the so-called flavour changing neutral current processes (FCNC) which
have been observed in $B$ decays, are induced as higher order effects (loops)
as the SM does not allow direct couplings of the
form $bsX$ or $bdX$, where $X=\gamma, Z, H^0, q\bar{q}$ or a gluon.
The effective induced couplings in the SM are governed by the GIM 
mechanism \cite{GIM} and are dominated by the intermediate (virtual)
top quark contribution - the quark with the largest Yukawa coupling -
through the transitions $b \to tW \to s$ and $b \to tW \to d$. Their
quantitative measurements therefore provide information about the
properties of the top quark, such as its mass and its weak mixing matrix
elements $V_{td}, V_{ts}$, and $V_{tb}$. 

 To extract the CKM matrix elements from the hadronic transitions,
one needs to implement the QCD perturbative corrections and calculate
the hadronic decay form factors and decay functions 
for the inclusive and exclusive decays, respectively.
A lot of work has gone into calculating the perturbative QCD corrections
in $B$ decays and this will be discussed in some detail here. The
aspects having to do with non-perturbative physics are not yet under
quantitative control, though important advances have been made and
partial answers are available.  In principle,
non-perturbative aspects in $B$ decays can 
all be calculated in the Lattice-QCD framework. In practice, the impact of
this technique is limited due to the inadequacy of the present 
computer technology
which restricts direct computation of the $B$ decay properties
involving the $b$-quark with a typical mass of
$O(5 ~\mbox{GeV})$. However, useful information on some form factors
and coupling constants has been
obtained by simulation of the charmed hadron systems and extrapolating
to the $b$-quark mass, often also using the constraints from the
limiting behaviour of QCD in the $m_Q \to \infty$ 
limit. There exist other non-perturbative 
theoretical tools such as the heavy quark effective theory HQET, the
QCD sum rules, and the good old potential models, 
which have been put to good use in the quantitative analyses of experimental
results in $B$ decays. We shall
review here some representative applications of each of these methods. 
They, in particular the HQET techniques, have
enabled us to determine the two mentioned matrix elements $V_{cb}$ and 
$V_{ub}$.

 The CKM matrix elements $V_{ti};~i=d,s,b$ are, in
principle, also measurable in the production and decays of the top quark
\cite{top95}. We note that first measurements of
 $\Vtbabs$ have been reported by the CDF collaboration \cite{CDFvtb},
through the ratio $R_{tb}$,
\be\label{rtbcdf}
 R_{tb} \equiv \frac{{\cal B}(t \to   
bW)}{\sum_{q=d,s,b} {\cal B}(t \to q W)}=
 0.94 \pm 0.27(\mbox{stat}) \pm 0.13(\mbox{syst}) ~.
\ee
Assuming three generations, this yields
\be
\Vtbabs = 0.97 \pm 0.15 \pm 0.07~,
\ee
which is consistent with unity but
within experimental errors also consistent with a value which is
considerably less than unity, namely at $95\%$ C.L. one gets $\Vtbabs > 
0.58$. This measurement is expected to
improve significantly in future. A precision of
 $\delta \Vtbabs/\Vtbabs \simeq 12\%$ is
projected at the Fermilab Tevatron with an integrated luminosity of
$2 (fb)^{-1}$, expected to be collected at the turn of this century
\cite{CDFvtb}. Eventually, $\Vtbabs$
will be measured in experiments at the linear collider(s) with a precision
$\delta \Vtbabs /\Vtbabs =O(1-2)\%$ from the anticipated
accuracy of $\delta \Gamma(t)/\Gamma(t) \simeq 1 \%$ on the top quark decay 
width \cite{Fuji96}. However, it will be difficult in the foreseeable 
future to get quantitatively useful  information on
 $V_{td}$ and $V_{ts}$ from direct decays of the top quark,
both due to the anticipated small branching ratios involving these
matrix elements, 
\be
{\cal B} (t \to sW) = O(10^{-3}), ~~~{\cal B} (t \to dW) = O(10^{-4}),
\ee
and, more importantly, due to the (present) low efficiency of
tagging light-quark jets. This is somewhat discomforting as the direct
determination of the CKM matrix elements in top quark decays and their
inferred values from FCNC processes, such as the ones from $B$ decays being 
discussed in these lectures, would have given very stringent
constraints on possible new physics or perhaps would have established 
its existence. It is likely that the FCNC processes in $B$ (and to a
lesser extent in $K$) decays will remain the major source of information
on $V_{td}$ and $V_{ts}$. 
We will discuss the present quantitative determinations of these matrix
elements and their possible improved 
measurements at the forthcoming $B$ facilities, 
such as the $B$ factories,
HERA-B, and the hadron colliders (Tevatron and LHC), using rare $B$ decays,
$\delmd$ and $\delms$.

 The weak interaction phase responsible for CP 
violation in the SM resides dominantly in the matrix elements 
$V_{td}$ and $V_{ub}$. This is manifest in
the Wolfenstein parameterization \cite{Wolfenstein} of the CKM matrix.   
Hence $B$ decays and mixings involving one or both of these 
matrix elements are potentially the most promising means to measure CP
violation. Since the information on the CP violating phase is
rather sparse, essentially confined at present to the decay $K_L \to \pi 
\pi$, the CP violating asymmetries in $B$ decays will be very welcome
and perhaps decisive input in testing the CKM paradigm for CP violation.
We shall give a profile of these CP violating asymmetries
in $B$ decays and the underlying CKM unitarity triangle
based on fitting the present data \cite{AL96} and   
will discuss measurements at future facilities which 
will go a long way in reducing the present uncertainties in the CKM 
parameter space. These experiments (and the anticipated theoretical progress)
have the possibility of putting the quark flavour physics at a comparable 
precision level as the present electroweak physics in the post-LEPI era.

  This writeup is organized as follows: In section 2, we introduce 
the CKM matrix and the unitarity triangle(s) using the Wolfenstein 
parametrization for this matrix. In section 3, we discuss the 
dominant $B$ decay modes, which determine the bulk quantities such as
the semileptonic branching ratio
${\cal B}_{SL }(B)$, the average number of charmed particles per $B$ decay
 $\langle n_c\rangle$, and the individual $B$ hadron
lifetimes. The present determination of the 
matrix elements
$\Vcbabs$ and $\Vubabs$ from semileptonic $B$ decays is also discussed 
in this section, 
using the HQET methods for the former. In section 4, we take up the 
discussion of the 
electromagnetic penguins and rare $B$ decays in the SM and make comparison
with data in terms of the branching ratio and the photon energy spectrum.
This measurement determines the ratio of the CKM matrix elements 
$\Vtsabs/\Vcbabs$ which we quantify.
The CKM-suppressed radiative rare decays $\BGAMAXD$
and the corresponding exclusive decays $B \to (\rho,\omega) \gamma$
are discussed in section 5. Their
role in determining the CKM matrix element $\Vtdabs$ (equivalently the
CKM-Wolfenstein parameters $\rho$ and $\eta$) is reviewed.  The
success of this proposal depends in a crucial way on reliable calculations 
of the so-called long distance (LD) contributions and we discuss some 
existing estimates of the same. In this section, we also take up  
the FCNC semileptonic decay $B \to X_s \ell^+ \ell^-$ in the SM model,
discussing first the QCD-improved rates and distributions from the 
short-distance (SD) contribution, including leading power corrections.
A quantitative understanding of these decays requires reliable estimates
of the LD  and non-perturbative effects which we also discuss. In
section 6, we give an update of the CKM matrix and the unitarity triangle 
(UT), taking into account the present measurements and theoretical estimates 
in a 
number of $B$ decays and $\abseps$, the CP violating parameter in the 
kaon sector.
The constraints on the CKM parameters from the present LEP bound on
$\delms$ are also analyzed. In section 7 we briefly discuss some representative
CP-violating asymmetries in $B$ decays and summarize their expected ranges
and correlations in the SM. We conclude with a brief summary in 
section 8. Some of the topics
discussed here have also been reviewed in \cite{ALI96}.

\section{CKM Matrix and the Unitarity Triangle}

We start by discussing the flavour changing transitions in the SM.
Since QCD is manifestly flavour-diagonal, the only possibility of FCNC
transitions is in the electroweak sector.  
Writing in terms of the physical boson
$(W_\mu^{\pm}, ~Z_\mu^0, ~A_\mu )$ and fermion fields,
it is easy to show that the neutral current
part of the standard electroweak model is also manifestly flavour-diagonal.
Denoting the quarks and leptons by $f_i (i=1...6)$,
the neutral current in the SM is given by:
\begin{equation}
J_\mu^{NC} = \sum_i \bar{f}_i
\bigg[ \frac{e}{\sin \theta_W \cos \theta_W}
Z_\mu \left( {I_3}_L - Q \sw \right)_i
 +  e A_\mu Q_i \bigg] f_i,
\label{neutralc}
\end{equation}
where $(I_3)_L=(1-\gamma_5)/2 (I_3)$ with
 $I_3=+1/2$ for $u_i$ and $\nu_i$ and $-1/2$ for $d_i$
and charged leptons $\ell_i$, and $Q_i$ is 
the electric charge of the fermion $f_i$ in units of the electron charge,
i.e., $Q_e=+1$. The electroweak mixing
angle in $J_\mu^{NC}$, denoted by $\theta_W$, has its origin in the
diagonalization of the gauge boson mass matrix, and it has 
the usual definition $\cos \theta_W = g_2/\sqrt{g_1^2
+g_2^2}$, with the electric charge defined as
$e \equiv g_2 \sin \theta_W$. Concerning the Higgs Yukawa couplings - 
a potential source of FCNC transitions in general - it is known that the
unitary transformations which diagonalize the quark mass matrix 
also diagonalize the Higgs Yukawa sector in the SM. This is most easily 
seen by
writing the Yukawa sector of the SM Lagrangian, which after spontaneous
symmetry breaking has the form
\be
{\cal L}_{\mbox{Yukawa}} = -\left[\bar{u}_{iL}M^{u}_{ij}u_{jR}
+ \bar{d}_{iL}M^{d}_{ij}d_{jR} + \bar{\ell}_{iL}M^{\ell}_{ij}\ell_{jR}
\right] \left( 1 + \frac{H}{v_0}\right) + h.c.~,
\ee
where the absence of the neutrino mass matrix is conspicuous and 
represents the SM choice of treating the neutrino massless (equivalently,
the absence of the right-handed neutrinos $v_{iR}$). 
In the basis where the quark masses are diagonal, this takes the form
\be
{\cal L}_{\mbox{Yukawa}} = - \sum_i m_i\bar{f}_i f_{i} \left( 1 + 
\frac{H}{v_0}\right) ~,
\ee
where $H$ is the Higgs field and $v_0$ is the Higgs vacuum expectation value.
 This manifest flavour diagonal form of ${\cal L}_{\mbox{Yukawa}}$ 
in general is not maintained 
in multi-Higgs models and one has to impose discrete
symmetries on the Higgs and fermion fields to forbid FCNC couplings
in ${\cal L}_{\mbox{Yukawa}}$, as emphasized by Glashow and
Weinberg quite some time ago \cite{GW77}.
The absence of such couplings in the SM owes itself to
the choice of a single Higgs doublet.

The charged current $J_\mu^{CC}$ in the SM, which couples
to the $W^{\pm}$, is
\begin{equation}
J_\mu^{CC} =  \frac{e}{ \sqrt{2} \sin \theta_W }
 \left( \bar{u}, \bar{c}, \bar{t} \right)_L
\gamma_\mu V_{\mbox{\footnotesize CKM}}
\left( \begin{array}{c} d\\ s \\ b \end{array} \right)_L ,
\end{equation}
where $V_{\mbox{\footnotesize CKM}} \equiv V_L^{\rm up} V_L^{{\rm down}\dag}$
is a $(3 \times 3)$ unitary matrix in flavour space, first
written down by Kobayashi and Maskawa in 1973 \cite{KM}.
The matrices $V_L^{\rm up}$ and $V_L^{\rm down}$ diagonalize the
up-type and down type quark mass matrices, respectively.  The matrix
$V_{\mbox{\footnotesize CKM}}$ is a
generalization of the Cabibbo rotation \cite{Cabibbo} for the
three-quark-flavour $(u, d, s)$ case, invented to keep the universality
of weak interactions, which took the form of a
                                           $(2\times 2)$ matrix by the
inclusion of $c$-quark with the GIM construction \cite{GIM}, and is
called the Cabibbo-Kobayashi-Maskawa (CKM) matrix. There are no FCNC 
transitions in the SM at the tree level by construction. They 
are induced by higher order CC transitions and 
the resulting FCNC amplitudes are determined by the masses of
the intermediate quarks, i.e. they reflect the flavour dependence of the 
Higgs Yukawa couplings, weighted with the appropriate CKM prefactors.
 
 The charged current in the SM
 has a $(V-A)$ structure, hence it violates
P and C maximally, conserves the electric charge and the lepton-
and baryon-number separately, but otherwise there are no restrictions
on it except that $V_{\mbox{\footnotesize CKM}}^{\dag}
 V_{\mbox{\footnotesize CKM}} = 1$. In general,
${\cal L}^{CC}$ violates CP due to the
possibility of a non-trivial phase in $V_{\mbox{\footnotesize CKM}}$.\\   
 Symbolically the matrix $V_{\mbox{\footnotesize CKM}}$ can be written as:
\begin{equation}
V_{\mbox{\footnotesize CKM}} \equiv \left( \begin{array}{lll}
V_{ud} & V_{us} & V_{ub}\\
V_{cd} & V_{cs} & V_{cb}\\
V_{td} & V_{ts} & V_{tb}
\end{array} \right).
\end{equation}
For quantitative discussions we need a parametrization of the CKM matrix.
The 
original parametrization due to Kobayashi and Maskawa \cite{KM}
was constructed from the rotation matrices in the flavour space
involving the angles $\theta_i ~(i = 1,2,3)$ and a
phase $\delta$,
\begin{equation}
V_{\mbox{\footnotesize KM}} = R_{23} ( \theta_3, \delta) R_{12} (\theta_1, 0)
R_{23} (\theta_2,0),
\end{equation}
where $0 \le \theta_i \le \pi/2,\,\, 0 \le \delta \le 2 \pi$, and
$R_{ij}(\theta, \phi)$ denotes a unitary rotation in the $(i,j)$
plane by the angle $\theta$ and the phase $\phi$.
The resulting representation is:
\begin{equation}
V_{\mbox{\footnotesize KM}} = \left( \begin{array}{lll}
c_1 & -s_1c_3 & -s_1 s_3 \\
s_1 c_2 & c_1c_2c_3-s_2s_3e^{i \delta} & c_1c_2s_3+
s_2c_3e^{i\delta} \\
s_1s_2 & c_1 s_2c_3+c_2s_3 e^{i \delta} & c_1s_2s_3-c_2c_3
e^{i\delta}
\end{array} \right),
\end{equation}
with $ c_i = \cos \theta_i, s_i = \sin \theta_i $.
This reduces to the usual Cabibbo form for $ \theta_2 = \theta_3 = 0$,
with the angle $\theta_1$, identified
(up to a sign) with the Cabibbo angle. In the PDG review \cite{PDG96}, 
however, another parametrization is advocated which differs
from $V_{\mbox{\footnotesize KM}}$ in assigning the
complex phases (dominantly) to the (1,3) and (3,1)
matrix elements of $V_{\mbox{\footnotesize CKM}}$.
An approximate but very useful form of the matrix
$V_{\mbox{\footnotesize CKM}}$ is due to Wolfenstein \cite{Wolfenstein}:
 \begin{equation}\label{Vwolf}
V_{\mbox{\footnotesize Wolfenstein}} = \left( \begin{array}{lll}
1-\frac{1}{2} \lambda^2 & \lambda &
A\lambda^3 (\rho - i \eta) \\
- \lambda & 1-\frac{1}{2} \lambda^2 
 & A \lambda^2 \\
A\lambda^3 (1-\rho-i \eta) & -A \lambda^2 & 1
\end{array} \right),
\label{Vwolfenstein}
\end{equation}
with $\lambda \equiv \sin \theta_c \simeq 0.221$.
Like other representations, $V_{\footnotesize{\mbox{Wolfenstein}}}$
 has also three
real parameters called $A$, $\lambda$ and $\rho$, and a phase $\eta$.
Since we shall be making extensive use of this parametrization, we
write some relations involving
the matrix elements of interest in this representation:
\begin{equation}
\frac{|\Vbu|}{|\Vbc|} = \lambda \sqrt{\rho^2 + \eta^2},
~~~\frac{|\Vtd|}{|\Vbc|} = \lambda \sqrt{(1-\rho)^2 + \eta^2},
\nonumber \\
\end{equation}
\begin{equation}
\frac{|\Vtd|}{|\Vbu|} =  \sqrt{\frac{(1-\rho)^2 + \eta^2}{\rho^2+ \eta^2}},
~~~\frac{|\Vts|}{|\Vbc|} = 1 ~,
\end{equation}
and the dominant phases are:
\begin{equation}
\Im (\Vbu ) = \Im (\Vtd ) = - A \lambda^3 \eta .
\end{equation}
 It should be recalled  
that the Wolfenstein parameterization given in Eq.~(\ref{Vwolfenstein})
is an approximation and in certain 
situations in the future it may become mandatory to specify  the matrix 
by taking into account the dropped terms in $O(\lambda^4)$ 
in $V_{\footnotesize{\mbox{Wolfenstein}}}$. For the present experimental and 
theoretical accuracy, the representation (\ref{Vwolf}) is
entirely adequate and we shall restrict ourselves to this form.
Further discussions on this point and suggestions on improved 
treatment to include higher order terms in $\lambda$
can be seen in \cite{BuBu94}. \\

\subsection{The CKM unitarity triangles}
\par
\begin{figure}
\epsfig{file=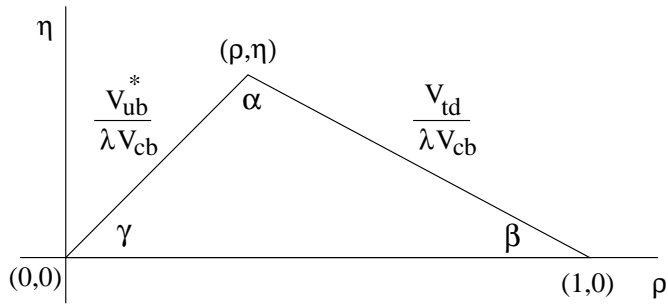,bbllx=30pt,bblly=285pt,bburx=390pt,bbury=494pt,
width=10cm}

\caption{The unitarity triangle. The angles $\alpha$, $\beta$ and $\gamma$
can be measured via CP violation in the $B$ system and the sides
from the CC- and FCNC-induced $B$ decays.}
\label{triangle}
\end{figure}

\indent
The CKM matrix elements obey unitarity constraints, which state that
any pair of rows, or any pair of columns, of the CKM matrix are
orthogonal. This leads to six orthogonality conditions which can be
depicted as six triangles in the complex plane of the CKM parameter space
\cite{AKL94}. The constraint 
stemming from the orthogonality condition on the first and third
row of $V_{\mbox{\footnotesize CKM}}$,
\begin{equation}
V_{ud} V_{td}^* + V_{us} V_{ts}^* + V_{ub} V_{tb}^* = 0 
\label{tdunit}
\end{equation}
 has received considerable attention. Since, as discussed in the 
introduction, present measurements are consistent with 
$V_{ud} \simeq 1, ~V_{tb} \simeq 1$ and $V_{ts}^* \simeq - V_{cb}$,
the unitarity relation (\ref{tdunit}) simplifies to:
\begin{equation}
V_{ub} + V_{td}^* = V_{us} V_{cb},
\end{equation}
which can be conveniently depicted as a triangle relation in
the complex plane, as shown in Fig.~\ref{triangle}. We shall refer to 
it as the unitarity triangle (UT).
 Thus, knowing the
sides of the UT, the three
angles  of this triangle $\alpha, \beta$ and $\gamma$ are determined.
These angles are all related to the Kobayashi-Maskawa phase
$\delta$ (equivalently the phase $\delta_{13}$
 in $V_{\mbox{\footnotesize PDG}}$ or the
phase $\eta$ in $V_{\mbox{\footnotesize Wolfenstein}}$),
 and they can, in principle, be independently measured in various
CP-violating $B$ decays. Restricting to the Wolfenstein representation in
which the dominant phases reside in the $(13)$ and $(31)$ matrix elements,
these angles are defined as follows:
\bea
\sin 2 \alpha &=& \arg \left( \frac{V_{ub}V_{td}}{V_{ub}^* V_{td}^*} \right)
=\frac{2 \eta (\eta^2 + \rho^2 - \rho)}{(\rho^2 + \eta^2)((1-\rho)^2 + 
\eta^2)} ~, \nonumber \\
\sin 2 \beta &=& \arg \left( -\frac{V_{td}}{V_{td}^*} \right)
= \frac{2 \eta (1-\rho)}{(1-\rho)^2 + \eta^2} ~, \nonumber\\
\sin 2 \gamma &=& \arg \left( -\frac{V_{ub}}{V_{ub}^*} \right)
= \frac{2 \eta \rho}{\rho^2 + \eta^2} ~.
\eea
Some estimates of the angles $\alpha, \beta$ and $\gamma$, and hence
CP asymmetries in $B$ decays, can be obtained at present by constraining the
parameters of the CKM matrix $\rho$ and $\eta$. Conversely, knowing
the CP asymmetries, the parameters $\rho$ and $\eta$ of the CKM matrix 
can be determined.
 
As already stated, the matrix elements
$V_{cb}$ and $V_{ub}$ are  known from the  CC $B$ decays.
         With more data and improved theory (in particular for the so-called
heavy-to-light transitions $b \to u \ell \nu_\ell$) 
one would be able to determine these matrix elements rather
precisely. The matrix element
           $V_{td}$ can, in principle, be determined from the
rare decays $b \to d + ~\gamma$, $b \to d + ~l^+ l^-$,
$ b \to d + \nu \bar{\nu}$ (and the corresponding exclusive decays),
and $B_d^0$--$B_d^0$ mixing. The mass difference $\delmd$
already provides a first measurement of $\Vtdabs$.
This set of measurements, which involves decay rates and mixing frequencies
but not CP-violating asymmetries,
then provides another way of determining the triangle, namely by measuring
its sides. The lengths of these sides  in the Wolfenstein
approximation are
\bea
R_b &\equiv & \frac{1}{\lambda}|\frac{V_{ub}}{V_{cb}}| ~, \nonumber\\
R_t &\equiv & \frac{1}{\lambda}|\frac{V_{td}}{V_{cb}}| ~.
\eea
The CP asymmetries in $B$ decays related to the angles $\alpha, \beta$ 
and $\gamma$ and the sides of the unitarity triangle obey the geometric
relations:
\bea
R_b &=& \frac{\sin \beta}{\sin \alpha} = \frac{\sin (\alpha + 
\gamma)}{\sin \alpha} = \frac {\sin \beta}{\sin (\alpha + \beta)} 
~,\nonumber\\
R_t &=& \frac{\sin \gamma}{\sin \alpha} = \frac{\sin (\alpha +
\beta)}{\sin \alpha} = \frac {\sin \gamma}{\sin (\gamma + \beta)}
~.
\eea
 By measuring both the sides and the angles, the UT
will be overconstrained which is one of the principal goals of the current
and forthcoming experiments in $B$  physics.
Before leaving this topic, we note that including the $O(\lambda^5)$ terms
in the imaginary part of the CKM matrix,
an additional phase emerges in the matrix element combination
$\arg ( -V_{cs}^*V_{cb}/V_{cb}^* V_{cd} )$. This phase, being equal to
$ \lambda^2 \eta$ is
bounded from the CKM fits to be less than $ 0.025$ and hence very small. 
 However, showing that this CP
violating phase is indeed small is both a test of the SM and 
belongs on the agenda of the forthcoming experiments in $B$ decays.
 
\par
Obviously,  there exists a large number of 
different parametrizations of the CKM matrix.
However, since the phases of the quark fields are unphysical
quantities, the different parametrizations, emerging from
specific choices of these phases, must all be equivalent.
The parametrization independent quantities are the
absolute values of the matrix elements $|V_{ij}|$ (hence also the angles 
of the unitarity triangles) and the area of the unitarity triangles, which is
the same for all six triangles and is an invariant measure of  CP violation.
This can be expressed as
 \be
 \mbox{area}[\Delta \mbox{(CKM)}] = \frac{1}{2}A\lambda^{6} \eta
\ee
for the Wolfenstein parametrizations of $ V_{\mbox{\footnotesize CKM}}$.
The Jarlskog invariant denoted by the symbol $J$ \cite{Jarlskog} is 
twice this area, which  
in the standard model is typically of $O(10^{-5})$. It is being 
debated if the intrinsic smallness of $J$ in the standard model is a 
serious problem in 
explaining the measured baryon asymmetry of the universe (BAU),
whose quantitative measure is the ratio of the baryon number density to
entropy density,
\begin{equation}
  \Delta_B= \frac{\rho(B)}{s},
\end{equation}
with $\Delta_B=(0.6 - 1) \times 10^{-10}$ \cite{PDG96}. Relating the Jarlskog
constant to $\Delta B$ is a profound problem and a topic of intense 
theoretical
research. It is an interesting  question if $B$ decays will lead
to some helpful clues towards understanding this connection. 
Recent studies indicate
that baryogenesis in the SM at the electroweak scale is unlikely 
\cite{Jansen96} due to the LEP constraints on the Higgs boson mass.
Very probably, the baryon number violation (as well as the lepton number 
violation) is governed by the physics at the grand unification scale,
which has then little direct influence on $B$ decays.
Experiments in $B$ physics will, however, provide an answer to the question if 
additional CP violating phases in the flavour changing sector are present.
These experiments, together with the searches of CP violation in the
flavour diagonal sector such as the electric dipole moment of the
neutron, will determine the effective low energy theory of CP violation.
\section{Dominant $B$ Decays in the Standard Model}

\par
We now turn to the mainstream $B$ physics and discuss the dominant decay 
rates which determine the lifetimes of the $B$ hadrons $\tau_B$,
 the semileptonic branching
ratios ${\cal B}_{SL }$, and the charm quark multiplicity in $B$ decays
$\langle n_c\rangle$ - a quantity which has become an important ingredient
in understanding the semileptonic branching ratios.
 
 The effective lowest-order
weak interaction Hamiltonian can be expressed in terms of
 $J_{\mu}^{CC}$, introduced earlier,\\
\begin{equation}
{\cal{H}}_{W}
      = \frac{G_{F}}{2 \sqrt{2}} \left( J_{\mu}^{CC}
J^{\mu\dag CC} + \mbox{h.c.} \right) ,
\end{equation}
where $G_{F}$ is the Fermi coupling constant. The calculational
framework that is used is QCD and we concentrate first on
perturbative QCD improvements of the decay rates and distributions
in $B$ decays. The leading order (in $\as$)
perturbative QCD improvements using ${\cal{H}_W}$ have been worked out in
semileptonic processes in \cite{CM} - \cite{JK89},
 which are modeled on the electromagnetic radiative
corrections in the decay of the $\mu$-lepton \cite{Behrends}.
For the non-leptonic decays, perturbative QCD corrections are
calculated in the effective Hamiltonian approach  
using the renormalization group techniques \cite{AM74}$ - $\cite{BW90}. The 
underlying theoretical framework and its numerous 
applications in weak decays of the $K$ and $B$ mesons have been  
reviewed in a comprehensive paper by
Buchalla, Buras and Lautenbacher \cite{BBL95}, to which we refer for 
details and confine ourselves here to some selected topics.

Apart from these perturbative QCD improvements, resulting in the so-called 
QCD-improved quark-parton model, one could also improve the quark-parton 
model itself by including power corrections in $1/m_Q$.
The method that is used in discussing such corrections is based 
on the heavy quark limit of QCD which allows one to
do a systematic expansion of decay amplitudes in $1/m_Q$, where
$m_Q \gg \Lambda_{QCD}$, and $\Lambda_{QCD}$ is the QCD scale parameter
which is typically of $O(200$ MeV) \cite{PDG96}. 
This technique \cite{Chayetal} - \cite{AHHM96} has the satisfying
feature that the parton model for heavy quark inclusive decays emerges as 
the leading term in the expansion of the decay amplitudes.
These methods can also be applied to calculate the energy-momentum spectra
of the decay products except in the end-point region, where the
heavy quark expansion breaks down. Here, one has at present little choice
other than smearing the (theoretical and experimental) spectra with weight 
functions to make meaningful comparison or modeling the 
non-perturbative effects. We shall return to the discussions of these
topics later.
\subsection{Inclusive semileptonic decay rates of the $B$ hadrons}
\par
We start with the assumption
that the inclusive decays of $B$ hadrons can
be modeled on the QCD-improved quark model decays. More specifically,
while calculating rates, we shall be equating the partial and total decay
rates of the
$B$ hadrons to the corresponding expressions obtained in the parton model,
relying on the heavy quark expansion \cite{Chayetal} - \cite{AHHM96}:
\begin{equation}
\Gamma(B \to X) = \Gamma (b \to x) + O(1/m_b^2) ~. 
\end{equation}
 For $b$ quark semileptonic
decays, one has two partonic CC transitions:
\begin{eqnarray}
  b & \longrightarrow & c\ell^{-}\bar{\nu}_{\ell}~,  \\
    & \longrightarrow & u\ell^{-}\bar{\nu}_{\ell}~. \nonumber
\label{bsemilept}
\end{eqnarray}   
There exists a  close analogy between the $b$ quark decays
  and $\mu$ decay,
$\mu^{-}\longrightarrow e^{-}\bar{\nu}_{e}\nu_{\mu}$,
with the identification:\\
\begin{equation}
[b,(c,u),\bar{\nu}_{\ell},\ell^{-}]\leftrightarrow [\mu^{-},e^{-},
\bar{\nu}_{e},\nu_{\mu}].
\end{equation}
 This analogy holds also at the one loop
level; $O(\alpha$) QED corrections to $\mu^{-}$ decay and
$O(\alpha_{s}$) QCD corrections to $b$ semileptonic
decays are related by simply replacing \cite{CM,Suzuki,Alipiet}\\
\begin{equation}
\alpha\longrightarrow\  \frac{1}{3}\alpha_{s} Tr\sum_{i=1}^{8}
\lambda_{i}\lambda_{i}   = \frac{4}{3}\alpha_{s},
\end{equation}
where $\lambda_{i}$ are the Gell-Mann $SU(3)$ matrices, and
$\as$ is the lowest order QCD effective coupling constant,
\begin{equation}
\alpha_{s} = \frac{12\pi}{(33-2n_{f})\ln(  \frac{m_{b}^{2}}
{ \Lambda_{\small{\mbox{QCD}}}^{2} }  )} \;\;,
\end{equation}
where $n_f$ is the number of effective quarks.
                     The semileptonic decay rates can then be
read off the expression for the $O(\alpha$) radiatively
corrected $\mu$-decay rate \cite{Behrends}.
The rates for $b\longrightarrow (u,c)\ell\nu_{\ell}$ decays,
setting $m_{\ell}=m_{\nu_{\ell}}=0$, are given by the expression: 
\begin{equation}
\Gamma_{SL }(b\longrightarrow (u,c)\ell\nu_{\ell})=
\Gamma_{0} f(r_{i})\left[1-\frac{2}{3}\frac{\alpha_{s} (m_{b}^{2})}
{\pi} g(r_{i})\right],
\end{equation}
with $\Gamma_{0}$ being the normalization factor in the lowest-order
rate
 \begin{equation}
\Gamma_{0} = \frac{G_{F}^{2}}{192\pi^{3}}
\mid V_{ib}\mid^{2} m_{b}^{5},
\end{equation}
 $r_{i} = m_{i}/m_{b} ~(i=u,b)$, and
\begin{equation}
f(r) = 1-8r^{2}+8r^{6}-r^{8}-24r^{4} \ln r .
\end{equation}
The function $g(r)$
              has the normalization $g(0)=\pi^{2}-\frac{25}{4}$,
and numerically $g(0.3)\simeq 2.51$, relevant for the $b \to u $ and $b 
\to c$ transitions, respectively \cite{CM,Suzuki,Alipiet}.
With $\Lambda_{\mbox{\small QCD}} \simeq 200$ MeV and $n_f=5$, this gives 
about 
$(15)\%$ corrections to the semileptonic decay widths involving $\ell =e,
\mu$, reducing $\Gamma_{SL }$ compared to the lowest order result
$ \Gamma_{SL }^{(0)} =\Gamma_{0} f(r)$.
The corresponding decrease in the decay width for the semileptonic
decay $b \to c \tau \nu_\tau$ is obtained by an expression very  
similar to the above one in which the $\tau$-mass effects are included
in the phase space and in the QCD corrections.
 \begin{equation}\label{bcln}
\Gamma(b\to c \tau \nu_\tau)=\Gamma_0 P(x_c,x_\tau,0)\left[
 1+\frac{2\as(\mu)}{3\pi} g(x_c,x_\tau,0)\right]
\end{equation}
where $P(x_1,x_2,x_3)$ is the well known three-body phase space factor
given for arbitrary masses $x_i=m_i/m_b$ in \cite{BKbook}.
The function $g(x_1,x_2,x_3)$ has been calculated for arbitrary arguments
in \cite{PHam83} in terms of a one-dimensional integral.
The functions $P(x_1, 0, 0)$ and $g(x_c, 0,0)$ go over to the
functions $f(r)$ and $(-)g(r)$, respectively, given above for the massless
lepton case.  The numerical
values for $g(x_c, x_\tau,0)$ and $g(x_c,0, 0)$ are tabulated in
\cite{Bagan94}. For the default value $x_c=0.3$, one has $g(x_c,
x_\tau,0) = -2.08$, yielding about a 12 \% decrease in $\Gamma (b \to c
\tau \nu_\tau)$ compared to $\Gamma_{SL }^{(0)} (b \to c \tau \nu_\tau)$
as a result of the leading order QCD corrections \cite{PHam83}. For 
more modern calculations of the decay rate
$\Gamma_{SL } (b \to c \tau \nu_\tau)$, see \cite{FCz95}.

\subsection{Inclusive non-leptonic decay rates of the $B$ hadrons}
\par
  The dominant CC-induced non-leptonic and semileptonic decays of $B$
hadrons are governed by the effective Lagrangian,
\begin{eqnarray}\label{Leff1}
{\cal L}_{eff} = -4\frac{G_F}{\sqrt{2}}\Vud^*\Vbc \left[ C_1(\mu){\cal 
O}_1(\mu)
 + C_2(\mu){\cal O}_2(\mu)\right] \nonumber\\
 -4\frac{G_F}{\sqrt{2}}\Vus^*\Vbc \left[ C_1(\mu){\cal O}_1^\prime(\mu)
+ C_2(\mu){\cal O}_2^\prime(\mu)\right] \nonumber\\
-4\frac{G_F}{\sqrt{2}}\Vbc \left[\sum_{\ell=e,\mu,\tau} \bar{\ell}_L 
\gamma_\mu \nu_\ell \bar{c}_L\gamma^\mu b_L \right] + h.c.~,
\end{eqnarray}
and we have just discussed the $O(\as)$ renormalization effects
to the matrix elements of the semileptonic piece in ${\cal L}_{eff}$.
Here ${\cal O}_1$ and ${\cal O}_2$ are the colour-octet and colour-singlet
four-Fermi operators, respectively  ($\alpha$ and $\beta$ are colour 
indices),
 \begin{eqnarray}
{\cal O}_1 &=& (\bar{d}_\alpha u_\beta)_L(\bar{c}_\beta b_\alpha)_L, 
\nonumber\\
{\cal O}_2 &=& (\bar{d}_\alpha u_\alpha)_L(\bar{c}_\beta b_\beta)_L,
\end{eqnarray}
and $q_L=1/2(1-\gamma_5)$ denotes a left-handed quark field. The operators
${\cal O}_{i}^\prime$ are related to the corresponding fields ${\cal O}_i$
by the relacement $\bar{d} \to \bar{s}$. The octet-octet $({\cal O}_1)$
and singlet-singlet $({\cal O}_2)$ operators emerge due to a single
gluon exchange between the weak current lines (quark fields) and follow from
the colour charge matrix $(T^{a}_{ij})$ algebra:
\begin{equation}
T_{ik}^{a} T_{jl}^{a} = -\frac{1}{2N_c} \delta_{ik} \delta_{jl} + \frac{1}{2}
 \delta_{il} \delta_{jk}~.
\end{equation}
Here, $N_c=3$ for QCD. The Wilson coefficients 
$C_i(\mu)$ are calculated at the scale $\mu =m_W$ and then scaled down 
to the scale typical for $B$ decays, $\mu =O(m_b)$, using the renormalization
group equations, which brings to the fore the influence of strong 
interactions
on the dynamics of weak non-leptonic decays. Without QCD corrections, the
two Wilson coefficients 
have the values $C_1(m_W)=0, ~C_2(m_W)=1$.
 Since the operators
${\cal O}_1$ and ${\cal O}_2$ mix under QCD renormalization, it is convenient
to introduce the operators ${\cal O}_\pm \equiv ({\cal O}_2 \pm {\cal 
O}_1)/2$ having the Wilson coefficients $C_\pm$
 which renormalize multiplicatively \cite{AM74}. The results
are now known to two-loop accuracy \cite{BW90}:
\begin{equation}\label{Cpm2loop}
C_\pm (\mu) = L_\pm(\mu) \left[1 + \frac{\as(m_W) - \as(\mu)}{4 \pi}
\frac{\gamma_\pm^{(0)}}{2 \beta_0} 
\bigg(\frac{\gamma_\pm^{(1)}}{\gamma_\pm^{(0)}}
-\frac{\beta_1}{\beta_0}\bigg) + \frac{\as(m_W)}{4 \pi} B_\pm \right],
\end{equation}
where the multiplicative factor in this expression represents the
solution of the RG equations in the leading 
order QCD \cite{AM74},
\begin{equation}\label{lpmmu}
L_\pm(\mu)=\left[\frac{\as(M_W)}{\as(\mu)} 
\right]^{d_\pm}~,
\end{equation}
and the exponents have the values $d_+=\gamma_+^{(0)}/(2\beta_0)$, 
$d_-=\gamma_-^{(0)}/(2 \beta_0)$. The
quantities $\gamma_\pm^{(i)}$ are the coefficients of the anomalous 
dimensions involving the operators ${\cal O}_\pm$ (and ${\cal 
O}_\pm^\prime$),
\begin{equation}
\gamma_\pm =\gamma_\pm^{(0)} \frac{\as}{4 \pi} + \gamma_\pm 
^{(1)}(\frac{\as}{4\pi})^2 + O(\as^3) ,
\end{equation}
 with
\begin{equation}
\gamma_+^{(0)}=4, ~~\gamma_-^{(0)}=-8, ~~\gamma_+^{(1)}=-7+\frac{4}{9}n_f,
~~\gamma_-^{(1)}=-14 -\frac{8}{9} n_f ,
\end{equation}
in the naive dimensional regularization (NDR) scheme, i.e., with 
anticommuting $\gamma_5$.
 The $\beta_i$ are the
first two coefficients of the QCD $\beta$-function, and they have the values
\begin{equation}
\beta_0=11 -\frac{2}{3} n_f, ~~~\beta_1=102-\frac{38}{3}n_f~.
\end{equation}
Finally, the functions $B_\pm$ are the matching conditions obtained by
demanding the equality of the matrix elements of the effective Lagrangian
calculated at the scale $\mu=m_W$ and in the full theory (i.e., SM)
up to terms of $O(\as(m_W^2))$. They
have the values:
\begin{equation}
B_\pm=\pm B \frac{N_c\mp 1}{2N_c},
\end{equation}
 The constant $B$ and the two-loop
anomalous dimension $\gamma_\pm^{(1)}$ are both regularization-scheme 
dependent. In the NDR scheme one has $B=11$. Following 
\cite{BW90}, we define a scheme-independent quantity $R_\pm$,
\begin{equation}
R_\pm = B_\pm + \frac{\gamma_\pm^{(0)}}{2 \beta_o}
\bigg(\frac{\gamma_\pm^{(1)}}{\gamma_\pm^{(0)}}-\frac{\beta_1}
{\beta_0}\bigg),
\end{equation}
in terms of which the Wilson coefficients read
\begin{equation}
C_\pm (\mu) = L_\pm(\mu) \left[1 + \frac{\as(m_W) - \as(\mu)}{4 \pi}
 R_\pm + \frac{\as(\mu)}{4 \pi} B_\pm \right].
\end{equation}
In this form all the scheme-dependence resides in the coefficients 
$B_\pm$ which is to be cancelled by the scheme-dependence of the matrix
elements of the corresponding operators.

  In addition to the decays $b \to c + \bar{u}d, ~b\to c + 
\bar{u}s$ and
$b \to c + \ell \nu_\ell$, which are described by the 
effective Lagrangian (\ref{Leff1}), there are other decays
involving the CC transitions $b \to u X$, $b \to (c,u) + 
\bar{c} s$ and $b \to (c,u) + \bar{c}d$,
 which are not included in this 
Lagrangian. In a systematic treatment involving QCD renormalization, 
one has to enlarge the operator basis to include these transitions and the 
so-called
penguin operators. We shall return to a discussion of this part of
the Lagrangian later in these lectures as we discuss rare $B$-decays,
where the operator basis
will be enlarged and the corresponding Wilson coefficients calculated
in the leading logarithmic approximation.

\par
   We now discuss the semileptonic branching ratio
 ${\cal B}_{\mbox{\small SL }}$ for the
$B$ mesons and to be specific will consider the case $\ell=e, \mu$. This
branching ratio is to a large extent free of the CKM matrix element
uncertainties but requires a QCD-improved calculation of the inclusive
decay rates, $\Gamma_{\mbox{SL }}$, discussed above, and 
$\Gamma_{\mbox{tot}}$,
\begin{equation}\label{bslr}
{\cal B}_{\mbox{SL }} \equiv \frac{\Gamma (B \to X e 
\nu_e)}{\Gamma_{\mbox{tot}}(B)}, \end{equation}
with 
\bea\label{gamatot}
\Gamma_{\mbox{tot}} (B) &=& \sum_{\ell =e, \mu, \tau} \Gamma (B \to X \ell 
\nu_\ell) + \Gamma(B \to X_cX)  + \Gamma(B \to 
X_{c\bar{c}}X) \nonumber\\
 &+& \Gamma (B \to X_uX) +  \Gamma (B) (\mbox{Penguins})~.
\eea
In the spirit of the parton model, we shall equate $ \Gamma(B \to X_cX) =
\Gamma (b \to c \bar{u} d) +~\Gamma (b \to c \bar{u} s)$, noting that
the so-called $W$-annihilation and $W$-exchange two-body decays are
expected to be small in inclusive $B$ decays. This will be quantified 
later as we discuss the lifetime differences among $B$ hadrons which arise
from the matrix elements of the operators representing these 
contributions.
 The corrections for the  decay widths
 $\Gamma (b \to c \bar{u} d)$ and $\Gamma (b \to c \bar{u} s)$
 are identical neglecting
$m_u$ and $m_s$, and so their contributions can be described by
similar functions. The
resulting next-to-leading order QCD corrected sum can be expressed as:
\begin{eqnarray}\label{bcud}
\Gamma(b\to c\bar ud) + \Gamma(b \to c\bar us )&=&\Gamma_0 
P(x_c,0,0)\nonumber\\
&\times&\left[2L(\mu)^2_++L(\mu)^2_-+
 \frac{\as(M_W)-\as(\mu)}{2\pi}(2L(\mu)^2_+ R_++L(\mu)^2_- R_-)\right.  
\nonumber\\
 &+&\frac{2\as(\mu)}{3\pi}\left(\frac{3}{4}(L(\mu)_+-L(\mu)_-)^2
c_{11}(x_c)+\frac{3}{4}(L(\mu)_++L(\mu)_-)^2c_{22}(x_c)\right.  
\nonumber\\
&&+\left.\left.\frac{1}{2}
(L(\mu)^2_+-L(\mu)^2_-)(c_{12}(x_c,\mu)-12 \ln\frac{\mu}{m_b})\right)\right]
\nonumber\\
&&\equiv 3\Gamma_0 \eta(\mu) J(x_c,\mu) ~,
\end{eqnarray}
with $\eta(\mu)$ representing the leading 
order QCD corrections.
 The scheme independent $R_\pm$ come from the NLO   
renormalization group evolution and are given by \cite{BW90}
\begin{eqnarray}\label{R+-}
R_{+} &=& \frac{10863 -1278n_f +80n_f^2}{6(33-2n_f)^2} , \nonumber\\
R_{-} & =& -\frac{15021 -1530 n_f + 80 n_f^2}{3(33-2n_f)^2}
\end{eqnarray}
For $n_f=5$,  $R_+=6473/3174$,
$R_-=-9371/1587$.  Note that the leading dependence of $L(\mu)_\pm$ on the
renormalization scale $\mu$ is canceled to ${\cal O}(\as)$ by the
explicit $\mu$-dependence in the $\as$-correction terms.  Virtual
gluon and Bremsstrahlung corrections to the matrix elements of four
fermion operators are contained in the mass dependent functions $c_{ij}(x)$.
 The analytic expressions for the functions $c_{11}(x), 
c_{12}(x), c_{22}(x)$ are given in \cite{Bagan94} where also their
numerical values are tabulated.
Lumping together all the perturbative and finite charm quark corrections in
a multiplicative factor $\Delta_{c}(m_b, x_c, \as(m_Z))$, the 
perturbatively corrected decay width can be expressed as:
\begin{equation}
\Gamma(b\to c\bar{u}d) + \Gamma(b\to c\bar{u}s)
=3  \Gamma_0 P(x_c,0,0)\left[1 +\Delta_{c}(x_c,
m_b, \as (m_Z)) \right].
\end{equation}
 For the central values of the parameters used here
($m_b=4.8$ GeV, $x_c=0.3$, $\mu=\mb$ and $\as(m_Z)=0.117$), the QCD 
corrections lead to an enhancement \cite{Bagan94}:
\begin{equation}
\Delta_{c}(m_b, x_c, \as (m_Z)) = 0.17 .
\end{equation}
Out of this, the bulk is contributed by the leading log factor
\begin{equation}
\eta(\mu)-1=\frac{1}{3} \left(2 L_{+}^2 + L_{-}^2 \right) -1= 0.10 ~.
\end{equation}

Next, we equate
 $\Gamma (B \to X_{c\bar{c}}) =\Gamma (b \to c \bar{c} s) + \Gamma 
(b \to c \bar{c} d)$ and 
 discuss the perturbative QCD corrections to the decay width
$\Gamma (b \to c \bar{c} s)$ and $\Gamma (b \to c \bar{c} d)$. Neglecting
$m_d$ and $m_s$, an assumption which has been found to be valid to a high
accuracy in \cite{Bagan95a}, the corrections in the two decay widths are
identical and the result can be written in close analogy with the ones
for the decay widths  $\Gamma (b \to c \bar{u} s)$ discussed above.  
\begin{eqnarray}\label{bccs}
\Gamma(b\to c\bar cs) + \Gamma(b\to c\bar cd)
&=&\Gamma_0 P(x_c,x_c,x_s) \nonumber\\
&\times&\left[2L(\mu)^2_++L(\mu)^2_-+   
 \frac{\as(M_W)-\as(\mu)}{2\pi}(2L(\mu)^2_+ R_++L(\mu)^2_- R_-)\right.
\nonumber\\
 &+&\frac{2\as(\mu)}{3\pi}\left(\frac{3}{4}(L(\mu)_+-L(\mu)_-)^2
k_{11}(x_c,\mu)+\frac{3}{4}(L_++L_-)^2k_{22}(x_c)\right.
\nonumber\\
&&+\left.\left.\frac{1}{2}
(L^2_+-L^2_-)(k_{12}(x_c)-12 \ln\frac{\mu}{m_b})\right)\right].
\end{eqnarray}
The functions $k_{ij}(m_b,x_c,\as (m_Z))$ have been calculated and 
their numerical values are tabulated in \cite{Bagan95b}. Again,
lumping together all the perturbative and finite charm quark corrections in
a multiplicative factor $\Delta_{cc}(m_b, x_c, \as(m_Z))$, the
perturbatively corrected decay width can be expressed as:
\begin{equation}\label{Deltac}
\Gamma(b\to c\bar{c}s)=3  \Gamma_0 P(x_c,x_c,x_s)\left[1 
+\Delta_{cc}(x_c, m_b, \as (m_Z)) \right].
\end{equation}
 With the values of the parameters used above, the QCD corrections lead
to the following enhancement \cite{Bagan95a,Bagan95b}:
\begin{equation}
\Delta_{cc}(m_b, x_c, \as (m_Z)) = 0.37 .
\end{equation}
This is by far the largest correction to the inclusive rates we have
discussed so far. Using pole quark masses and the renormalization scale
$\mu=m_b$, one gets \cite{Bagan95a}:
\begin{equation}
\frac{\Gamma (b \to c\bar{c}s)(NLO)}{\Gamma (b \to c \bar{c}s)(LO)}= 1.32 
\pm 0.07 ~.
 \end{equation}
The NLO corrections go in the right direction in bringing theoretical
estimates closer to the experimental value for the semileptonic branching
ratio. However, this will also lead to enhanced charmed quark multiplicity
$\langle n_c \rangle$ in $B$ decays, as discussed a little later.

 The CKM-suppressed and penguin transitions
contribute at a smaller rate to $\Gamma_{\mbox{tot}}(B)$. They are of
two kinds:
\begin{itemize}
\item $\Gamma(B \to X_u+X)$,
 which is suppressed due to the CKM matrix
element $\vert V_{ub} \vert$,
 with the rate depending on $\vert V_{ub} \vert^2$, and
\item $\Gamma(B)(\mbox{Penguin})$: The so-called penguin transitions $b 
\to s +X$, where $X=c\bar{c}$
and $X=g$ (QCD penguins), $X=\gamma$ (electromagnetic penguins),
$X=\ell^+ \ell^-, \nu\bar{\nu}$ (electroweak penguins).
\end{itemize}
 There are also transitions involving $b \to d +X$, as well as
a host of other rare decays, which can all be neglected.
The dominant contributions in the SM add up to \cite{ALI96}:
\begin{equation}
\Gamma(B \to X_u+X) + \Gamma(B)(\mbox{Penguins)} \simeq 1.25 \times 
10^{-2} \Gamma_0 ~,
\end{equation}
and hence not of much consequence for
the semileptonic branching ratio or the  $B$ hadron lifetime estimates.
\subsection{Power corrections in $\Gamma_{SL}(B)$ and $\Gamma_{NL}(B)$}
 Before we discuss the numerical results for ${\cal 
B}_{\mbox{\small SL}}$, we include the
$O(1/m_b^2)$ power corrections in the inclusive partonic decay widths.
 They constitute the first non-trivial corrections to the parton model 
results and have been calculated using the operator product expansion 
techniques \cite{Chayetal}- \cite{Bigi}.

\par
 In HQET, the $b$-quark field is represented by a 
four-velocity-dependent field, denoted by $b_v(x)$. To first 
order in $1/m_b$, the $b$-quark field in  QCD $b(x)$
 and the HQET-field
$b_v(x)$ are related through:
\begin{equation}\label{hqetb}
b(x) = e ^{-im_b v.x} \left[ 1 + i\frac{\not\!\! D}{2m_b} \right] b_v(x)
\end{equation}
The QCD Lagrangian for the $b$ quark in HQET in this order is:
\begin{equation}\label{hqetlang}
 {\cal L}^{\mbox{\small HQET}} = \bar{b}_v iv.\not\!\! D b_v + \bar{b}_v
 \frac{i(\not\!\! D)^2}{2m_b} b_v
   -Z_b \bar{b}_v\frac{gG_{\alpha\beta}\sigma^{\alpha\beta}}{4 m_b} b_v
+ O\left[ \frac{1}{m_b^2} \right],
\end{equation}  
where $Z_b$ is a renormalization factor, with $Z_b(\mu=m_b)=1$
and $\not\!\! D = D_\mu \gamma^{\mu}$, with $D_\mu$ being the covariant
derivative. The 
operator
$\bar{b}_v(i\not\!\! D)^2b_v/2m_b$ is not renormalized due to the 
symmetries of
HQET. (In technical jargon, this is termed as a consequence of the
reparametrization invariance of ${\cal L}^{\mbox{\small HQET}}$.)
 With this Lagrangian, it has been shown in \cite{Chayetal} -
\cite{Bigietal} that in
the heavy quark expansion in order $(1/m_b^2)$, the
hadronic corrections can be expressed in terms of two matrix elements
\begin{equation}\label{lambda1}
\langle B^{(*)} \vert \bar{b}_v (iD)^2 b_v \vert B^{(*)}\rangle = 2 
m_{B^{(*)}} \lambda_1 , \nonumber 
\end{equation}
\begin{equation}\label{lambda2} 
\langle B^{(*)} \vert \bar{b}_v \frac{g}{2}\sigma_{\mu \nu} F^{\mu \nu} b_v 
\vert B^{(*)}\rangle = 2 d_{B^{(*)}} m_{B^{(*)}} \lambda_2 ,
\end{equation}
where $F^{\mu \nu}$ is the gluonic field strength tensor,
and the constants $d_{B^{(*)}}$ have the value 3 and $-1$ for $B$ and $B^*$,
respectively. The
constant $\lambda_2$ can be related to the hyperfine splitting in the $B$ 
mesons, which gives:
\begin{equation}
\lambda_2 \simeq \frac{1}{4} (m_{B^*}^2 - m_B^2) = 0.12 ~\mbox{GeV}^2.
\end{equation}
The other quantity $\lambda_1$ is  the average kinetic energy of the
$b$ quark inside a $B$ meson and has been estimated in various ways, using
 the QCD sum rule 
approach \cite{kinetic}, the virial theorem \cite{Neubert96},
lattice QCD \cite{Martinelli96} and data \cite{GKLW96}. A range
\be
 \lambda_1= -(0.5 \pm 0.3) ~\mbox{GeV}^2
\ee
is compatible with most estimates. (For a recent compilation
of $\lambda_1$ estimates, see \cite{Martinelli96}.) 
 Taking into account these corrections, the semileptonic
and non-leptonic decay rates of a $B$ meson $B \to X\ell \nu_\ell$ and
$B \to X_cX$ can be written as \cite{Bigietal,MW94}:
\begin{eqnarray}\label{pcsld}
\Gamma(B\longrightarrow X_c\ell\nu_{\ell}) &=&
\Gamma^{(0)} f(r_{c})\bigg[\left(1-\frac{2}{3}\frac{\alpha_{s} (m_{b}^{2})}
{\pi} g(r_{c})\right) \left( 1 + \frac{\lambda_1}{2 m_b^2}
+ \frac{3 \lambda_2}{2m_b^2} - 
\frac{6(1-r_c)^4}{f(r_c)}\frac{\lambda_2}{m_b^2}\right) \nonumber\\
& & \mbox{}
+ O(\as^2,\frac{\as}{m_b^2},\frac{1}{m_b^3})\bigg],
\end{eqnarray}
and
\begin{eqnarray}\label{pcnld}
\Gamma(B\longrightarrow X_cX) &=&
3 \Gamma^{(0)} \bigg[\eta(\mu)J(\mu)
 \left( 1 + \frac{\lambda_1}{2 m_b^2}
+ \frac{3 \lambda_2}{2m_b^2} - 
\frac{6(1-r_c)^4}{f(r_c)}\frac{\lambda_2}{m_b^2}\right) \nonumber\\
& & \mbox{}
-\left(L_+(\mu)^2 -L_-(\mu)^2\right) \frac{4(1-r_c)^3}{f(r_c)} 
\frac{\lambda_2}{m_b^2} +
 O(\as^2,\frac{\as}{m_b^2},\frac{1}{m_b^3})\bigg],
\end{eqnarray}
where the product $\eta(\mu)J(\mu)$ denotes the NLO corrected result for
the partonic decay discussed above in (\ref{bcud}), to which Eq. 
(\ref{pcnld}) reduces in the limit $\lambda_1=\lambda_2=0$.

\par
The decay rates depend on the quark masses, which unlike lepton masses,
do not appear as poles in the $S$-matrix nor do the quarks  
exist as asymptotic states. They are parameters of an interacting theory 
and hence subject to renormalization effects. Consequently, they require 
a regularization scheme, such as the $\overline{MS}$ scheme, and a scale,
where they are normalized, to become well-defined quantities.
 For example, the quark masses
in the so-called $\overline{MS}$ scheme and the pole masses (OS scheme) are 
related in the leading order \cite{mtmsbar},
\begin{equation}\label{polmsbar}
\overline{m}_Q(m_Q)=m_Q\left[1- 4 \frac{\as (m_Q)}{(3\pi)} +...\right] ~.
\end{equation}

 In HQET, quark masses can be expressed in terms of the heavy meson 
masses $m_M$ and the parameters $\lambda_1, ~\lambda_2$ and a quantity 
called $\bar{\Lambda}$, where
\begin{equation}\label{biglambda}
m_M=m_Q + \bar{\Lambda} - \frac{\lambda_1 + d_M \lambda_2}{2m_Q}+...
\end{equation}
This yields
\begin{equation}
m_b-m_c=m_B-m_D + \frac{\lambda_1 + 3 \lambda_2}{2} (\frac{1}{m_b} - 
\frac{1}{m_c}) + O(\frac{1}{m^2}) ,
\end{equation}
and the quark mass differences can then be calculated knowing $\lambda_1$ 
and $\lambda_2$, giving $(m_b - m_c)= (3.4 \pm 0.03 \pm 0.03)$ GeV 
\cite{Neubert95}.
 This difference, which determines the inclusive
rates and shape of the lepton energy spectrum in semileptonic decays,  
has also been determined from an analysis of the experimental lepton energy 
spectrum in $B$ decays, yielding  $(m_b-m_c)=3.39\pm 0.01$ GeV
for the pole masses \cite{GKLW96}, in excellent agreement with the QCD 
sum rule based estimates.

\subsection{Numerical estimates of ${\cal B}_{SL}(B)$ and $\langle n_c 
\rangle$}
\par
\begin{figure}[htb]
\vspace{0.10in}
\centerline{
\epsfig{file=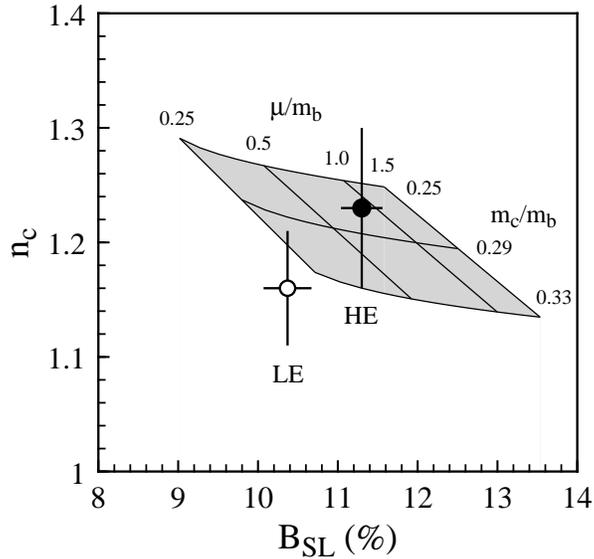,height=3in,angle=0}}
\vspace{0.08in}
\caption[]{The charm content $n_c$ vs. $B_{\small SL}$ in $B$ decays,
shown as a function of the quark mass ratio $m_c/m_b$ and the
renormalization scale $\mu/m_b$ in the OS scheme. The data show the 
experimental
averages from the $\Upsilon(4S)$ (LE) and $Z^0$ (HE) measurements. (Figure
taken from \protect\cite{NS96}).}
\label{ncslBB}
\end{figure}

The theoretical framework described in the previous section
can now be used to predict two important quantities
in $B$ decays ${\cal B}_{SL}(B)$ and
$\langle n_c \rangle $, which have been measured. Concerning 
${\cal B}_{SL}(B)$, there is some discrepancy between the two set of 
experiments
performed at the $\Upsilon(4S)$ and at the $Z^0$ resonance, although it
must be stressed that these experiments measure a different mixture
of $B$ hadrons. The 
present measurements give:
\begin{eqnarray}\label{slbr95}
{\cal B}_{SL}(B) &=& (10.37 \pm \pm 0.30)\% ~~~\mbox{at} ~\Upsilon(4S) 
~\cite{Tomasz95}, \nonumber\\
{\cal B}_{SL}(B) &=& (11.11 \pm 0.23)\% ~~~~~~\mbox{at} ~Z^0 
~\cite{Perret95}, \nonumber\\
\langle n_c \rangle &=& 1.16 \pm 0.05 ~~~\mbox{at} ~\Upsilon(4S) 
~\cite{Tomasz95}\, \nonumber\\
\langle n_c \rangle &=& 1.23 \pm 0.07 ~~~\mbox{at} ~Z^0 ~\cite{Calderini96},
\end{eqnarray}
where the number for $\langle n_c \rangle$ at the $Z^0$ is from the 
ALEPH collaboration. We use the following average  in which the error 
on ${\cal B}_{SL}(B)$ is inflated to bridge the gap in
the experimental measurements \cite{NS96}:
\bea
{\cal B}_{SL}(B) &=& (10.90 \pm \pm 0.46)\% ~\, \nonumber\\
\langle n_c \rangle &=& 1.18 \pm 0.04 ~.
\eea 
   The theoretical predictions for these quantities 
have been updated by Bagan et al. \cite{Bagan95a}, and more recently by
Neubert and Sachrajda \cite{NS96}, using the same theoretical input.
We shall use here the numerical results from \cite{NS96}
where the following ranges of parameters have been used:
\begin{equation}\label{BBparameters1}
m_b(\mbox{pole})=4.8 \pm 0.2 ~\mbox{GeV}; ~~~\as (m_Z)=0.117 \pm 0.004,
~~~m_b/2 < \mu < 2 m_b,\nonumber\\
\end{equation}
and $0.25 \leq m_c/m_b \leq 0.33$. Here $m_b$ is the pole mass
defined to one-loop order in perturbation theory.
 At order $1/m_b^2$ in
the heavy quark expansion, non-perturbative effects are described by
the parameter $\lambda_2$, as the dependence on the parameter $\lambda_1$
cancels out in calculating ${\cal B}_{SL}(B)$ and $\langle n_c \rangle$. 
This analysis leads to the following values using the pole 
masses (OS scheme)\cite{NS96}: \bea
\label{bslrpole}
{\cal B}_{SL} &=& (12.0 \pm 1.0)\% ~~(\mbox{for} ~\mu =m_b) ,
 ~~(10.9 \pm 1.0)\% ~~(\mbox{for} ~\mu =m_b/2)~, \nonumber\\
\langle n_c \rangle &=& 1.20 \mp 0.06 ~~~~~(\mbox{for} ~\mu =m_b) ,
~~~1.21 \mp 0.06 ~~~~(\mbox{for} ~\mu =m_b/2) . 
\eea
 One could also use,
following Bagan et al. \cite{Bagan95a}, the $\overline{\mbox{MS}}$ scheme
and the results in this scheme are as follows \cite{NS96}
\bea
\label{bslrmsbar}
{\cal B}_{SL}(\overline{\mbox{MS}}) &=& (10.9 \pm 0.9)\% ~~(\mbox{for} ~\mu 
=m_b) ,
 ~~(10.3 \pm 0.9)\% ~~(\mbox{for} ~\mu =m_b/2)~, \nonumber\\
\langle n_c \rangle &=& 1.25 \mp 0.05 ~~~~~(\mbox{for} ~\mu =m_b) ,
~~~1.24 \mp 0.06 ~~~~(\mbox{for} ~\mu =m_b/2) ,
\eea
 The numbers in the $\overline{\mbox{MS}}$ scheme correspond to
using the two-loop anomalous dimension matrix in the running of the quark
masses and the errors from various sources have been added in quadrature.
The estimates (\ref{bslrpole}) and (\ref{bslrmsbar}) show that both
$\langle n_c\rangle $ and ${\cal B}_{SL}$ are scheme-dependent; in addition
${\cal B}_{SL}$ also depends on the scale $\mu$.
A comparison of the theoretical estimates in the OS scheme (eqs. 
(\ref{bslrpole})) and data on
 $\langle n_c\rangle $ and ${\cal B}_{SL}$ is shown in Fig. ~\ref{ncslBB}.
 Given the parametric dependence on the scale 
$\mu$ and the ratio $m_c/m_b$, the agreement between theory and experiment
is reasonably good. In the $\overline{\mbox{MS}}$ scheme, the semileptonic
branching ratio is generally smaller and  $\langle n_c \rangle $
somewhat higher (the two are anti-correlated).
%
%
To make more precise predictions, one has to calculate the missing
$O(\alpha_s^2)$ corrections, which, as the experience has it, will
considerably reduce the scale dependence. However, equally important is
to reduce the present theoretical uncertainty in the ratio $m_c/m_b$.
Here, precise data on the lepton and hadron energy distribution
in semileptonic $B$ decays will
help. So, while there is certainly much room for improvement,
it is fair to conclude that
within existing uncertainties the current theoretical estimates
for ${\cal B}_{SL}$ and $\langle n_c \rangle$ in the SM do not disagree
significantly from the corresponding experimental values.
%
\subsection{ $B$-Hadron Lifetimes in the Standard Model}

\par
   A matter closely related to the semileptonic branching ratios is that 
of the
individual $B$ hadron lifetimes. The QCD-improved spectator model gives
almost equal lifetimes. Power corrections will split the
$B$-baryon lifetime from those of $B_d^0, B^{\pm}$ and $B_s$. However,
first estimates of these differences 
are at the few per cent level \cite{Bigietal}.
 The experimental situation has been 
summarized as of  summer 1996 in \cite{Richman96}:
\begin{equation}
\frac{\tau(B^-)}{\tau(B_d)}=1.04 \pm 0.04;
~~\frac{\tau(B_s)}{\tau(B_d)}=0.98 \pm 0.05; 
~~\frac{\tau(\Lambda_b)}{\tau(B_d)}=0.78 \pm 0.04 ~.
\end{equation}

    The subject of exclusive $B$ hadron lifetimes has received renewed 
theoretical attention lately
 \cite{NS96,Uraltsev96,Rosner96}, in which the possibly enhanced roles of 
the four-Fermion operators involving baryonic states (as compared to the
mesonic state) has 
been studied. We recall that such operators enter at 
$O(1/m_b^3)$ in the heavy quark expansion discussed above \cite{Bigietal}. 
In this order, there are
four such operators, which using the notation of \cite{NS96}, can be
expressed as:
\begin{eqnarray}\label{4quarkNS}
{\cal O}^q_{V-A} &=& (\bar{b}_L\gamma_\mu q_L)(\bar{q}_L\gamma^{\mu}b_L),
\nonumber\\
{\cal O}^q_{S-P} &=& (\bar{b}_R q_L)(\bar{q}_L b_R),
\nonumber\\
{\cal T}^q_{V-A} &=& (\bar{b}_L\gamma_\mu t_a 
q_L)(\bar{q}_L\gamma^{\mu}t_a b_L), \nonumber\\
{\cal T}^q_{S-P} &=& (\bar{b}_Rt_a q_L)(\bar{q}_Lt_a b_R),
\end{eqnarray}
where $t_a$ are generators of colour $SU(3)$. The matrix elements of these
operators between various $B$-meson and $\Lambda_b$-baryons
are in general different and this contribution will thus split the decay 
widths
of the various $B$ hadrons. In general, the operators
(\ref{4quarkNS}) introduce eight new 
parameters corresponding to the matrix elements of these operators.
In the large-$N_c$ limit, however, it has been
argued in \cite{NS96} that the $B$-mesonic matrix elements of the operators
$\langle B_q \vert {\cal O}^q_{V-A} \vert B_q \rangle$ and
$\langle B_q \vert {\cal O}^q_{S-P} \vert B_q \rangle$ are the dominant 
ones. While accurate numerical estimates require a precise knowledge of
these matrix elements, one expects that they give rise 
typically 
to  the spectator-type effects (using the parton model language):
\begin{equation}
\frac{\Gamma_{\mbox{spec}}}{\Gamma_{\mbox{tot}}} \simeq (\frac{2 \pi f_B}
{m_B})^2 \simeq 5\% ~,
\end{equation}
with $f_B$ of order 200 MeV.
In the case of $\Lambda_b$ baryons, one can use the heavy quark spin symmetry
to derive two relations involving the operators
given above taken between the $\Lambda_b$ 
states. The problem is then reduced to the estimate of two matrix elements
which in \cite{NS96} are taken to be the following:
\begin{equation}
 \frac{1}{2m_{\Lambda_b}}\langle \Lambda_b \vert {\cal O}^q_{V-A} \vert 
\Lambda_b \rangle \equiv -\frac{f_B^2 m_B}{48} r(\frac{\Lambda_b}{B_q}) ,
\end{equation}
and
\begin{equation}
\langle  \Lambda_b \vert \tilde{ {\cal O}}^q_{V-A} 
\vert \Lambda_b \rangle = -\tilde{B}
\langle \Lambda_b \vert  {\cal O}^q_{V-A}
\vert \Lambda_b \rangle ~,  
\end{equation}
The operator $\tilde{\cal O}_{V-A}$ is a linear combination of the
operators ${\cal T}_{V-A}$ and ${\cal O}_{V-A}$ introduced earlier,
$\tilde{{\cal O}}_{V-A}= 2{\cal T}_{V-A} + 3 {\cal O}_{V-A}$, following from
colour matrix algebra \cite{NS96}, and
$r(\Lambda_b/B_q)$ is the ratio of the squares of the wave functions  
which can be expressed in terms of the probability of finding a light quark
at the location of a $b$ quark inside $\Lambda_b$ baryon and the $B$ meson,
i.e.
\begin{equation}
r(\frac{\Lambda_b}{B_q})  = \frac{\vert \Psi^{\Lambda_b}_{bq} \vert^2}
        {\vert \Psi^{B_q}_{b\bar{q}} \vert^2} ~.
\end{equation}
 One expects $\tilde{B} =1$ in the valence-quark approximation.
However, the ratio $r(\Lambda_b/B_q)$ has a large uncertainty on it, ranging 
from $r(\Lambda_b/B_q) \simeq 0.5$ in the non-relativistic quark model 
\cite{Guberinaetal79}
to $r(\Lambda_b/B_q) =1.8 \pm 0.5$ if one uses the ratio of the spin 
splittings 
between $\Sigma_b$ and $\Sigma_b^*$ baryons and $B$ and $B^*$ mesons, as
advocated by Rosner \cite{Rosner96} and using the preliminary data 
from DELPHI, $m(\Sigma_b^*) - m(\Sigma_b) = (56 \pm 16)$ MeV 
\cite{DELPHISIGMA}.

Using the ball-park estimates that $\tilde{B} $ and
$r(\Lambda_b/B_q) $ are both of order unity yields
for the lifetime ratio  $\tau(\Lambda_b)/\tau(B_d) > 0.9$ \cite{NS96},
significantly larger than the present world average.
 Reliable estimates of these constants can be got, in principle, using 
lattice-QCD and QCD sum rules. Very recently, QCD sum rules have been used
to estimate
$\langle  \Lambda_b \vert \tilde{ {\cal O}}^q_{V-A}
\vert \Lambda_b \rangle$ and $\tilde{B}$, yielding 
$\langle  \Lambda_b \vert \tilde{ {\cal O}}^q_{V-A}
\vert \Lambda_b \rangle= (0.4 - 1.2) \times 10^{-3} ~\mbox{GeV}^3$
and $\tilde{B}=1.0$ \cite{CD96}. This corresponds to the parameter
$r(\Lambda_b/B_q)$ having a value in the range
$r(\Lambda_b/B_q) \simeq 0.1 - 0.3$, much too small to explain the
observed lifetime difference. We  mention here the possibility
of linear power corrections in the inclusive decay rates, 
which are not encountered in the explicit power corrections discussed 
above but may enter via the breakdown of the parton-hadron duality.
Phenomenological parametrizations presented in \cite{AMPR96}
in support of such a scenario are interesting but not persuasive.
 One must conclude that the lifetime ratio
$\tau(\Lambda_b)/\tau(B_d)$ remains a puzzle. New and improved measurements
are needed, which we trust will be forthcoming from HERA-B and the Tevatron
experiments in not-too-distant a future.

   Before leaving this section, we mention that from a theoretical point
of view one expects measurable lifetime differences between the two mass
eigenstates of the $\BS$-$\BSB$ complex \cite{Bigietal}. Recently, leading 
order 
perturbative $(O(\Lambda_{\mbox{\small QCD}}/m_b))$ and power corrections
$(O(m_s/m_b))$ to the differences in the decay rates $\Gamma_1^{s}$ and 
$\Gamma_2^{s}$ of the two mass eigenstates have been analyzed in 
\cite{BBD96}. The perturbative
corrections go along very much the same lines as discussed earlier. Power
corrections bring in the four-quark operators already mentioned.
Quantitative estimate of $\Delta \Gamma$ requires the knowledge of 
non-perturbative quantities, 
bag factors called $B$ and $B_S$, involving the expectation values of the 
operators $\langle {\cal O}^{s}_{V-A} \rangle$ and $\langle {\cal 
O}^{s}_{S-P} \rangle$, and the pseudoscalar meson coupling constant 
$f_{B_s}$. The resulting expression for the ratio $(\Delta 
\Gamma/\Gamma)_{B_s}$ can be expressed as \cite{BBD96}:
 \be
\left(\frac{\Delta \Gamma}{\Gamma}\right)_{B_s} = \left[aB + bB_S + c\right] 
\left(\frac{f_{B_s}} {210 ~\mbox{MeV}}\right)^2 ~,
\ee
where the constants $B$ and $B_S$ are the mentioned bag factors and the
coefficients $a,b$ and $c$ depend on the parameters such as $m_b$ and  
$\mu$; $c$ incorporates the explicit $1/m_b$ corrections. 
For the choice $B=B_S=1$ (corresponding to the vacuum  
insertion approximation), $m_b= 4.8$ GeV, $\mu=m_b$ and $f_{B_s}=210$
MeV, one gets $a=0.009, ~b=0.211, ~c=-0.065$, yielding \cite{BBD96}:
\be
\left(\frac{\Delta \Gamma}{\Gamma}\right)_{B_s} = 0.155 ~.
\ee 
This difference is large enough to be measured in the forthcoming
experiments. If accurately measured, 
$\Delta \Gamma_{B_s}$ has the potential of providing an alternative estimate
of the mass difference in the $\BS$ - $\BSB$ complex, $\Delta M_{B_s}$, as 
the ratio $\Delta \Gamma_{B_s}/\Delta M_{B_s}$, as opposed to the mass
difference itself, does not depend on $f_{B_s}$. However, there is still
some dependence in this ratio on the unknown bag constants
and the coefficients. The present 
determination  of this ratio is: $\Delta \Gamma_{B_s}/\Delta 
M_{B_s}=(5.6 \pm 2.6) \times 10^{-3}$ \cite{BBD96}, which is in need of 
substantial improvement if it has to make any impact on the CKM 
phenomenology. We 
shall return to the estimates of $\delms$ and a discussion of the related 
issues later.   
\subsection{Determination of $\Vcbabs$ and $\Vubabs$}
\par
 The CKM matrix element $V_{cb}$ can
be obtained from semileptonic decays of $B$ mesons. We shall restrict
ourselves to the methods based on HQET \cite{HQET,IW} to calculate the 
exclusive 
semileptonic decay rates and use the heavy quark expansion to estimate the
inclusive rates. Concerning exclusive decays, we recall that
 in the heavy quark limit $(\mb \to \infty)$, it has been
observed that all hadronic form factors in the semileptonic decays $B \to
(D,D^*) \ell \nu_\ell$ can be expressed in terms of a single function, the
Isgur-Wise function \cite{IW}. It has been shown that the HQET-based
method works best for $B\to D^*l\nu$ decays, since these are unaffected by
$1/m_Q$ corrections \cite{Luke,Boyd,Neubert}.
Using HQET, the differential decay rate in $B \to D^* \ell \nu_\ell$ is
\begin{eqnarray}
\frac{d\Gamma (B \to D^* \ell \bar{\nu})}{d\omega }
&=& \frac{G_F^2}{48 \pi^3} (m_B-m_{D^*})^2 m_{D^*}^3 \eta_{A}^2
 \sqrt{\omega^2-1} (\omega + 1)^2 \\ \nonumber
&~& ~~~~~~~~~~~~~~\times [ 1+ \frac{4 \omega}{\omega + 1}
 \frac{1-2\omega r + r^2}{(1-r)^2}] \Vcbabs ^2 \xi^2(\omega) ~,
\label{bdstara1}
\end{eqnarray}
where $r=m_{D^*}/m_B$, $\omega=v\cdot v'$ ($v$ and $v'$ are the
four-velocities of the $B$ and $D^*$ meson, respectively), and $\eta_{A}$
is the short-distance correction to the axial vector form factor. In the
leading logarithmic approximation, this was calculated by Shifman and
Voloshin some time ago -- the so-called hybrid anomalous dimension
\cite{hybrid}. In the absence of any power corrections, $\xi (\omega=1)=1$.
Estimating
the size of the $O(1/\mb^2)$ and $O(1/m_c^2)$ corrections to the Isgur-Wise
function, $\xi (\omega)$ has generated some lively theoretical debate
\cite{Vainshtein95,Neubert95} but it seems that a convergence
has now emerged on their magnitude. We take \cite{Neubert95}:
\be
\label{neubertxiold}
\xi (1) = 1+ \delta (1/m^2)= 0.945 \pm 0.025 ~.
\ee
The quantity $\eta_{A}$, and its counterpart for the vector
current matrix element renormalization, $\eta_{V}$, have now been calculated
in the complete next-to-leading order by Czarnecki \cite{Cz96}, getting
\begin{eqnarray}
\label{Czetas}
\eta_{A} &=& 0.960 \pm 0.007~, \nonumber \\
\eta_{V} &=& 1.022 \pm 0.004 ~.
\end{eqnarray}
 The error on ${\cal F}(1)$ is now 
dominated by the power corrections in $\xi (1)$, yielding 
\cite{Cz96}:     
\be
{\cal F}(1)=\xi \cdot \eta_{A}=0.907 \pm 0.026~.
\label{alxi}
\ee
 Since the rate is zero at
the kinematic point $\omega=1$, one uses data for $\omega >1$ and an
extrapolation procedure to determine $\xi (1) 
\Vcbabs$. As the range of accessible energies in the decay
$B \to D^* \ell \bar{\nu}$ is rather small $(1 ~< \omega < ~1.5)$, the
extrapolation to the symmetry point can be done using a Taylor expansion
around $\omega =1$,
\be
{\cal F}(\omega) = {\cal F}(1) \left[1- \hat{\rho}^2(\omega -1) + \hat{c}
          (\omega -1)^2+...\right].
\ee
The present experimental input from the exclusive
semileptonic channels is based on the data from
 CLEO, ARGUS, ALEPH, DELPHI and OPAL,
which is summarized by Gibbons  at the Warsaw conference \cite{Gibbons96}. 
He obtains
 \be
  \vert V_{cb}\vert \cdot {\cal F}(1) = 0.0357 \pm 0.0020 \pm 0.0014 ~.
\ee
Using ${\cal F}(1)$ from Eq.~(\ref{alxi}), gives the following
value:
 \be
  \vert V_{cb} \vert= 0.0393 \pm 0.0021 ~(\rm{expt}) \pm 0.0015 (\rm{curv})
 \pm 0.0017 ~(\rm{th}),
\label{Vcbhqet95}
\ee
where the error from the curvature of the Isgur-Wise function has also
been indicated.
Combining the errors quadratically gives
\be
\vert V_{cb} \vert = 0.0393 \pm 0.0028~.
\label{vcbhqet}
\ee
\par
 A value of $\vert V_{cb}
\vert$ has also been obtained from the inclusive semileptonic $B$
 decays using heavy quark expansion. The inclusive analysis has the 
advantage of having very small statistical error.
However, as discussed previously,  there is
about $2\sigma$ discrepancy between the semileptonic branching ratios
at the $\Upsilon(4S)$ and in $Z^0$ decays.   
Using an averaged value for the semileptonic decay width from these two 
sets of measurements and inflating the error as before 
to take into account the disagreement leads to a value \cite{Tomasz95}:
\be
 \vert V_{cb} \vert = 0.0398 \pm 0.0008 ~\mbox{ (expt)} \pm 0.004 
~\mbox{ (th)} ~.
 \ee
where the theoretical error estimate ($\pm 10 \%)$ has been taken from  
Neubert \cite{Neubert95}.
For further discussion of these matters we refer to
\cite{Vainshtein95,Neubert95}. The agreement in the values
of $\Vcbabs$ obtained from the exclusive and inclusive semileptonic 
decays is remarkably good. This can be taken as a quantitative check of
the notion of quark-hadron duality in semileptonic decays.
 We shall use  the following values
for $\Vcbabs$ and the Wolfenstein parameter $A$ in the CKM fits discussed
later:
\be
 \vert V_{cb} \vert = 0.0393 \pm 0.0028 \Longrightarrow ~A = 0.81 \pm 0.058~.
\label{Avalueal}
\ee

\par
 The knowledge of the CKM matrix
element ratio $|V_{ub}/V_{cb}|$ is based on the analysis of the end-point
lepton energy spectrum in semileptonic decays $B \to X_{u} \ell 
\nu_\ell$  
and the measurement of the exclusive semileptonic decays $B \to (\pi, 
\rho)
\ell \nu_\ell$ reported by the CLEO collaboration \cite{Tomasz95}. As 
noted
in \cite{Bartelt93}, the inclusive measurements suffer from a large
extrapolation factor from the measured end-point rate to the total
branching ratio, which is model dependent. The exclusive measurements 
allow
a discrimination among a number of models \cite{Gibbons96}, all of which
were previously allowed from the inclusive decay analysis alone. It is
difficult to combine the exclusive and inclusive measurements to get a
combined determination of $\Vubabs/\Vcbabs$. However, it has been noted
that the disfavoured models in the context of the exclusive decays are 
also
those which introduce a larger theoretical dispersion in the 
interpretation
of the inclusive $B \to X_u \ell \nu_\ell$ data. Excluding them from
further consideration, measurements in both the inclusive and exclusive
modes are compatible with \cite{Gibbons96}:
\beq
\left\vert \frac{V_{ub}}{V_{cb}} \right\vert = 0.08\pm 0.016~.
\label{vubvcbn}
\eeq
This gives
\beq
\sqrt{\rho^2 + \eta^2} = 0.363 \pm 0.073~.
\eeq

\par
   We summarize this section by observing that the bulk properties of
$B$ decays are largely accounted for in the standard model. On the
theoretical front, parton model estimates of the earlier epoch have been
replaced by theoretically better founded calculations with controlled
errors.
 In particular, methods based on  HQET and heavy
quark expansion have led to a quantitative determination of $\Vcbabs$ at
$\pm 7 \%$ accuracy, which makes it after $\Vudabs$ and $\Vusabs$, the
third best measured CKM matrix element. The 
matrix element $\Vubabs$ has still 
large uncertainties ($\pm 20 \%)$ and there is every need to reduce this,
as this error is one of the two handicaps at present in testing the 
unitarity of the CKM matrix precisely.
The quantities ${\cal B}_{SL}$, $\langle n_c \rangle$, and the individual
$B$-hadron lifetimes are now in reasonable agreement with data.
A completely quantitative comparison requires the missing NLL corrections and
in the case of lifetime differences better evaluations of the matrix
elements of four-quark operators, which we hope will be forthcoming.
Finally, we stress that it will be very helpful to measure the semileptonic
branching ratios $B_{SL}$ for the $\Lambda_b$ baryons. With the lifetimes
of the $B$ hadrons now well measured, such a measurement would allow to
compare $\Gamma_{SL}(B_d), \Gamma_{SL} (B^\pm)$ and 
$\Gamma_{SL}(\Lambda_b)$, to check the pattern of power corrections 
in semileptonic decays.
%
%
%
 \section{Electromagnetic Penguins and Rare $B$ Decays in the Standard Model}

\par
We now discuss the FCNC transitions SM which in the SM
are induced through the exchange of $W^\pm$ bosons
in loop diagrams. We shall discuss  representative examples
from several such transitions involving $B$ decays,
 starting with the decay $\BGAMAXS$, which has been measured by CLEO
 \cite{CLEOrare2}. This was preceded by the measurement of the exclusive 
decay $\BGAMAKSTAR$ \cite{CLEOrare1}:
\begin{eqnarray}
{\cal B}(\BGAMAXS) &=& (2.32\pm 0.57\pm 0.35)\times 10^{-4} ~,\\
{\cal B}(\BGAMAKSTAR) &=& (4.2\pm 0.8\pm 0.6)\times 10^{-5}~,
\end{eqnarray}
yielding an exclusive-to-inclusive ratio:
\begin{equation}
R_{K^*} = \frac{\Gamma(\BGAMAKSTAR)}{\Gamma(\BGAMAXS)}=(18.1\pm 6.8)\% ~.
\end{equation}
These decay rates test the SM and the models for decay form factors
and we shall study them quantitatively.

 The leading contribution to $b \to s +\gamma$ arises
at one-loop from the so-called penguin diagrams and
the matrix element in the lowest order can be written as:
\begin{equation} \label{e1}
  {\cal M} (b \to s ~+\gamma)
    = \frac{G_F}{\sqrt{2}} \,\frac{e}{2 \pi^2}
 \sum_{i} V_{ib} V_{is}^*
      F_2 (x_i)\, q^\mu \epsilon^\nu \bar{s} \sigma_{\mu \nu}
      (m_bR ~+ ~m_sL)b ~,
 \end{equation}
where $x_i= ~m_i^2/m_W^2$,
$q_\mu$  and $\epsilon_\mu$ are, respectively, the photon four-momentum
and polarization vector,
the sum is over the quarks, $u, ~c$, and $t$, and $V_{ij}$ are the
CKM matrix elements.
The (modified) Inami-Lim function $F_2(x_i)$ derived from the (1-loop) 
penguin diagrams is given by \cite{InamiLim}:
\begin{equation}
F_{2}(x) = \frac{x}{24 (x-1)^{4}} \ \left[6 x (3 x -2 )
\log x - (x-1) (8 x^{2} +5 x -7 ) \right], \nonumber \\
\end{equation}
where in writing the expression for $F_2(x_i)$ above we have
left out a constant from the function derived by Inami and Lim, since on 
using the unitarity constraint these sum to zero.
It is instructive to write the unitarity constraint for the decays
$\BGAMAXS$ in full:
\begin{equation} \label{e4}
 V_{tb} V_{ts}^* + V_{cb}V_{cs}^* + V_{ub}V_{us}^* =0 ~.
 \end{equation}
Now, since the last term in this sum is completely negligible compared to 
the others
(by direct experimental measurements), one could set it to zero enabling
us to express the one-loop electromagnetic penguin amplitude as follows:
\begin{equation} \label{e2}
 {\cal M }(b \to s ~+\gamma)
    = \frac{G_F}{\sqrt{2}}\,\frac{e}{2 \pi^2} \,\lambda_{t}
   \,(F_2 (x_t)-F_2(x_c))\, q^\mu \epsilon^\nu \bar{s} \sigma_{\mu \nu}
      (m_bR ~+ ~m_sL)b ~.
 \end{equation}
The GIM mechanism \cite{GIM} is manifest in this amplitude and the
CKM-matrix element dependence is factorized in $\lambda_t\equiv V_{tb} 
V_{ts}^*$. The measurement of the branching ratio for $\BGAMAXS$ can then be 
readily interpreted in terms of the CKM-matrix element product
$\lambda_t/\Vcbabs$ or equivalently $\Vtsabs/\Vcbabs$. In the approximation
we are using (i.e., setting $\lambda_u=0$), this is equivalent to
measuring $\Vcsabs$.
For a quantitative determination
of $\Vtsabs/\Vcbabs$, however,  QCD radiative
corrections have to be computed and
the contribution  of the so-called long-distance effects estimated.
 We proceed to
discuss them below.
 \subsection{The effective Hamiltonian for $B \to X_s \gamma$}
\par
  The appropriate framework to incorporate
QCD corrections is that of an effective theory obtained by integrating 
out the
heavy degrees of freedom, which in the present context are the top quark 
and $W^\pm$ bosons. This effective theory
is an expansion in $1/m_W^2$ and  involves a tower of
increasing higher dimensional operators
built from the  quark fields $(u,d,s,c,b)$, photon, gluons and leptons. The 
presence
of the top quark and of the $W^\pm$ bosons is reflected through the
effective coefficients of these operators which become functions of
their masses. The operator basis depends on the underlying theory and 
in these lectures we shall concentrate on the standard model. The
basis that we shall use is restricted to dimension-6 operators and 
the operators which vanish on using the
equations of motion are not included.
The effective Hamiltonian ${\cal H}_{eff}$ given below
 covers not only the decay $b \to s + \gamma$, in which we are
principally interested in this section, but also other processes
such as $b \to s + g$ and $b \to s + q \bar{q}$.
				 
 It is to be expected in general that due to
 QCD corrections, which induce
 operator-mixing, 
 additional contributions with  different CKM pre-factors have to be
included in the amplitudes.
Thus, QCD effects alter the CKM-matrix element dependence of the
decay rates for both $\BGAMAXS$ and (more importantly) $\BGAMAXD$. 
 However, with the help of the unitarity condition given above,
the CKM matrix
dependence in the effective Hamiltonian incorporating the QCD
corrections for the decays $\BGAMAXS$ factorizes, and
one can write this Hamiltonian as \footnote{Note
that in addition to the penguins with the $u$-quark intermediate state
there are also non-factorizing contributions due to the
operators $ (\bar{u}_{L \alpha} \go{\mu} b_{L \alpha})
(\bar{s}_{L \beta} \gu{\mu} u_{L \beta})$, which like the $u$-quark
contribution to the 1-loop electromagnetic penguins are proportional
to the CKM-factor $\lambda_u \equiv V_{us} V_{ub}^*$,
                              and hence are consistently set to zero.}:
\begin{equation}\label{heffbsg}
{\cal H}_{eff}(b \to s +\gamma) = - \frac{4 G_F}{\sqrt{2}} V_{ts}^* V_{tb}
        \sum_{i=1}^{8} C_i (\mu) {\cal O}_i (\mu) ,
\end{equation}
where the operator basis is chosen to be (here $\mu$ and $\nu$ are 
Lorentz indices and $\alpha$ and $\beta$ are colour indices)
\begin{eqnarray}
{\cal O}_1 &=& (\bar{s}_{L \alpha} \gamma_\mu b_{L \alpha})
               (\bar{c}_{L \beta} \gamma^\mu c_{L \beta}),    \\
{\cal O}_2 &=& (\bar{s}_{L \alpha} \gamma_\mu b_{L \beta})
               (\bar{c}_{L \beta} \gamma^\mu c_{L \alpha}),    \\
{\cal O}_3 &=& (\bar{s}_{L \alpha} \gamma_\mu b_{L \alpha})
               \sum_{q=u,d,s,c,b}
               (\bar{q}_{L \beta} \gamma^\mu q_{L \beta}),    \\
{\cal O}_4 &=& (\bar{s}_{L \alpha} \gamma_\mu b_{L \beta})
                \sum_{q=u,d,s,c,b}
               (\bar{q}_{L \beta} \gamma^\mu q_{L \alpha}),    \\
{\cal O}_5 &=& (\bar{s}_{L \alpha} \gamma_\mu b_{L \alpha})
               \sum_{q=u,d,s,c,b}
               (\bar{q}_{R \beta} \gamma^\mu q_{R \beta}),    \\
{\cal O}_6 &=& (\bar{s}_{L \alpha} \gamma_\mu b_{L \beta})
                \sum_{q=u,d,s,c,b}
               (\bar{q}_{R \beta} \gamma^\mu q_{R \alpha}),    \\
{\cal O}_7 &=& \frac{e}{16 \pi^2} m_b
               (\bar{s}_{L \alpha} \sigma_{\mu \nu} b_{R \alpha})
                F^{\mu \nu},                                   \\
{\cal O}_7^\prime &=& \frac{e}{16 \pi^2} m_s
               (\bar{s}_{R \alpha} \sigma_{\mu \nu} b_{L \alpha})
                F^{\mu \nu},                                    \\
{\cal O}_8 &=& \frac{g}{16 \pi^2} m_b
(\bar{s}_{L \alpha} T_{\alpha \beta}^a \sigma_{\mu \nu} b_{R \beta})
                G^{a \mu \nu},                                   \\
{\cal O}_8^\prime &=& \frac{g}{16 \pi^2} m_s
(\bar{s}_{R \alpha} T_{\alpha \beta}^a \sigma_{\mu \nu} b_{L \beta})
                G^{a \mu \nu},                                   
\end{eqnarray}
where $e$ and $g_s$ are the electromagnetic and the strong
coupling constants, and
 $F_{\mu \nu}$ and $G^A_{\mu \nu}$
denote the electromagnetic and the gluonic field strength
tensors, respectively.
We call attention to  the explicit mass factors in
 ${\cal O}_7 ({\cal O}_7^\prime)$
and ${\cal O}_8({\cal O}_8^\prime)$, which will undergo renormalization 
just as the Wilson coefficients.
                The dominant contributions
in the radiative decays $\BGAMAXS$
arise from the operators  ${\cal O}_2$, ${\cal O}_7$ and ${\cal O}_8$,
whereas the operators ${\cal O}_3,..., {\cal O}_6$
get coefficients through operator mixing only, which numerically
are negligible. Historically, the anomalous
dimension matrix was calculated in a truncated basis \cite{BSGAM}
 and this basis is
still often used for the sake of ease 
 in calculating the real and virtual corrections, though
as we discuss below, now the complete anomalous dimension matrix is
available \cite{Ciuchini}.

  The perturbative QCD corrections to the decay rate $\GGAMAXS$ have two 
distinct contributions:
\begin{itemize}
\item Corrections to the Wilson coefficients
$C_i(\mu)$, calculated with the help of the 
 renormalization group equation, whose solution requires the
knowledge of the anomalous dimension matrix in a given order in $\as$.

\item Corrections to the matrix elements of the operators
${\cal O}_i$ entering through the effective Hamiltonian
at the scale $\mu=O(m_b)$.
\end{itemize}
The anomalous dimension matrix is needed in order 
to use the renormalization group and sum up large logarithms, 
i.e., terms 
like $\as^{n}(m_W)\log^{m}(m_b/M)$, where $M=m_t$ or $ m_W$ and $m\leq n$
(with $n=0,1,2,...)$. Until recently, only the leading logarithmic 
corrections $(m=n)$ have been calculated systematically in the complete basis
given above \cite{Ciuchini}. Very recently,  the next-to-leading order
anomalous dimension has also been calculated and reported this summer by
Misiak at the Warsaw conference \cite{Misiak96}. 
 
Next-to-leading 
order corrections to the matrix elements are now available completely. 
They are of two kinds: 
 \begin{itemize}
 \item QCD Bremsstrahlung corrections $b \to s \gamma + g$, which are
needed both to cancel the infrared divergences
 in the decay rate for
$\BGAMAXS$ and in obtaining a non-trivial QCD contribution to the
photon energy spectrum in the inclusive decay $\BGAMAXS$.
\item Next-to-leading order virtual corrections to the matrix elements
in the decay $b \to s +\gamma$. 
\end{itemize}
The Bremsstrahlung corrections were calculated in \cite{ag1} - 
\cite{ag3} in the truncated basis and last year also in the complete 
operator basis \cite{ag95}, which have been checked in 
\cite{Pott95}.
The higher order matching conditions, i.e., $C_i(m_W)$, are known up to
the desired accuracy, i.e., up to $O(\as(M_W)$ terms \cite{Yao94}. 
 The next-to-leading order virtual
corrections have been completed by Greub, Hurth and Wyler recently 
\cite{GHW96}.
We discuss the presently available pieces in the SM calculation
of $\BBGAMAXS$ in the NLO accuracy.

\par
We recall that the Wilson coefficients obey the 
renormalization group equation
\begin{equation} \label{RGE}
\left[\mu \frac{\partial}{\partial \mu}
+ \beta(g) \frac{\partial}{\partial g} \right]
C_i \left(\frac{M^2_W}{\mu^2},g \right)
= \hat{\gamma}_{ji} (g) C_j \left(\frac{M^2_W}{\mu^2},g \right) .
\end{equation}
The  QCD beta function $\beta(g)$ has been defined earlier
and $\hat{\gamma}(g)$ is the anomalous dimension matrix, which,
to leading logarithmic accuracy, is given by
\begin{equation}  \label{gam}
\hat{\gamma}(g) =  \gamma_0 \frac{g^2}{16 \pi^2} .
\end{equation}
Here $\gamma_0$ is a $8 \times 8$ matrix given in \cite{Ciuchini,Buras94}.
The non-zero initial conditions in the SM are given at the scale $M_W$ and 
read \cite{InamiLim}
\begin{eqnarray}
C_2 &=& 1 \\
C_7 (M_W) &=& -\frac{1}{2} x \left[ \frac{2x^2/3 + 5x/12-7/12}{(x-1)^3} -
                  \frac{3x^2/2-x}{(x-1)^4} \ln x \right],
\\
C_8(M_W) &=& -\frac{1}{2} x \left[ \frac{x^2/4 - 5x/4-1/2}{(x-1)^3} +
                  \frac{3x/2}{(x-1)^4} \ln x \right],
\end{eqnarray}
and
$x = m^2_t / M_W^2$.
Also, for subsequent discussion it is useful to define two
effective Wilson coefficients $C_7^{\mathit{eff}}(\mu)$ and
 $C_8^{\mathit{eff}}(\mu)$
\cite{Buras94}:
 \begin{eqnarray}
\label{C78eff}
C_7^{\mathit{eff}} &\equiv & C_7 - \frac{C_5}{3} -  C_6 \quad , \nonumber\\
C_8^{\mathit{eff}} &\equiv & C_8 + C_5 \quad .
\end{eqnarray}
 The numerical values for the Wilson coefficients 
at $\mu=M_W$ (``Matching Conditions") and at three other scales
$\mu= 10.0$ GeV, $5.0$ GeV and $10.0$ GeV can, for example, be seen in 
\cite{ALI96} and will not be given here. In LO, one gets \cite{Buras94}: 
$${\cal B} (\BGAMAXS )= (2.8 \pm 0.8) \times 10^{-4},$$
reflecting the parametric uncertainties of the underlying framework.

\par
 Now, we discuss the explicit $O(\alpha_s)$ improvement of the decay rate.
The real and virtual $O(\as)$ corrections to the matrix
element for $b \to s + \gamma$ at the scale $\mu \approx m_b$  form a 
well-defined gauge invariant albeit scheme-dependent set of corrections.
This scheme dependence will be cancelled against the one in the anomalous
dimension $\gamma^{(1)}$, as discussed in the context of the dominant
contributions to the non-leptonic decays of the $B$ hadron  
earlier. The results presented here correspond to the NDR scheme.

\par
The Bremsstrahlung corrections  in $b \to s \gamma + g$, calculated 
in \cite{ag1} - \cite{ag3} and \cite{ag95}, were aimed at
 getting a non-trivial photon energy spectrum at the partonic level.
In these papers, the virtual corrections to $b \to s \g$ in $O(\as)$
were included only partially by taking into account those virtual diagrams 
which are needed to cancel the infrared
singularities (and also the collinear ones in the limit $m_s \to 0$)
 generated by the 
Bremsstrahlung diagram. The emphasis was on deriving the photon energy 
spectrum
in $B \to X_s + \gamma$ away from the end-point $x_\gamma \to 1$ and the
Sudakov-improved photon energy spectrum in the region $x_\gamma \to 1$.
The left-out virtual diagrams,
however, do contribute to the overall decay rate in $B \to X_s + \gamma$.
Recently, these virtual correction have been evaluated in
\cite{GHW96}, neglecting the small contributions from the operators 
$O_3$--$O_6$. The additional contribution reduces substantially
the scale dependence of the leading order (or partial next-to-leading order)
 decay 
width $\Gamma(B \to X_s + \gamma)$, which previously was found to be 
substantial 
and constituted a good fraction of the theoretical uncertainty in the
inclusive decay rate
\cite{AGM92,Buras94,Ciuchini94,ag95}.

 Concentrating on the dominant operators
$O_2, ~O_7$ and $O_8$,  the contribution of the next-to-leading order  
correction to the matrix element part in $b \to s+\gamma$ can be
expressed as follows: 
 \begin{equation}
{\cal M} = {\cal M}_2 + {\cal M}_7 + {\cal M}_8
\end{equation}
and the various terms (including appropriate counter term 
contributions) can be summarized as \cite{GHW96}:
\be
\label{m2ren}
{\cal M}_2 = \bra s \g |O_7| b \ket _{tree} \, \frac{\a_s}{4 \p} \,
\left( \ell_2 \log \frac{m_b}{\mu}  + r_2 \right) \quad ,
\ee
with
\be
\label{l2}
\ell_2 = \frac{416}{81}.
\ee
\begin{eqnarray}
\label{rer2ndr}
\Re r_2 &=& \frac{2}{243} \, \left\{- 833 + 144 \pi^2 z^{3/2}
\right. \nonumber \\
&& \hspace{0.3cm}
+ \left[ 1728 -180 \pi^2 -1296 \zeta (3) + (1296-324 \pi^2) L +
108 L^2 + 36 L^3 \right] \, z \nonumber \\
&& \hspace{0.3cm}
+ \left[ 648 + 72 \pi^2 + (432 - 216 \pi^2) L + 36 L^3 \right] \, z^2
\nonumber \\ 
&& \hspace{0.3cm}        \left.                 +
\left[ -54 - 84 \pi^2 + 1092 L - 756 L^2 \right] \, z^3 \, \right\}
\end{eqnarray}
\begin{eqnarray}
\label{imr2ndr}
\Im r_2 &=& \frac{16 \p}{81} \, \left\{- 5
+ \left[ 45-3 \pi^2 + 9 L +
9 L^2 \right] \, z
+ \left[ -3 \pi^2 + 9 L^2 \right] \, z^2 +
\left[ 28 - 12 L  \right] \, z^3 \, \right\}~.
\end{eqnarray}
Here, $\Re r_2$ and $\Im r_2$ denote the real and the imaginary part
of $r_2$, respectively, $z=(m_c/m_b)^2$ and $L=\log (z)$. 

The real and virtual corrections associated with the
operator $O_7$, calculated in \cite{ag1,ag2,ag95} can be combined
into a {\it modified matrix element} for
$b \to s \g$,
in such a way that its square reproduces
the result derived in these papers. This modified 
matrix element ${\cal M}_7^{mod}$ reads \cite{GHW96}:
\be
\label{m7lr}
{\cal M}_7^{mod} = \bra s \g|O_7| b \ket _{tree}
\, \left( 1+ \frac{\a_s}{4\p} \left( \ell_7 \, \log \frac{m_b}{\mu}
+r_7
\right) \right)
\ee
with
\be
\label{l7r7}   
\ell_7 = \frac{8}{3}  \quad , \quad
r_7 = \frac{8}{9} \, (4 - \pi^2) \quad .
\ee

Finally, the result for ${\cal M}_8$ is \cite{GHW96}:
\be
\label{m8lr}
{\cal M}_8 = \bra s \g |O_7| b \ket _{tree} \, \frac{\a_s}{4 \p} \,
\left( \ell_8 \log \frac{m_b}{\mu}  + r_8 \right) \quad ,
\ee
with
\be
\label{l8r8}
\ell_8 = - \frac{32}{9} \quad , \quad
r_8 = - \frac{4}{27} \,
\left( -33 +2\p^2  -6 i \p   
\right)
\quad .
\ee
\par
With the results given above,
one can write down the amplitude ${\cal M}(b \to s \g )$ 
by summing the various contributions already mentioned.
Since the relevant scale for a $b$ quark decay is expected to
be $\mu \sim m_b$,  the matrix elements of the
operators may be expanded around
$\mu=m_b$ up to order $O(\a_s)$  
and the next-to-leading order result can be written as:
\be
\label{amplitudevirtuell}
{\cal M}(b \to s \g ) = -\frac{4 G_F \l_t}{\sqrt{2}} \, D \,
\bra s \g|O_7(m_b)|b \ket _{tree}
\ee
with
\be
\label{d}
D = C_7^{\mathit{eff}}(\mu) + \frac{\a_s(m_b)}{4\p} \left(
C_i^{(0)eff}(\mu) \gamma_{i7}^{(0)eff} \log \frac{m_b}{\mu} +
C_i^{(0)eff} r_i
\right)         \quad ,
\ee
where the quantities $\gamma_{i7}^{(0)eff}=\ell_i + 8 \, \delta_{i7}$
are  the entries of the (effective) leading order anomalous
dimension matrix 
and  the quantities $\ell_i$ and $r_i$ are given
for $i=2,7,8$ in eqs. (\ref{l2},\ref{rer2ndr}),
(\ref{l7r7}) and (\ref{l8r8}), respectively.
The first term, $C_7^{\mathit{eff}}(\mu)$,
on the r.h.s. of Eq.~(\ref{d})  has to be
taken up to  next-to-leading logarithmic precision in order
to get the full next-to-leading logarithmic result, whereas
it is sufficient to use the leading logarithmic values of
the other Wilson coefficients in Eq.~(\ref{d}).

\begin{figure}[htb]
\vspace{0.10in}
\centerline{\epsfig{file=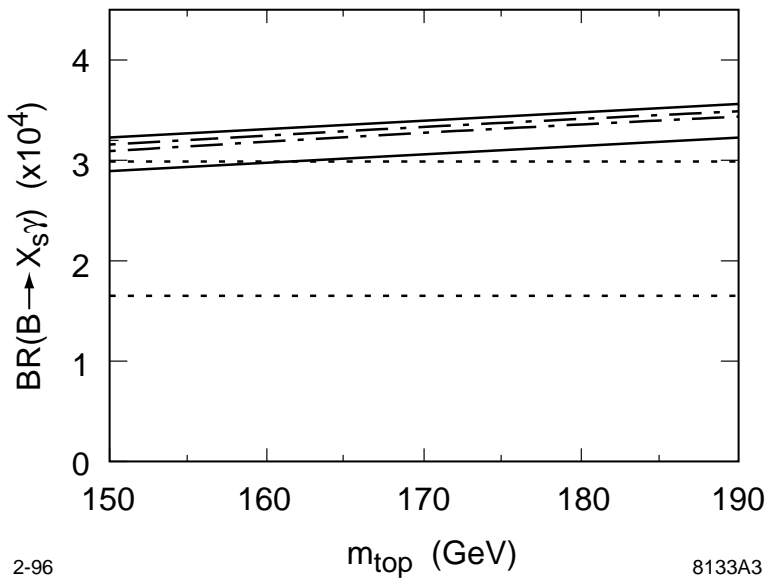,height=3in}}
\vspace{0.08in}
\caption[]{Branching ratio for $b \to s \g (g)$
calculated in \protect\cite{GHW96} with the parameters
$\Vtsabs/\Vcbabs =1,~\Vtbabs =1, ~m_b^{pole}=4.8$ GeV and $m_c/m_b=0.29$.
The different curves are explained in the text.
\label{GHRfig3}}
\end{figure}

The decay width $\G^{virt}$ which follows
from ${\cal M}(b \to s \g)$ in Eq.~(\ref{amplitudevirtuell}) reads 
\be
\label{widthvirt}
\G^{virt} = \frac{m_{b,pole}^5 \, G_F^2 \l_t^2 \a_{em}}{32 \p^4}
\, F \, |D|^2 \quad ,
\ee
where the terms of $O(\a_s^2)$
in $|D|^2$ have been discarded.
The factor $F$ in Eq.~(\ref{widthvirt}) is
\be
F = \left( \frac{m_b(\mu=m_b)}{m_{b,pole}} \right)^2 =
1- \frac{8}{3} \,  \frac{\a_s(m_b)}{\p} \quad ,  
\ee
and its origin lies in the explicit presence of $\mb$ in the operator
$O_7$.
To get the inclusive decay width for $b \to s \g (g)$, also
the Bremsstrahlung corrections (except the part
already absorbed above) must be added. The contribution of the operators
${\cal O}_2$ and ${\cal O}_7$ was  calculated already in \cite{ag1}.
As pointed out by Buras et al. \cite{Buras94},
the explicit logarithms of the form $\a_s(m_b) \log(m_b/\mu)$ in Eq.
(\ref{d}) are cancelled by the $\mu$-dependence of
$C_7^{(0)\mathit{eff}}(\mu)$. Therefore, the scale dependence is
significantly reduced by including the virtual corrections
completely to this order.
 This is shown in
Fig. ~\ref{GHRfig3} (solid curves), which 
 yields an error of $\pm 6 \%$ on ${\cal B}(B \to X_s + \gamma)$ 
 varying $\mu$ in the range $m_b/2 \leq \mu \leq 2 m_b$.
We recall that in the LO calculations, this scale dependence was $\pm 
20\%$. The CLEO result is shown as dashed lines. The two other curves
(dashed-dotted) represent more stringent assumption on the $\mu$-dependence
and we refer to \cite{GHW96} for further details.
%
%
%

\subsection{Estimating long-distance effects in $\BGAMAXS$}

\par
 In order to get the complete amplitude for $\BGAMAXS$ one has to include
also the effects of the long-distance contributions, which arise from 
the matrix elements of the four-quark operators  in ${\cal H}_{eff}$,
$\langle X_s \gamma \vert {\cal O}_i \vert B \rangle$. It is 
usual to invoke the hypothesis of factorization, which is then
combined with the additional assumption of  vector
meson dominance, involving the decays
 $ B \to \sum_{i} V_{i} + X_s \to \gamma +X_s$, where $V_{i}= J/\psi,
\psi^\prime,...$ \cite{bsgamld} - \cite{Ricciardi}.
One has to ensure that the
resulting amplitude  ${\cal M}(\BGAMAXS)$  remains
manifestly gauge invariant. In practice,
this amounts to discarding the longitudinal polarization contribution in
the non-leptonic decays $B \to (J/\psi, \psi^\prime,...) +X_s$, which
in fact dominates the decay widths \cite{BH95}, and keeping only the smaller
contribution from the transverse polarization of $J/\psi, \psi^\prime,...$.
 Following
\cite{DHT95,GP95}, one can write the decay amplitude as:
\begin{eqnarray}
{\cal M}(b\rightarrow s J/\psi )_T
=  {G_F\over \sqrt{2}}a_2V_{cb}V_{cs}^* {g_{J/\psi}(m_{J/\psi}^2)\over m_b
m_{J/\psi}^2}
\bar s \sigma^{\mu\nu}(1+\gamma_5)b q_\nu \epsilon^\dagger_\mu(q)\;,
\end{eqnarray}
where $g_\psi$ is defined as $\langle \psi(q) \vert \bar{c}\gamma_\mu c\vert
0 \rangle =-i g_\psi(q^2) \epsilon_\mu^\dagger (q)$.
 For the
decays under consideration one needs the value of the
 BSW coefficient $a_2$ \cite{BSW}, 
which has been determined to be $\vert a_2 \vert = 0.24 \pm 0.04$ 
\cite{BH95}.
 One also needs to evaluate the coupling constant
$g_V(q^2)$ at the point $q^2=0$. From leptonic decays of vector mesons, one
gets, however,  $g_V(q^2=M_V^2)$.
As noted in \cite{DHT95,GP95}, using $g_V(q^2=0)=g_V(q^2=M_V^2)$ would
substantially overestimate the long-distance contribution due to
the expected dynamical suppression of the effective coupling 
$g_V(q^2)$, as one extrapolates to the point $q^2=0$. In fact, such a
suppression is supported by HERA data on photoproduction of $J/\psi$.
Including all the ($c\bar c$) resonances and the short distance contribution
${\cal M}_{SD}$, the two-body  decay amplitude ($b \to s \gamma)$ can be
written as \begin{eqnarray}
{\cal M}(b\rightarrow s\gamma)
=  -{eG_F\over 2\sqrt{2}}V_{tb}V_{ts}^*[{1\over 4\pi^2}m_b D(\mu)
-a_2{2\over 3} \sum_i{g_{\psi_i}^2(0)\over m_{\psi_i}^2m_b} ]
\bar s \sigma^{\mu\nu}(1+\gamma_5)b F_{\mu\nu}\;,
\end{eqnarray}
where $\psi_i$ represents the following  vector $c\bar c$ resonant
states:
$\psi(1S)$, $\psi(2S)$, $\psi(3770)$, $\psi(4040)$, $\psi(4160)$, and
$\psi(4415)$, and $D$ is the function given earlier.
 Taking this estimate
 as giving the right order of magnitude for the long-distance
contribution, Deshpande et al. \cite{DHT95} conclude that such
long-distance effects can be as large as 10\%. Other estimates, in
particular by Golowich and Pakvasa \cite{GP95}, lead to an even smaller
long-distance contribution. Clearly, one can not argue very conclusively
if such estimates are completely quantitative due to the assumptions
involved. In future, one
could improve them by  using  data from HERA on elastic $J/\psi
-$, and $\psi^\prime$-photoproduction
to get $g_{J/\psi}(0)$ and $g_{\psi^\prime}(0)$ directly, reducing at 
least the extrapolation uncertainties involved in the presently adopted 
procedure of extracting these coupling constants from the leptonic decay 
widths of each state and extrapolating to the point $q^2=0$ using an Ansatz.
In conclusion, LD effects in $\BGAMAXS$ are dynamically suppressed.
%
\subsection{Estimates of $\BBGAMAXS$ in the Standard Model}
 In the quantitative
estimates of the SM branching ratio $\BBGAMAXS$ given below we have 
neglected the LD-contributions. It is 
theoretically preferable to calculate this quantity in terms of the
semileptonic decay branching ratio
\begin{equation}
\label{brdef}
{\cal B} ( B \ra  X_{s} \g) = [\frac{\Gamma(B \ra  
\gamma + X_{s})}{\Gamma_{SL}}]^{th}
\, {\cal B} (B \to X\ell \nu_\ell)  \qquad ,
\end{equation}  
where, the leading-order QCD corrected
 $\G_{SL}$ has been given earlier. The leading order power 
corrections in the heavy quark expansion, discussed in the context of the 
semileptonic decay rate, are identical 
in the inclusive decay rates for  $\BGAMAXS$ and $B \to X \ell \nu_\ell$, 
entering in the
numerator and denominator in the square bracket, respectively, and hence 
drop out \cite{Chayetal,Bigietal}.

   The error on the branching ratio ${\cal B} ( B \ra  X_{s} \g)$ comes
from the following sources. 
\begin{enumerate}
\item $\Delta m_t$ and $\Delta \mu$:  The present value of $m_t$ is $m_t=175 
\pm 9$ GeV \cite{CDFvtb}, which is usually interpreted as the pole mass.
 With this
the running top quark mass in the $\overline{MS}$ scheme is
$\overline{m_t} = 166 \pm 9$ GeV. This leads to an error of $\pm 4\%$
in ${\cal B} ( B \ra  X_{s} \g)$. The combined error on $\Delta m_t$ and  
$\Delta \mu$ is about $\pm 9\%$ as can be seen in Fig. ~\ref{GHRfig3}.
\item Errors from the extrinsic parameters ($\Delta (m_b)$, 
$\Delta(\as(m_Z))$, 
and $\Delta(BR_{SL})$, the experimental uncertainty on the semileptonic
 branching ratio): This gives an uncertainty of $\pm 12\%$  
on ${\cal B}(B \to X_s + \gamma)$ as estimated in \cite{ag95}, of which 
half is due to the assumed uncertainty $\Delta(\as(m_Z))=0.006$. 
\end{enumerate}
As mentioned already, the Wilson coefficient
 $C_7^{\mathit{eff}}$ has been calculated in the 
next-to-leading order \cite{Misiak96}. The NLO corrections are found to
be small but the numerical difference is still being worked out. 
Replacing $C_7^{\mathit{eff}}$ by its leading log 
value yields the branching ratio \cite{ALI96}:
\begin{equation}\label{smbsgbr}
{\cal B} (\BGAMAXS )= (3.20 \pm 0.30 \pm 0.38) \times 10^{-4}
\end{equation}
where the first error comes from the combined error on $\Delta m_t$ and
$\Delta \mu$, as can be seen in Fig. ~\ref{GHRfig3},
and the second from the extrinsic source.
Combining the theoretical errors in quadrature gives
\begin{equation}\label{smbsgbrf}
{\cal B} (\BGAMAXS )= (3.20 \pm 0.48) \times 10^{-4}.
\end{equation}
Using the same input, a branching ratio ${\cal B} (\BGAMAXS )= (3.25 \pm 
0.50) \times 10^{-4}$ has been calculated in \cite{GH96}. The error on
the NLO branching ratio is now $\pm 15\%$, reduced by a factor 2 from the
corresponding LO value.
 The SM branching ratio
${\cal B} (\BGAMAXS )$ is compatible with the present measurement
${\cal B} (\BGAMAXS )= (2.32 \pm 0.67) \times 10^{-4}$ \cite{CLEOrare2}.
On its face value,
the electroweak penguin rate in the SM is nominally larger than the present
experimental value, but due to the large errors this difference is not
significant. Nevertheless, this comparison suggests that there is little
room for an additive beyond-the-SM contribution, which, for example, is
the case in multi-Higgs doublet models. For constraints on such models, see
\cite{Hewett96}. 
Expressed in terms of the CKM matrix element ratio, one gets
\begin{equation}\label{vtscb}
\frac{\Vtsabs}{\Vcbabs} = 0.85 \pm 0.12 (\mbox{expt}) \pm 0.08 (\mbox{th}),
\end{equation}
which is within errors consistent with unity, as expected from the
unitarity of the CKM matrix.

  Finally, we note that the ratio $R_{K^*}$ has been calculated in a 
large number of models. Not surprisingly, taken together they give rise to a 
large 
dispersion for this quantity. However, one should stress that QCD sum rules
and models based on quark-hadron duality give values which are in good
agreement with the CLEO measurements. Some representative results are:
\bea
R_{K^*} &=& 0.20 \pm 0.06 ~~[\mbox{Ball}~\protect\cite{bksnsr}], 
\nonumber\\
R_{K^*} &=& 0.17 \pm 0.05 ~~[\mbox{Colangelo et al.}
~\protect\cite{bksnsr}], \nonumber\\
R_{K^*} &=& 0.16 \pm 0.05 ~~[\mbox{Ali, Braun and Simma}
~\protect\cite{abs93}], \nonumber\\
R_{K^*} &=& 0.16 \pm 0.05 ~~[\mbox{Narison}             
~\protect\cite{bksnsr}], \nonumber\\
R_{K^*} &=& 0.13 \pm 0.03 ~~[\mbox{Ali \& Greub}             
~\protect\cite{ag3}].
\eea 
\subsection{Photon energy spectrum in $\BGAMAXS$}
The two-body partonic process $b \to s \gamma$
yields a photon energy spectrum which is just a discrete line, $1/(\Gamma) d 
\Gamma (b \to s \gamma) = \delta(1-x)$, where 
the scaled photon energy $x$ is defined as
$ E_\g = (m_b^2-m_s^2)/(2 \, m_b )\, x $. The physical photon energy spectrum
is built up by convoluting the non-perturbative effects due to 
the hadronic states involved in the decay and the
perturbative QCD corrections, such as $b \to s \gamma + g$, which give a
characteristic Bremsstrahlung spectrum in $x$ in the interval $[0,1]$
peaking near the end-points, $E_\gamma \to
E_\gamma ^{max}$ (or $x \to 1$) and $E_\gamma \to 0$ (or $x \to 0$), arising 
from the soft-gluon and soft-photon configurations, respectively.
Near the end-points, one 
has to improve the spectrum obtained in fixed order perturbation 
theory. This is usually done in the region $x \to 1$ by isolating and 
exponentiating the leading behaviour in $\alpha_{em}\alpha_s(\mu)^m 
\log^{2n} (1-x)$  with $m\leq n$,
where $\mu$ is a typical momentum in the decay $\BGAMAXS$. The running of 
$\alpha_s$ is a non-leading effect, but  as it is characteristic of QCD it
modifies the Sudakov-improved end-point photon energy spectrum
 \cite{KS94,Shifmangamma} 
compared to its  analogue in QED \cite{Sudakov}.
 As long as the $s$-quark mass
is non-zero, there is no collinear singularity in the spectrum.
However, parts of the spectrum have large logarithms of the form
$\as \log (m_b^2/m_s^2)$, which are important near the end-point
$x \to 0$ but their influence persists also in the
intermediate photon energy region and they have to be resummed
\cite{ag95,klp95}. Implementation of non-perturbative effects is at present
model dependent.

\par
  We shall confine ourselves 
to the discussion of the photon energy spectrum calculated in 
\cite{ag1,ag3,ag95}, incorporating    
the perturbatively computed spectrum, discussed in the previous section, 
and non-perturbative smearing for which a model is used. 
 In this model 
\cite{Alipiet}, which admittedly is simplistic but has received some
theoretical support in the HQET approach subsequently \cite{MW94, Bigietal2},
  the $b$ quark in $B$ hadron is assumed to have a Gaussian
distributed Fermi motion determined by a non-perturbative parameter, $p_F$,
\begin{equation}
\label{lett13}
 \phi(p)= \frac {4}{\sqrt{\pi}{p_F}^3} \exp (\frac {-p^2}{{p_F}^2})
\quad , \quad p = |\vec{p}|
\end{equation}
with the wave function normalization 
$ \int_0^\infty \, dp \, p^2 \, \phi(p) = 1.$
The photon energy spectrum from
the decay of the $B$-meson at rest is then given by
\begin{equation}
\label{lett15}
 \frac{d\Gamma}{dE_\gamma}= \int_0^{p_{max}} \, dp \, p^2 \, \phi(p)
  \frac {d\Gamma_b}{dE_\gamma}(W,p,E_\g) \quad ,
\end{equation}
where $p_{max}$ is the maximally allowed value of $p$ and
$ \frac{d\Gamma_b}{dE_\g}$
 is the photon energy spectrum from the decay of the $b$-quark in
flight, having a momentum-dependent mass $W(p)$. This is calculated
in perturbation theory taking into account the appropriate Sudakov
behaviour in the $E_\gamma$ end-point region at the partonic level.

\begin{figure}[htb]
\vspace{0.10in}
\centerline{
\epsfysize=3in
\rotate[r]{
\epsffile{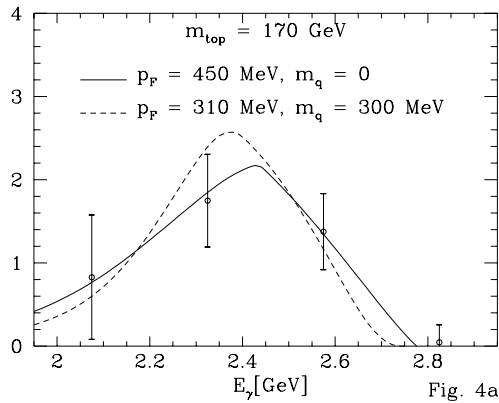}
}
}
\vspace{0.08in}
\caption[]{Comparison of the normalized photon energy distribution
using the
 CLEO data \protect\cite{CLEOrare2} corrected for detector effects and 
theoretical
distributions from \protect\cite{ag95} , both  normalized to unit area in
the photon energy interval between 1.95 GeV and 2.95 GeV. The solid 
 curve corresponds to the values with the minimum $\chi^2$,
 $(m_q,p_F)$=(0, ~450 MeV), and the dashed curve to the values
 $(m_q,p_F)$=(300 MeV, 310 MeV).
\label{agfig4}}
\end{figure}

An analysis of the CLEO photon 
energy spectrum has been undertaken in \cite{ag95}  to determine 
the non-perturbative parameters of this model, namely  
$m_b(pole)$ and  $p_F$. The latter is related to the
kinetic energy parameter $\lambda_1$ defined earlier in the HQET approach. 
The experimental
errors are still large and the fits result in relatively small $\chi^2$ 
values; the minimum, $\chi^2$ is obtained for $p_F=450$ MeV and 
$m_b(pole)=4.77$ GeV. While the value of the kinetic energy parameter 
$\lambda_1$  is at present in a state of flux
\cite{Martinelli96} and hence no
quantitative conclusions can be drawn, the value of the $b$-quark pole mass
determined from the photon energy spectrum is in good agreement with 
theoretical estimates of the same, namely
$m_b(pole)= 4.8\pm 0.15$ GeV \cite{Neubert94,bqmass}.
In Fig. \ref{agfig4} we have plotted the photon energy spectrum normalized
to unit area in the interval between 1.95 GeV and 2.95 GeV for
the parameters which correspond to the minimum $\chi^2$ (solid curve)
and for another set of parameters that lies near the
$\chi^2$-boundary defined by $\chi^2=\chi^2_{min} +1$.
(dashed curve). Data from CLEO \cite{CLEOrare2} are also 
shown. Further details of this analysis can be seen in \cite{ag95}.

\par
It is a very desirable goal to calculate the photon energy spectrum in
$\BGAMAXS$ in a theoretically more robust framework. In this context we note
that attempts to calculate the photon and lepton energy spectra
in the heavy quark expansion method lead to formal expressions which near
the end-point are divergent \cite{MW94,JR94,Bigietal2}.
Near the end-point, the energy released for the light   
quark system in the final state is not of $O(\mb)$ but of the order of   
the parameter
 $\bar{\Lambda}=M_B-m_b \sim O(\Lambda_{\mbox{\small QCD}})$. Thus, the
expansion parameter is no longer $1/\mb^2$
but rather $1/\mb \bar{\Lambda} =O(1)$ and the operator product expansion 
breaks down. This divergent series in the effective theory has to be
 cleverly resummed and the distributions averaged over momentum bins
\cite{neubertbsg}. The resummation allows us to define, in principle,
 an effective non-perturbative shape function
\cite{neubertbsg,Shifmangamma}, which though can not be calculated in the
effective theory but one could use this concept advantageously to relate
the energy spectra in the semileptonic decays $B \to X_u \ell \nu_\ell$  
and $B \to X_s + \gamma$. Since the perturbative corrections are 
process-dependent, they have to be included accordingly.
 This programme remains to be implemented in a
phenomenological analysis of data.

%
%
%
\section{Inclusive radiative decays \bgamaxd }
\par
The theoretical interest in studying the 
(CKM-suppressed) inclusive radiative decays
\bgamaxd\ lies in the first place in the 
 possibility of determining the parameters of the CKM
matrix.
With that goal in view, one of the interesting quantities in the
decays $B \to X_d + \gamma$ is the end-point photon energy spectrum,
which has to be measured requiring that
 the hadronic system $X_d$ recoiling against the
photon does not contain strange hadrons to separate the large-$E_\g$
photons from the decay $\BGAMAXS$. Assuming that this is feasible,
one can determine 
 from the ratio of the decay rates
$\BBGAMAXD/\BBGAMAXS$ the CKM-Wolfenstein parameters $\rho$ and $\eta$.
 This measurement was first proposed in \cite{ag2}, where
the photon energy spectra were also worked out.

\indent
 In close analogy
with the \bgamaxs\ case discussed earlier,
the complete set of dimension-6 operators relevant for
the processes $b \to d \gamma$ and $b \to d \gamma g$ 
can be written as:
\begin{equation}
\label{heffd}
{\cal H}_{eff}(b \to d)=
 - \frac{4 G_{F}}{\sqrt{2}} \, \xi_{t} \, \sum_{j=1}^{8}
C_{j}(\mu) \, \hat{O}_{j}(\mu),\quad
\end{equation}
where $\xi_{j} = V_{jb} \, V_{jd}^{*}$ with $j=t,c,u$. The operators
 $\hat{O}_j, ~j=1,2$, have implicit in them CKM factors. In the
Wolfenstein parametrization \cite{Wolfenstein}, one can express these
factors as :
\begin{equation} 
\xi_u = A \, \lambda^3 \, (\rho - i \eta),
~~~\xi_c = - A \, \lambda^3 ,
~~~\xi_t=-\xi_u - \xi_c.
\end{equation}
We note that all three CKM-angle-dependent quantities
$\xi_j$ are of the
same order of magnitude, $O(\lambda^3)$. It is calculationally convenient 
to define the operators $\hat{O}_1$ and 
$\hat{O}_2$ entering in ${\cal H}_{eff}(b \to d)$ as follows \cite{ag2}:
\begin{eqnarray}
\label{basis}
&&\hat{O}_{1} =
 -\frac{\xi_c}{\xi_t}(\bar{c}_{L \beta} \go{\mu} b_{L \alpha})
(\bar{d}_{L \alpha} \gu{\mu} c_{L \beta})
 -\frac{\xi_u}{\xi_t}(\bar{u}_{L \beta} \go{\mu} b_{L \alpha})
(\bar{d}_{L \alpha} \gu{\mu} u_{L \beta}) ,\nonumber \\
&& \hat{O}_{2} =
-\frac{\xi_c}{\xi_t}(\bar{c}_{L \alpha} \go{\mu} b_{L \alpha})
(\bar{d}_{L \beta} \gu{\mu} c_{L \beta}) 
 -\frac{\xi_u}{\xi_t}(\bar{u}_{L \alpha} \go{\mu}
b_{L \alpha}) (\bar{d}_{L \beta} \gu{\mu} u_{L \beta}) ,
\end{eqnarray}
with the rest of the operators $\hat{O}_j$ 
defined like their
counterparts ${O}_j$ in ${\cal H}_{eff}(b \to s)$, with the obvious 
replacement
$s \to d$. With this choice, the matching conditions $C_j(m_W)$
 and the solutions
of the RG equations yielding $C_j(\mu)$ become
identical for the two operator bases $O_j$ and $\hat{O}_j$.
The essential difference between  $\GGAMAXS$ and $\GGAMAXD$ 
lies in the matrix elements of the first two operators $O_1$ and $O_2$
(in ${\cal H}_{eff}(b \to s)$) and $\hat{O}_1$ and $\hat{O}_2$ (in 
${\cal H}_{eff}(b \to d)$).
The branching ratio  $\BBGAMAXD$ in the SM  can be written as:
\begin{equation}
\label{branstruc}
\BBGAMAXD = D_1 \lambda^2 \{
(1-\rho)^2 + \eta^2 -(1-\rho) D_2 - \eta D_3 +D_4  \} , \quad
\end{equation}
where the functions $D_i$ depend on the parameters $\mt,\mb,m_c,\mu$,
as well as the others we discussed in the context of ${\cal B}(\BGAMAXS)$.
These functions were first calculated in \cite{ag2} in the leading 
logarithmic 
approximation. Recently, these estimates have been improved in 
\cite{aag96}, making use of the NLO calculations in \cite{GHW96} discussed
in the context of the decay $\BGAMAXS$ earlier.
 To get the inclusive branching 
ratio, the CKM parameters $\rho$ and $\eta$ have to be constrained from the
unitarity fits.  Present data and theory 
restrict the parameters $\rho$ and $\eta$ to lie in the following
range (at 95\% C.L.) \cite{AL96}:
\begin{eqnarray}
 0.20 &\leq & \eta \leq 0.52 , \nonumber \\
 -0.35 &\leq & \rho \leq 0.35 ~,
\label{rhoetarange}
\end{eqnarray}
which, on using the current lower bound from LEP on
 the $B_s^{0}$ - $\overline{B_s^{0}}$ mass difference  
 $\delms > 9.2$ (ps)$^{(-1)}$ \cite{Gibbons96},
restricts $\rho$ to lie in the range $ -0.25 \leq \rho \leq 0.35$, with 
$\eta$ not changed significantly. This is based on assuming 
$\xi_s =1.1$, where $\xi_s$ is the $SU(3)$-breaking parameter $\xi_s = 
f_{B_s} \sqrt{\hat{B}_{B_s}}/f_{B_d} \sqrt{\hat{B}_{B_d}}$.
The preferred CKM-fit values are \cite{AL96}
\beq
(\rho,\eta) = (0.05,0.36) ~,
\eeq
for which one gets \cite{aag96}   
\begin{equation}
 \BBGAMAXD = 1.62 \times 10^{-5},
\end{equation}
whereas $\BBGAMAXD =8.0 \times 10^{-6}$ and $2.8 \times 10^{-5}$ for the 
other 
two extremes $\rho=0.35, ~\eta=0.50$ and $\rho=-\eta=-0.25$, respectively.
In conclusion, we note that  
the functional dependence of $\BBGAMAXD$ on the Wolfenstein parameters   
$(\rho,\eta)$ is mathematically different than that of $\delms$. However,
since the non-factorizing terms represented by the coefficients $D_2$ - $D_4$
in the expression for $\BBGAMAXD$ are numerically small \cite{aag96}, the 
resulting constraints from this decay mode and $\delmd/\delms$ are
qualitatively very similar. From the experimental point of
view, the situation $\rho <0$ is favourable for both these measurements as
in this case one expects (relatively) smaller values for $\delms$ and 
larger values for the branching ratio $\BBGAMAXD$, as compared to the
$\rho > 0$ case which would yield larger $\delms$ and smaller $\BBGAMAXD$. 
\vspace*{3.0ex}
\subsection{${\cal B}(B \to V + \gamma )$ and constraints on the CKM 
parameters}

\par
Exclusive radiative
 $B$ decays $B \to V + \gamma$, with $V=K^*,\rho,\omega$, are also 
potentially
very interesting from the point of view of determining the CKM parameters
\cite{abs93}. The extraction of these parameters would, however,  involve a 
trustworthy 
estimate of the SD- and LD-contributions in the decay amplitudes.
\par
  The SD-contributions in the 
 exclusive decays $(B^\pm, B^{0}) \to (K^{*\pm}, K^{* 0})+ \gamma$,
$(B^\pm, B^{0}) \to (\rho^\pm,\rho^{0}) + \gamma$,
$B^{0} \to \omega + \gamma$  and the
corresponding $B_s$ decays, $B_s \to \phi + \gamma $, and
$B_s \to K^{* 0} + \gamma $,
involve the magnetic moment operator ${\cal O}_7$ and the related one 
obtained by the obvious change $s \to d$, $\hat{O}_7$.
The transition form factors governing these decays
can be generically  defined as:
\be
 \langle V,\lambda |\frac{1}{2} \bar \psi \sigma_{\mu\nu} q^\nu b
 |B\rangle  =
     i \epsilon_{\mu\nu\rho\sigma} e^{(\lambda)}_\nu p^\rho_B p^\sigma_V
F_S^{B\rightarrow V}(0).
\label{defF}
\ee
Here $V$ is a vector meson
with the polarization vector $e^{(\lambda)}$,
$V=\rho, \omega, K^*$ or $\phi$;
$B$ is a generic
$B$-meson $B^\pm, B^{0}$ or $B_s$, and $\psi$ stands for the
field of a light $u,d$ or $s$ quark. The vectors $p_B$, $p_V$ and
$q=p_B-p_V$
correspond to the 4-momenta of the initial $B$-meson and the
outgoing vector
meson and photon, respectively.
 Keeping only the SD-contribution 
 leads to obvious relations among the exclusive 
decay rates, exemplified here by the decay
rates for $(B^\pm,B^0) \to (\rho^\pm,\rho^0) + \gamma$ and $(B^\pm,B^0) \to 
(K^{*\pm}, K^{*0}) + \gamma$:
 \be
\frac{\Gamma ((B^\pm,B^{0}) \to (\rho^\pm,\rho^{0}) + \gamma)}
     {\Gamma ((B^\pm,B^{0}) \to (K^{*\pm},K^{* 0}) + \gamma)} 
  = \frac{\vert \xi_t \vert^2}{\vert\lambda_t \vert ^2}
      \frac{\vert F_S^{B \to \rho }(0)\vert^2}
          {\vert F_S^{B \to K^* }(0)\vert^2} \Phi_{u,d}
  \simeq \kappa_{u,d}\left[\frac{\Vtdabs}{\Vtsabs}\right]^2 \,,
\label{SMKR}
\ee
where $\Phi_{u,d}$ is a phase-space factor which in all cases is close to 1
and\\
 $\kappa_{i} \equiv [F_S^{B_i \to \rho}/F_S^{B_i \to K^*}]^2$.
The transition form factors $F_S$ are model dependent.
Estimates of $F_S^{B_i \to K^*}$ in the QCD sum rule approach  
are in good agreement with the CLEO data, as already discussed.
The ratios of the form factors, i.e. $\kappa_i$,
should therefore also be reliably calculable in this approach as they depend
essentially only on the SU(3)-breaking effects.

 If the SD-amplitudes were the only contributions, the measurements of the
 CKM-suppressed radiative decays $(B^\pm,B^0) \to (\rho^\pm,\rho^0) + 
\gamma , ~B^0 \to \omega + \gamma$ and $B_s \to K^* + \gamma$ could be
used in conjunction with the decays $(B^\pm,B^0) \to (K^{*\pm},K^{*0}) + 
\gamma$ to determine 
the CKM parameters. The present experimental upper limits on the CKM ratio
$\Vtdabs/\Vtsabs$ from radiative $B$ decays 
are indeed based on this assumption, yielding at 90\% C.L.\cite{CLEOwarsaw}:
\be
\left\vert {V_{td} \over V_{ts}} \right\vert \leq 0.45 - 0.56~,
\ee
depending on the models used for the $SU(3)$ breaking effects
in the form factors \cite{abs93,bksnsr}.

  The possibility of significant
LD-contributions in
radiative $B$ decays from the light quark intermediate states
has been raised in a number of papers
\cite{bsgamld}.
Their amplitudes necessarily involve other CKM matrix elements and hence the
simple factorization of the decay rates in terms of the CKM factors
involving $\Vtdabs$ and $\Vtsabs$ no longer holds thereby
 invalidating the relation (\ref{SMKR}) given above. As we already discussed,
the LD-contributions are small in the exclusive
decays $B \to K^* + \gamma$ and so this issue hinges sensitively upon the
LD-contributions in the CKM-suppressed decays,
$B^\pm \to \rho^\pm \gamma$ and $B^0 \to (\rho^0,\omega) \gamma$.

The LD-contributions in $B \to V + \gamma$,
induced by the matrix elements of the
four-Fermion operators $\hat{O}_1$ and $\hat{O}_2$ (likewise $O_1$ and 
$O_2$), have been investigated in \cite{wyler95,ab95} using 
 a technique \cite{BBK89}
which treats the photon emission from the light quarks in a theoretically
consistent and model-independent way. This has been combined
with the light-cone QCD sum rule approach to calculate both the SD and LD
--- parity conserving and parity violating --- amplitudes
in the decays $(B^\pm, B^{0}) \to (\rho^\pm,\rho/\omega) + \gamma$.
To illustrate this, we concentrate on the $B^\pm$ decays
$B^\pm \to \rho^\pm + \gamma$, and take up the neutral $B$ decays
$B^{0} \to \rho (\omega) + \gamma$ at the end.

\par
The LD-amplitude of the four-Fermion operators $\hat{O}_1$, $\hat{O}_2$
is dominated by  the
 contribution of the weak annihilation
of valence quarks in the $B$ meson and it is color-allowed for the
decays of charged $B^\pm$ mesons.
Using factorization, the LD-amplitude in the decay $B^\pm \to \rho^\pm + 
\gamma$ can be written in terms of the form factors $F_1^L$ and $F_2^L$,
\begin{eqnarray}\label{Along}
{\cal A}_{long} &=&
-\frac{e\,G_F}{\sqrt{2}} V_{ub}V_{ud}^\ast
\left( C_2+\frac{1}{N_c}C_1\right) m_\rho
\varepsilon^{(\gamma)}_\mu \varepsilon^{(\rho)}_\nu
\nonumber\\&&{}\times
 \Big\{-i\Big[g^{\mu\nu}(q\cdot p)- p^\mu q^\nu\Big] \cdot 2 F_1^{L}(q^2)
  +\epsilon^{\mu\nu\alpha\beta} p_\alpha q_\beta
 \cdot 2 F_2^{L}(q^2)\Big\}\,.
\end{eqnarray}
 Estimates from the light-cone QCD sum rules give
\cite{ab95}:
\begin{equation}\label{result}
 F^L_1/F_S = (1.25\pm 0.10)\times 10^{-2}\,,\quad F^L_2/F_S = (1.55\pm 
0.10) \times 10^{-2} ~,
\end{equation}
where the errors correspond only to the variation of the
Borel parameter in the QCD sum rules. Including other possible 
uncertainties, 
 one expects an accuracy  of order 20\% for the ratios in (\ref{result}).
 The parity-conserving and parity-violating amplitudes turn out
to be numerically close to each other in the QCD sum rule approach, 
$F_1^L\simeq F^L_2 \equiv F_L$,
hence the ratio of the LD- and the SD- contributions reduces to a number 
\cite{ab95}
 \begin{equation}\label{ratio2p}
{\cal A}_{long}/{\cal A}_{short}=
R_{L/S}^{B^\pm\to\rho^\pm\gamma}
\cdot\frac{V_{ub}V_{ud}^\ast}{V_{tb}V_{td}^\ast} ~.
\end{equation}
Using $C_2=1.10$, $C_1=-0.235$, $C_7^{\mathit{eff}}=-0.306$
(corresponding to the scale $\mu=5$ GeV) gives:
\begin{equation}\label{result2}
R_{L/S}^{B^\pm\to\rho^\pm\gamma} \equiv
 \frac{4 \pi^2 m_\rho(C_2+C_1/N_c)}{m_b C_7^{\mathit{eff}}}
\cdot\frac{F_L^{B^\pm \to \rho^\pm \gamma}}{F_S^{B^\pm \to \rho^\pm 
\gamma}}=-0.30\pm 0.07 ~, \end{equation}
which is not small.
 To get a ball-park estimate of the ratio
${\cal A}_{long}/{\cal A}_{short}$, we take the central value from 
the CKM fits,  $\Vubabs/\Vtdabs \simeq 0.33$ \cite{AL96}, yielding
\begin{equation}
|{\cal A}_{long}/{\cal A}_{short}|^{B^\pm\to\rho^\pm\gamma}
= |R_{L/S}^{B^\pm\to\rho^\pm\gamma}|
\frac{|V_{ub}V_{ud}|}{|V_{td}V_{tb}|} \simeq 10\% ~.
\label{bpmld}
\end{equation}
Thus, the CKM factors suppress the LD-contributions in
${B^\pm\to\rho^\pm\gamma}$. 

The analogous LD-contributions in the neutral $B$ decays
$B^{0}\to\rho\gamma $ and $B^{0}\to\omega\gamma $ are
expected to be much smaller, a point
that has also been noted in the context of the VMD and quark model
based estimates \cite{bsgamld}. 
 The corresponding form factors for the decays
$B^{0} \to \rho^0(\omega)  \gamma$ are obtained from
the ones for the decay $B^\pm\to\rho^\pm \gamma$ discussed above by the
replacement of the light quark charges
 $e_u\to e_d$, which gives the factor $-1/2$; in addition,
and more importantly, the
LD-contribution to the neutral $B$ decays
is colour-suppressed, which reflects itself
through the replacement of the BSW-coefficient
$a_1$  by $a_2$ \cite{BSW}. This yields for the ratio
\begin{equation}
\frac{R_{L/S}^{B^{0}\to\rho\gamma}}{R_{L/S}^{B^\pm\to\rho^\pm\gamma}}=
\frac{e_d a_2}{e_u a_1} \simeq -0.13 \pm 0.05 ,
\end{equation}
where the numbers are based on using
$a_2/a_1 = 0.27 \pm 0.10$ \cite{BH95}. Thus, in this approach
$R_{L/S}^{B^{0}\to\rho\gamma} \simeq R_{L/S}^{B^{0}\to\omega\gamma}=0.05$,
which in turn gives
\begin{equation}
 \frac{{\cal A}_{long}^{B^{0}\to\rho\gamma}}{{\cal 
A}_{short}^{B^{0}\to\rho\gamma}}\leq 0.02.
\end{equation}
The above estimate, as well as the one in eq.~(\ref{bpmld}), should be 
taken only as indicative in view of the approximations made in 
\cite{wyler95,ab95}. That the LD-effects remain small
in ${B^{0}\to\rho\gamma}$ has been supported in a recent analysis   
based on the soft-scattering of the on-shell hadronic decay products
$B^{0} \to \rho^0 \rho^0 \to \rho \gamma$ \cite{DGP96},
though this paper estimates them somewhat higher (between $4 -8\%$).

The relations, which
follow from the SD-contribution and isospin invariance
\beq\label{ratio2}
\Gamma(B^\pm \to \rho^\pm \gamma)=2 ~\Gamma(B^{0}\to \rho^0  \gamma)
    = 2 ~\Gamma (B^{0} \to \omega  \gamma)~,
\eeq
on including the LD-contributions get modified to
\begin{eqnarray}\label{ratio5}
\lefteqn{\frac{\Gamma(B^\pm\to \rho^\pm\gamma)}{2\Gamma(B^{0}\to \rho\gamma)}
=\frac{\Gamma(B^\pm\to \rho^\pm\gamma)}{2\Gamma(B^{0}\to \omega\gamma)}
 =\left|1+R_{L/S}^{B^\pm\to\rho^\pm\gamma}
\frac{V_{ub}V_{ud}^\ast}{V_{tb}V_{td}^\ast}\right|^2 =
}
\nonumber\\&&{}
=1+2\cdot R_{L/S} V_{ud}\frac{\rho(1-\rho)-\eta^2}{(1-\rho)^2+\eta^2}
+(R_{L/S})^2 V_{ud}^2\frac{\rho^2+\eta^2}{(1-\rho)^2+\eta^2}\,.
\end{eqnarray}
where $R_{L/S}\equiv R_{L/S}^{B^\pm\to\rho^\pm\gamma}$.  
The ratio\\
$\Gamma(B^\pm\to \rho^\pm\gamma)/2\Gamma(B^{0}\to \rho\gamma)
(=\Gamma(B^\pm\to \rho^\pm\gamma)/2\Gamma(B^{0}\to \omega\gamma))$,
estimated to lie in the range $0.7$ - $1.2$ in \cite{ab95}, 
constrains the Wolfenstein parameters
$(\rho, \eta)$, with the dependence on $\rho$ more marked
than on $\eta$.

\par
\begin{figure}[htb]
\vskip -2.4truein
\centerline{\epsfysize=7in
{\epsffile{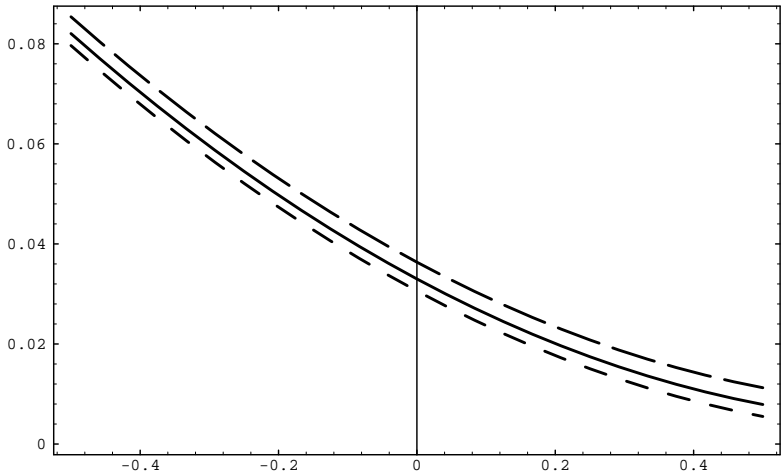}}}
\vskip -1.0truein
\caption[]{
 Ratio of the CKM-suppressed and CKM-allowed radiative $B$-decay
rates\\
$\Gamma (B^\pm \to \rho^\pm \gamma)/\Gamma (B \to K^* \gamma)$ (with 
$B=B^\pm$ or $B^0$) as a function of the Wolfenstein parameter $\rho$,
a) with $\eta =0.2$ (short-dashed curve), $\eta =0.3$ (solid curve), and
$\eta =0.4$ (long-dashed curve). (Figure taken from \protect\cite{ab95}.)
\label{abfig3}}
\end{figure}

The ratio of the CKM-suppressed and CKM-allowed
 decay rates  for charged $B$ mesons
 also gets modified due to the LD contributions. Following \cite{bsgamld},
we ignore the LD-contributions in $\Gamma(B \to K^*\gamma)$. The ratio of
the decay rates in question can therefore be written as:
\begin{eqnarray}\label{ratio3}
\lefteqn{\frac{\Gamma(B^\pm\to \rho^\pm\gamma)}{\Gamma(B^\pm\to 
K^{*\pm}\gamma)} = \kappa_u \lambda^2[(1-\rho)^2+\eta^2]
}
\nonumber\\&&{}
\times\Bigg\{
1+2\cdot R_{L/S} V_{ud}\frac{\rho(1-\rho)-\eta^2}{(1-\rho)^2+\eta^2}
+(R_{L/S})^2 V_{ud}^2\frac{\rho^2+\eta^2}{(1-\rho)^2+\eta^2}\Bigg\}\,,
\end{eqnarray}
 Using the central value from the estimates $\kappa_u=0.59 \pm 0.08$
  \cite{abs93}, we show the ratio (\ref{ratio3}) in Fig.~\ref{abfig3}
 as a function of $\rho$ for $\eta=0.2,0.3$, and $0.4$.
It is seen that the dependence of this ratio is  weak on $\eta$
but it depends on $\rho$ rather sensitively.
The effect of the LD-contributions is modest but not negligible, introducing
an uncertainty  
comparable to the $O( 15\%)$ uncertainty in the overall normalization
due to the $SU(3)$-breaking effects in the quantity $\kappa_u$.
%
%
%

\indent
Neutral $B$-meson radiative decays are less-prone to the LD-effects
as argued above, and hence one expects that to a good approximation
(say, of $O(10\%)$)
the ratio of the decay rates for neutral $B$ meson obtained in the
approximation of SD-dominance remains valid \cite{abs93}:
\begin{equation}
\frac{\Gamma(B^0\to \rho^0\gamma,\omega\gamma)}{\Gamma(B\to K^*\gamma)}
 = \kappa_d\lambda^2 [(1-\rho)^2+\eta^2]~,
\end{equation}
where this relation holds for each of the two decay modes separately.

 Finally, combining the estimates for the LD- and SD-form factors in
\cite{ab95} and
\cite{abs93}, respectively, and restricting the Wolfenstein
parameters in the range $-0.25 \leq \rho \leq 0.35$ and $ 0.2 \leq \eta
\leq 0.4$ from the CKM-fits \cite{AL96}, one gets the
following ranges for the branching ratios:
\begin{eqnarray}\label{ratio4}
{\cal B}(B^\pm\to \rho^\pm\gamma)
&=& (1.5 \pm 1.1) \times 10^{-6} ~,
\nonumber\\
{\cal B}(B^{0}\to \rho\gamma) &\simeq& {\cal B}(B^{0}\to \omega \gamma)
= (0.65 \pm 0.35) \times 10^{-6} ~,
\end{eqnarray}
where we have used the experimental value for the branching ratio
${\cal B} (B \to K^* + \gamma)$
\cite{CLEOrare1},
adding the errors in quadrature. The large error reflects the poor
knowledge of the CKM matrix elements and hence experimental determination
of these branching ratios will put rather stringent constraints on the
Wolfenstein parameter $\rho$.

In addition to studying the radiative penguin decays of the $B^\pm$ and
$B^0$ mesons discussed above, hadron machines such as HERA-B will be in a 
position to study the
corresponding decays of the $B_s^0$ meson and $\Lambda_b$ baryon, such as
$B_s^0 \to \phi + \gamma$ and $\Lambda_b \to \Lambda + \gamma$, which have
not been measured so far. Their estimates can be seen in \cite{Alihera}.

\subsection{Inclusive decays $B \to X_s \ell^+ \ell^-$ in the SM}

\par
The decays \bxslll, with $\ell=e,\mu,\tau$, provide a more sensitive search
strategy for finding new physics in rare $B$ decays
than for example the decay \bxsg , which constrains
 the magnitude of $C_7^{\mathit{eff}}$.
 The sign of $C_7^{\mathit{eff}}$, which
 depends on the underlying physics, is not
determined by the measurement of ${\cal B}(\BGAMAXS)$. This sign, which 
in our convention is negative in the SM, is in general model dependent.
It is known (see for example \cite{AGM94}) that
in supersymmetric (SUSY) models, both the negative and positive signs are 
allowed as one scans over the allowed SUSY parameter space.
We recall that for low dilepton masses, the differential decay 
rate for \bxsll is dominated by the contribution of the virtual photon 
to the charged lepton pair, which in turn  depends on the
effective Wilson coefficient $C_7^{\mathit{eff}}$.
However, as is well known, the \bxsll ~amplitude in the standard model
has two additional terms, arising from the two FCNC four-Fermi operators
 \footnote{This also
holds for a large class of models such as MSSM and the two-Higgs doublet
models but not for all SM-extensions. In LR symmetric models, for example, 
there
are additional FCNC four-Fermi operators involved \cite{LRsymmetry}.},
which are not constrained by the $\BGAMAXS$ data.  
Calling their coefficients $C_{9}$ and $C_{10}$, it has been argued in
\cite{AGM94} that the signs and
magnitudes of all three coefficients $C_7^{\mathit{eff}}$, $C_{9}$ and 
$C_{10}$
can, in principle,  be determined from the decays $\BGAMAXS$ and \bxsll .

\par
 The SM-based rates for the decay \bsll , calculated in the free quark decay
approximation, have been known in the LO approximation for some time
\cite{BSGAM}. The required NLO calculation is in the meanwhile
available, which reduces the scheme-dependence of the LO effects in these
decays \cite{MisiakBM94}. In addition,
long-distance (LD) effects, which are expected to be very important in the
decay \bxsll  \cite{long}, have also been estimated from data
 on the assumption that they arise dominantly due to
the charmonium resonances $ J/\psi$ and $\psi'$
and higher resonances through the decay chains
$B \rightarrow X_s J/\psi (\psi',...) \rightarrow X_s \ell^+ \ell^-$.
Likewise, the  leading $(1/{m_b}^2)$ power corrections
to the partonic decay rate and the dilepton invariant mass distribution
have been calculated with the help of the operator product expansion in the 
effective heavy quark theory \cite{falketalbsll}. The results of 
\cite{falketalbsll} have, however, not been confirmed in a recent calculation
\cite{AHHM96}, which finds that the power corrections in the branching
ratio ${\cal B}(B \to X_s \ell^+ \ell^-)$ are small. Moreover, the
end-point dilepton invariant mass spectrum in this order is not calculable 
in the heavy quark expansion, which requires either resummation
in the context of the HQE approach or a
non-perturbative model.
 We review the salient features of the decay \bxsll here.   
 
The amplitude for \bxsll is calculated in the effective theory
approach, which we have discussed earlier,  by
extending the operator basis of the effective Hamiltonian
introduced in Eq.~(\ref{heffbsg}):
\begin{equation}\label{heffbsll}
{\cal H}_{eff}(b \to s + \gamma ; b \to s + \ell^+\ell^- )
  = {\cal H}_{eff} (b \to s + \gamma) -\frac{4 G_F}{\sqrt{2}} V_{ts}^* V_{tb}
\left[ C_9 {\cal O}_9 +C_{10}{\cal O}_{10} \right],
\end{equation}
where the two additional operators are:
\begin{eqnarray}
{\cal O}_9 &=& \frac{\alpha}{4 \pi} \bar{s}_\alpha \gamma^{\mu} P_L b_\alpha 
\bar{\ell} \gamma_{\mu} \ell , \nonumber\\
{\cal O}_{10} &=& \frac{\alpha}{4 \pi} \bar{s}_\alpha \gamma^{\mu} P_L 
b_\alpha \bar{\ell} \gamma_{\mu}\gamma_5 \ell ~.
\end{eqnarray}

The analytic expressions for $C_{9}(m_W)$ and $C_{10}(m_W)$ can be seen
in \cite{MisiakBM94} and will not be given here.
 We recall that the
coefficient $C_9$ in LO is scheme-dependent. However, this is compensated
by an additional scheme-dependent part in the
(one loop) matrix element of ${\cal O}_9$. We call the
sum  $C_9^{\mathit{eff}}$, which is scheme-independent and enters in the 
physical decay amplitude given below,
\begin{eqnarray}
\lefteqn{{\cal M}(b \to s +\ell^+\ell^-) =
 \frac{4 G_F}{\sqrt{2}} V_{ts}^* V_{tb}\frac{\alpha}{\pi}} \nonumber\\
&\times &\left[ C_9^{\mathit{eff}}\bar{s} \gamma^{\mu} P_L b \bar{\ell} 
\gamma_{\mu} \ell
 +C_{10}\bar{s} \gamma^{\mu} P_L b \bar{\ell} \gamma_{\mu}\gamma_5 \ell
- 2C_7^{\mathit{eff}} \bar{s} i\sigma_{\mu \nu} 
\frac{q^\nu}{q^2}(m_bP_R+m_sP_L)b
\bar{\ell} \gamma^{\mu} \ell \right],\nonumber\\
&& {}
\end{eqnarray}
with
\begin{equation}
C_9^{\mathit{eff}} (\hat{s}) \equiv C_9\eta({\hat{s}}) + Y(\hat{s})~,
\end{equation}
where $\hat{s}=q^2/m_b^2$.
The function $Y(\hat{s})$ is the one-loop matrix element of ${\cal O}_9$
and can be seen in literature \cite{ALI96,MisiakBM94}.
The dilepton invariant mass distribution in \bxsll (ignoring lepton masses) 
is,
 \begin{eqnarray}
	{{\rm d}{\cal B}(\hat{s}) \over {\rm d}\hat{s}} & = &
		{\cal B}_{sl} \frac{\alpha^2}{4 \pi^2} \frac{ 
		\lambda_t^2}{\Vcbabs^2} \frac{1}{f(\hat{m}_c) \kappa(\hat{m}_c)}
		u (\hat{s}) \left[ \vphantom{\frac{1}{1}}
		\left( |C_9^{\mathit{eff}}(\hat{s})|^2
 		+ C_{10}^2 \right) \alpha_1 (\hat{s},\hat{m}_s)
		\right. \nonumber \\
& & \left. + \frac{4}{\hat{s}} (C^{eff}_7)^2 \alpha_2 (\hat{s},\hat{m}_s)
+ 12 \alpha_3 (\hat{s},\hat{m}_s) C^{eff}_7 {\cal 
\Re}(C_9^{\mathit{eff}}(\hat{s}))
		\right] ,
	\label{eqn:dbrs}
\end{eqnarray}
with $u(\hat{s})=\sqrt{\left[\hat{s}-(1+\hat{m_s})^2\right]
\left[\hat{s}-(1-\hat{m_s})^2 \right]} $,
and $f(\mc)$ has been given earlier; likewise 
$\kappa(\mc)=1-2\as(\mu)/3\pi \left[(\pi^2-31/4)(1-\mc)^2 + 3/2 \right]$
is the same as the corresponding function in $B \to X_c \ell \nu_\ell$, 
and the rest of the functions are defined as
 \begin{eqnarray}
	\alpha_1 (\hat{s},\hat{m}_s) & = &
		- 2 \hat{s}^2 + \hat{s} (1+ \hat{m}_s^2)
       		+(1-\hat{m}_s^2)^2 ,
		\label{eqn:alpha1} \\
	\alpha_2 (\hat{s},\hat{m}_s) & = &
		-(1+ \hat{m}_s^2) \hat{s}^2
           	- (1+14 \hat{m}_s^2+\hat{m}_s^4) \hat{s}
            	+ 2 (1+ \hat{m}_s^2)(1-\hat{m}_s^2)^2 ,
		\label{eqn:alpha2} \\
	\alpha_3 (\hat{s},\hat{m}_s) & = &
		(1-\hat{m}_s^2)^2 - (1+ \hat{m}_s^2) \hat{s}. 
		\label{eqn:alpha3}
\end{eqnarray}
Here $\hat{m_i}=m_i/m_b$ and
 ${\cal \Re}(C_7^{\mathit{eff}})$ represents the real part of 
$C_7^{\mathit{eff}}$.
A useful quantity is the  differential FB asymmetry in the c.m.s. of the
dilepton
defined in refs. \cite{amm91}:
\begin{equation}\label{FBasym}
\frac{d {\cal A}(\hat{s})}{d\hat{s}} = \int_0^1 \frac{d{\cal B}}{dz}
                                      -\int_0^{-1} \frac{d{\cal B}}{dz},
\end{equation}
where $z=\cos \theta$, with $\theta$ being the angle between the
$\ell^+$ direction and the $b$-quark direction in this system, 
 which can be expressed as:
 \begin{eqnarray}
	{{\rm d}{\cal A}(\hat{s}) \over {\rm d}\hat{s}} & = &
	- {\cal B}_{sl} \frac{3 \alpha^2}{4 \pi^2}
        \frac{1}{f(\hat{m}_c)} u^2 (\hat{s})
	C_{10} \left[ \hat{s}{\cal \Re} ( C_9^{\mathit{eff}}(\hat{s})) +
	2 C^{eff}_7 (1 + \hat{m}_s^2) \right] .
	\label{eqn:dasym}
\end{eqnarray}
 The Wilson coefficients
$C^{eff}_7$, $C^{eff}_9$ and $C_{10}$ appearing in the above equations
can be determined from data by solving the partial branching ratio
${\cal B}(\Delta \hat{s})$ and partial FB asymmetry
${\cal A}(\Delta \hat{s})$, where $\Delta \hat{s}$ defines an
interval in the dilepton invariant mass \cite{AGM94}.
 A third quantity, called energy asymmetry,
proposed by Cho, Misiak and Wyler \cite{CMW96}, defined as
\begin{equation}
{\cal A}=\frac{N(E_{\ell^-} > E_{\ell^+}) - N(E_{\ell^+} > E_{\ell^-})}
              {N(E_{\ell^-} > E_{\ell^+}) + N(E_{\ell^+} > E_{\ell^-})}~,
\end{equation}
where $N(E_{\ell^-} > E_{\ell^+})$ denotes the number of lepton pairs
where $\ell^+$ is more energetic than $\ell^-$ in the $B$-rest frame,
is directly proportional to the FB asymmetry discussed above. The relation
is \cite{AHHM96}:
\begin{equation}
\int {\cal A}(\hat{s})= {\cal B} \times A~.
\end{equation}

Taking into account the spread in the values of the input parameters,
$\mu, ~\Lambda, ~\mt$, and ${\cal B}_{SL}$
discussed in the previous section in the context of ${\cal B}(B \to X_s +
\gamma)$, following branching ratios for the SD-piece
(i.e., from the intermediate top quark contribution only)
have been estimated in \cite{AHHM96}:
 \begin{eqnarray}\label{brbsll}
{\cal B}(\bxsee) &=& (8.4 \pm 2.3) \times 10^{-6}, \nonumber\\
{\cal B}(\bxsmm) &=& (5.7 \pm 1.2) \times 10^{-6}, \nonumber\\
{\cal B}(\bxstt) &=& (2.6 \pm 0.5) \times 10^{-7}.
\end{eqnarray}
A good fraction of this uncertainty is contributed by the assumed
($\pm 9 $ GeV) uncertainty on $m_t$, reflecting this sensitivity as
first pointed out in \cite{HWS87}.
 The present experimental limit for the inclusive branching ratio in
\bxsll is actually still the one set by the UA1 collaboration some time
ago \cite{UA1R}, namely ${\cal B}(\bxsmm) > 5.0 \times 10^{-5}$. As far
as we know, there are no interesting limits on the other two modes,
involving $X_s e^+e^-$ and $X_s \tau^+ \tau^-$.

   The leading power corrections in $1/\mb$ in the Dalitz distribution
for the decay $B\to X_s \ell^+\ell^-$, and the resulting
dilepton invariant mass distribution and FB-asymmetry have been 
calculated in \cite{AHHM96} and discussed in detail including comparison
with the earlier calculation of the dilepton mass given in 
\cite{falketalbsll}.
We give here the simpler expressions  in the limit $m_s=0$.
 For the dilepton invariant mass distribution, the result is 
\cite{AHHM96}:
   \begin{eqnarray}
        \frac{{\rm d}{\cal B}}{{\rm d}\s} & = & 2 \; {\cal B}_0
                \left\{
                  \left[
                \frac{1}{3} (1-\s)^2 (1+2 \s) \; (2 + \lo)
                + ( 1 - 15  \s^2 + 10 \s^3\right) \lt
                        \right]
                \left( |C_9^{\mbox{eff}} |^2 + |C_{10}|^2 \right)   
                \nonumber \\
        & &
             +  \left[
                 \frac{4}{3} (1-\s)^2 (2+ \s) \; (2 + \lo)
                + 4  \left( -6 -3  \s + 5 \s^3 \right) \lt
                \right] \frac{|C_7^{\mbox{eff}}|^2}{\s}
                \nonumber \\
        & &     \left.
           +    \left[
                4 (1-\s)^2 (2+ \lo)
               + 4  \left( -5 -6  \s + 7 \s^2 \right) \lt
                                \right] Re(C_9^{\mbox{eff}}) \, 
C_7^{\mbox{eff}}
                \right\}\, .
\label{eqn:dbds0}
\end{eqnarray}  

 The (unnormalized) FB asymmetry reads as,
\begin{eqnarray}
        \frac{{\rm d}{\cal A}}{{\rm d}\s} & = &
                - 2 \; {\cal B}_0 
                \left\{
                \left[   
                2 (1 - \s)^2 \s
                + \frac{\s}{3} (3+2 \s +3 \s^2) \lo
                +  \s \, (-9 -14 \s + 15 \s^2) \, \lt \right]
                         \, Re(C_9^{\mbox{eff}} ) \, C_{10}
                \right.
                \nonumber \\
        & &     \left.
                + \left[ 4 (1-\s)^2
                + \frac{2}{3} (3+ 2 \s + 3\s^2) \lo
                + 2 (-7 -10 \s + 9 \s^2)\, \lt \right] \, Re(C_{10}) \, 
C_7^{\mbox{eff}}
                 \,  \right\}\, .
\end{eqnarray}
 
The normalization constant ${\cal B}_0$ now includes also the power 
corrections in $\Gamma_{\small SL}$,
 \begin{equation}
        {\cal B}_0 \equiv
                {\cal B}_{\small SL} \frac{3 \, \alpha^2}{16 \pi^2} 
\frac{
    {\vert V_{ts}^* V_{tb}\vert}^2}{\Vcbabs^2} \frac{1}{f(\mc) 
[\kappa(\mc)+ h(\mc)/2m_b^2]}
                \, ,
\label{eqn:seminorm}
\end{equation}
where the functions  $f(\mc)$ and $\kappa(\mc)$ have been defined earlier,
and the function $h(\mc)$ is defined as:
\be
h(\mc) = \lambda_1 + \frac{\lambda_2}{f(\mc)} \left[ -9 +24 
\mc^2
-72\mc^4 + 72\mc^6 -15\mc^8 -72 \mc^4 \ln \mc \right]\, .
\label{eqn:ghr}
\ee

 Doing the integration, one derives the (leading) power corrected 
branching ratio for $B \to X_s \ell^+ \ell^-$. 
The decay width itself may be written
in the numerical form \cite{AHHM96}:
\begin{equation}
\Gamma^{\mbox{\small HQE}}=\Gamma^{b} (1 + C_1\hat{\lambda}_1 + C_2
\hat{\lambda}_2 )\, ,
\label{gammahqet}     
 \end{equation}
where $\Gamma^{b}$ is the parton model decay width for \bsll and the
coefficients have the values
$$C_1 =0.50 ~\mbox{and} ~C_2 = -7.425~.$$
 This leads to a reduction in the
power-corrected decay
width by $-4.1\%$, using  $\lambda_1=-0.2$ GeV$^2$ and 
$\lambda_2 =0.12$ GeV$^2$.  Moreover, this reduction is mostly contributed by
the $\lambda_2$-dependent term. We recall that the coefficient of the
$\hat{\lambda}_1$ term above is universal, i.e., it is
 the same as in the inclusive widths
$\Gamma(B \to X_u \ell \nu_\ell)$ and  $\Gamma(\BGAMAXS)$,
but the coefficient of the $\hat{\lambda}_2$ term above is larger than
the corresponding coefficient $(=-9/2)$ in the semileptonic decay width.
Hence,  power corrections in $\Gamma(B \to X_u \ell \nu_\ell)$ and
$\Gamma(\mbox{\bxslll})$ are rather similar but not identical. Finally,
the effect of the power corrections leads to a reduction of about $1.5\%$
in the branching ratio ${\cal B}(B \to X_s \ell^+ \ell^-)$.

\begin{figure}[htb]
\vskip -2.0truein
\centerline{\epsfysize=7in
{\epsffile{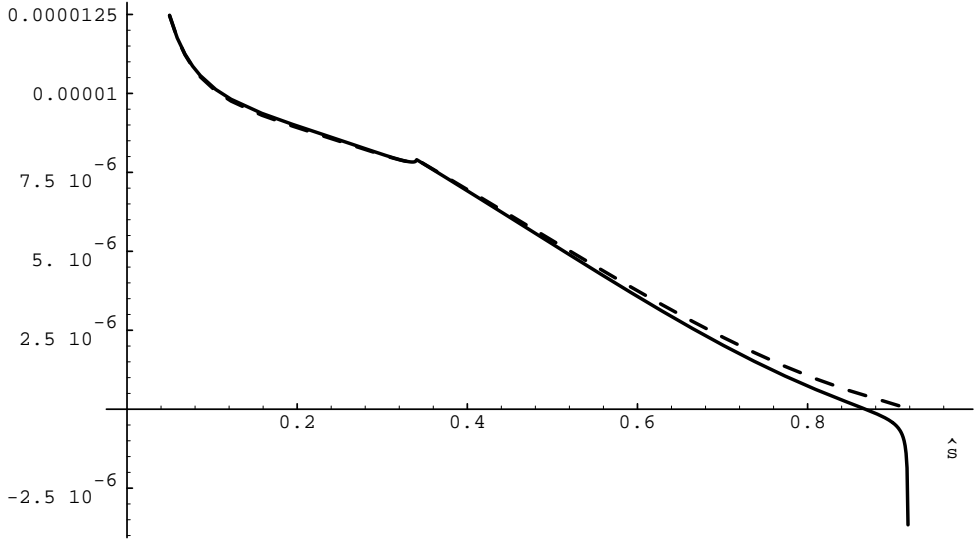}}}
\vskip -2.0truein
\caption[]{Dilepton invariant mass spectrum ${\rm d}{\cal B} (B \to X_s e^+
e^-)/{\rm d} \hat{s}$ in the parton model (dashed curve) and with
leading   
power corrections calculated in the 
HQE approach (solid curve). (Figure taken from \protect\cite{AHHM96}).} 
\label{fig:hillerfig1} 
\end{figure} 
Concerning power corrections to the dilepton invariant mass distribution 
and the FB asymmetry, we would like to make the following observations:
\begin{itemize}
\item The results given above in the HQE approach reproduce the 
parton model expressions for the dilepton invariant mass distribution
and FB asymmetry in the
limit $\lambda_1 \to 0$ and $\lambda_2 \to 0$.
\item The power-corrected dilepton invariant mass
distribution given above retains the characteristic
$1/\hat{s}$ behaviour following from the one-photon exchange
in the parton model.
\item Power corrections in the dilepton mass distribution are
found to be small over a good part of the dilepton mass $\s$.
However, these corrections become increasingly large
and negative as one approaches $\hat{s} \to \hat{s}^{max}$.
 Since the parton model spectrum falls steeply near the
end-point $\hat{s} \to \hat{s}^{max}$, this leads to the
uncomfortable result that the power corrected
dilepton mass distribution becomes negative for the high dilepton masses. 
 We show in
Fig.~\ref{fig:hillerfig1} this distribution in the parton model and the
HQE approach.
\item The normalized FB asymmetry, $d\bar{\cal A}(\s)/d\s$ is stable against
leading order power corrections up to $\s \leq 0.6$, but the corrections
become increasingly large and eventually uncontrollable due to the
unphysical
behaviour of the HQE-based dilepton mass distribution as $\s$ approaches
$\s^{max}$ \cite{AHHM96}.
\end{itemize}
Based on these investigations, we must conclude that the HQE-based 
approach  
has a  restrictive kinematic domain for its validity. In particular,
it breaks down for the high dilepton invariant mass region in \bxsll.
This behaviour is very similar to what has been observed earlier in the
context of the photon energy spectrum in the decay $\BGAMAXS$.  

As an alternative to the heavy quark expansion,
  the non-perturbative effects due to the $B$-hadron wave function on the
decay distributions in \bxsll have been evaluated in \cite{AHHM96} using the 
familiar Fermi motion model.
 The dilepton invariant mass distribution is found to be stable
against such effects over most part of the dilepton mass. However, the
end-point spectrum is sensitive to the model parameters. The FB asymmetry
turns out to be more sensitive to the model parameters.
Further details on these points can be seen in \cite{AHHM96}.
            
\begin{figure}[htb]
\vskip -0.5truein
\centerline{\psfig{figure=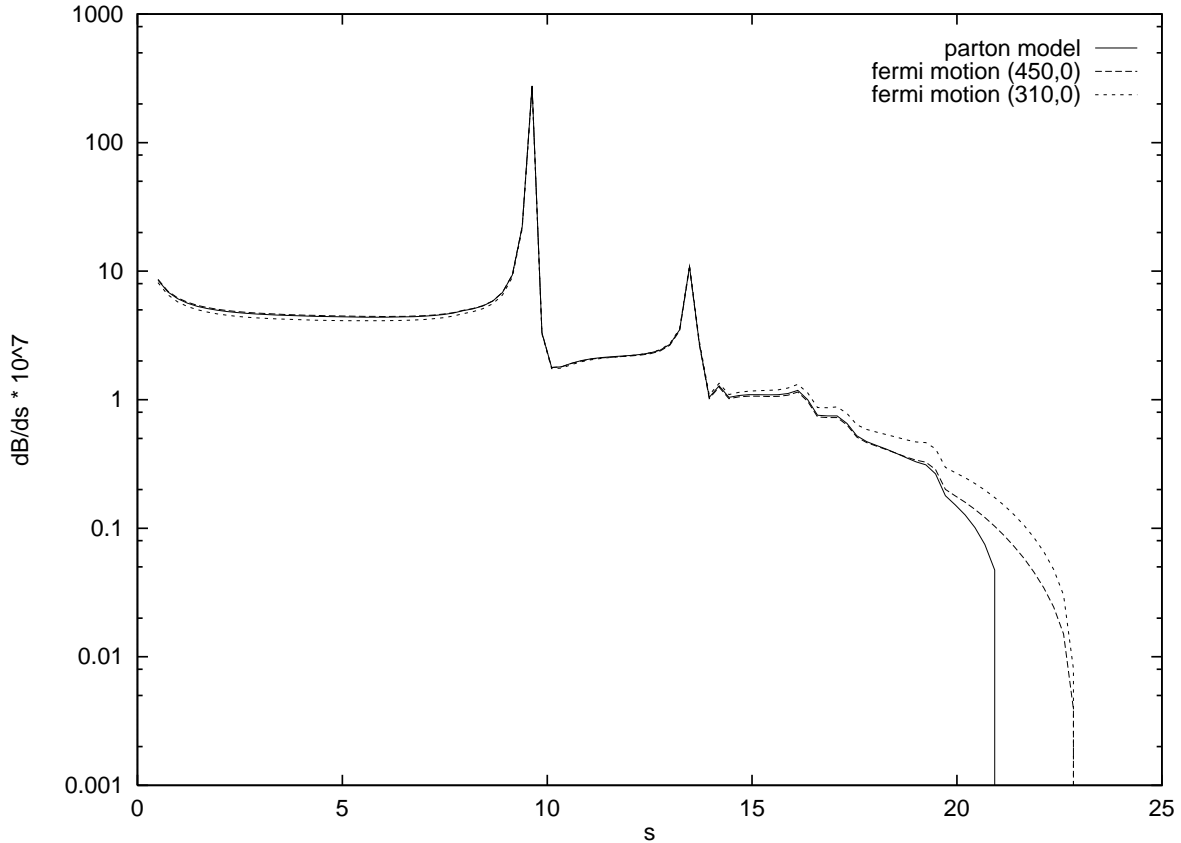,height=16.0cm,angle=270}}
\vskip -0.1truein
\caption[]{
Dilepton invariant mass distribution in $B \to X_s \ell^+ \ell^-$ in
 the SM including
next-to-leading order QCD correction and LD effects. The solid curve
corresponds to the parton model and the short-dashed and long-dashed
curves correspond to including the Fermi motion effects. The
values of the Fermi motion model are indicated in the figure.
(Figure taken from \protect\cite{AHHM96}).}
\label{fig:dbrnsm}
\end{figure}

\begin{figure}[htb]
\vskip -0.5truein
\centerline{\psfig{figure=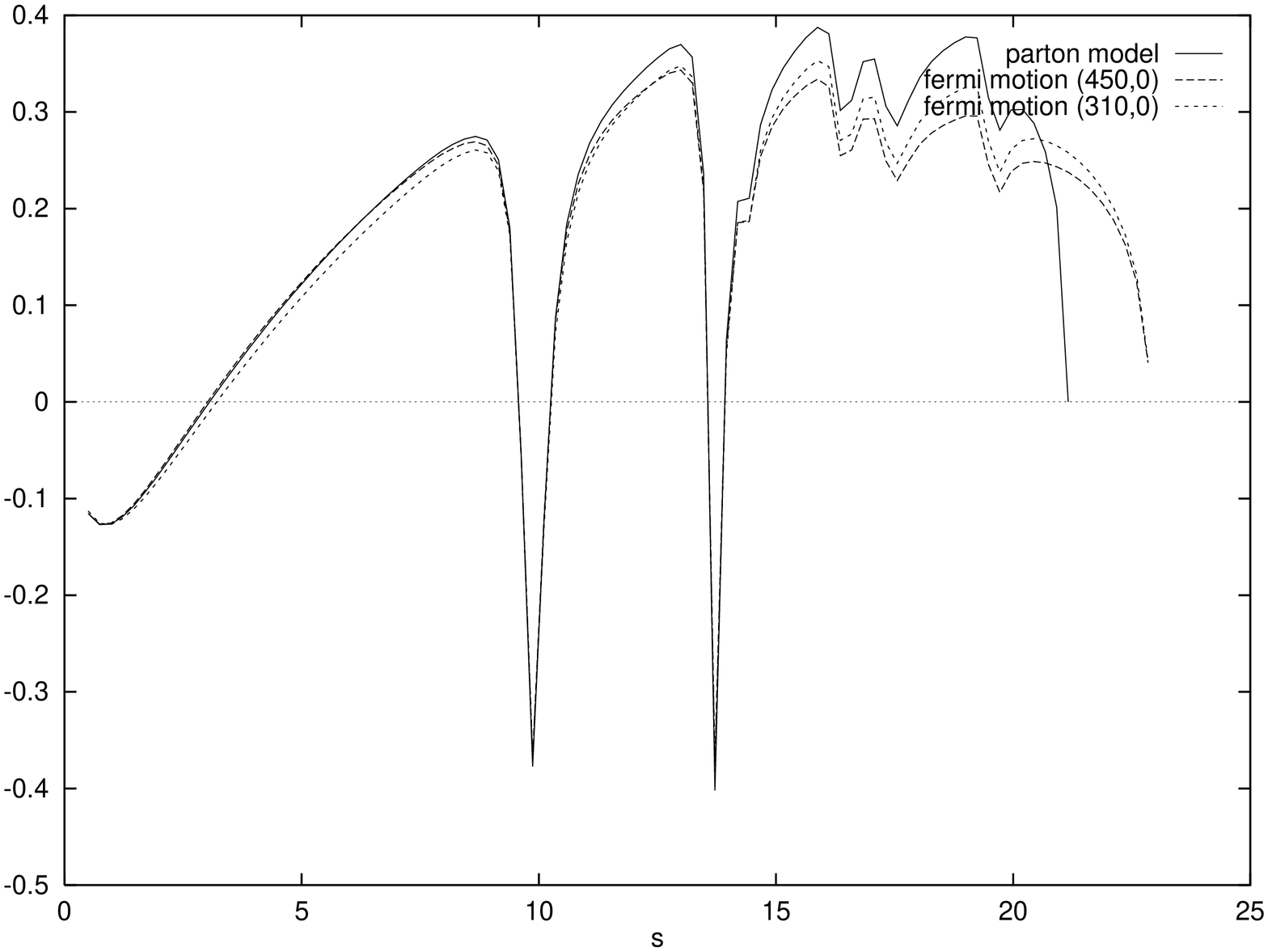,height=16.0cm,angle=270}}
\vskip -0.1truein
\caption[]{FB asymmetry for $B \to X_s \ell^+ \ell^-$ in the SM
as a function of the dimuon invariant mass  including the
                next-to-leading order QCD correction and LD effects.
 The solid curve
corresponds to the parton model and the short-dashed and long-dashed
curves correspond to including the Fermi motion effects. The
values of the Fermi motion model are indicated in the figure.
(Figure taken from \protect\cite{AHHM96}).}
\label{fig:asymmnsm}
\end{figure}

 Concerning the LD effects in  $B \to X_s \ell^+ \ell^-$, it is worth 
noting that such
 contributions ( due to the vector mesons such as $J/\psi$ and 
$\psi^\prime$ as well as the continuum $c\bar{c}$ contribution already
discussed) 
appear as an effective $(\bar{s}_L \gamma_\mu b_L)(\bar{\ell} \gamma^\mu 
\ell)$ interaction term only, i.e. in the operator ${\cal O}_9$.
 This implies that the LD-contributions should change
$C_9$ effectively. The LD-contribution from the 
matrix element of the four-quark operators
${\cal O}_1$ and ${\cal O}_2$ discussed earlier in the context of the
decay $\BGAMAXS$ can be absorbed in $C_7$. However, as we have discussed
earlier, to a good approximation 
$C_7$ is dominated by the SD-contribution. Finally,  
$C_{10}$ has no LD-contribution. In accordance with this, 
the function $Y(\hat{s})$ is replaced by,
\begin{equation}
	Y(\hat{s}) \rightarrow Y^\prime(\hat{s}) \equiv Y(\hat{s}) + 
		Y_{\mbox{res}}(\hat{s}),
\end{equation}
where $Y_{\mbox{res}}(\hat{s})$ is given as \cite{amm91},
\begin{equation}
	Y_{\mbox{res}}(\hat{s}) = \frac{3}{\alpha^2} \kappa 
		\left(3 C_1 + C_2 + 3 C_3 + C_4 + 3 C_5 + C_6 \right)
		\sum_{V_i = J/\psi, \psi^\prime,...}
		\frac{\pi \Gamma(V_i \rightarrow l^+ l^-) M_{V_i}}{
		M_{V_i}^2 - \hat{s} m_b^2 - i M_{V_i} \Gamma_{V_i}} ,
\end{equation}
where $\kappa$ is a fudge factor, which appears due to the inadequacy
of the factorization framework in describing data on $B \to J/\psi X_s$.
With 
$$\kappa \left( 3 C_1 + C_2 + 3 C_3 + C_4 + 3 C_5 + C_6 \right) = +0.88~, $$ 
one reproduces (in average) the measured 
branching ratios for  $B \to J/\psi 
X_s$ and $B \to \psi' X_s$, after the contributions from the $\chi_c$ states
have been subtracted.
 The long-distance effects lead to significant interference effects
in the  dilepton invariant mass
distribution and the FB asymmetry in \bxsll shown in Figs. \ref{fig:dbrnsm}
and \ref{fig:asymmnsm}, respectively. This can be used to
test the SM, as the signs of the Wilson coefficients in
general are model dependent.
%
%
%

It is obvious from Fig.~\ref{fig:dbrnsm} that   
only in the dilepton mass region far away from
the resonances is there a hope of extracting the Wilson coefficients
governing the short-distance physics. The region below the $J/\psi$ resonance
is well suited for that purpose as the dilepton invariant 
mass distribution here is dominated by the SD-piece.
Including the LD-contributions, following branching ratio has been
estimated for the dilepton mass range
 $2.1 ~\mbox{GeV}^2 \leq s \leq 2.9 ~\mbox{GeV}^2$ in 
\cite{AHHM96}:
\be \label{brbslld}
{\cal B}(\bxsmm) = (1.3 \pm 0.3) \times 10^{-6}, 
\ee
with ${\cal B}(\bxsee) \simeq {\cal B}(\bxsmm)$. The FB-asymmetry is 
estimated to be 
in the range $10\%$ - $27\%$, as can be seen in Fig.~\ref{fig:asymmnsm}. 
These branching ratios and the FB asymmetry
are expected to be measured within the next several years at
the forthcoming experiments.
 In the high invariant mass region, the short-distance contribution 
dominates. However, the rates are down by roughly an order of magnitude
compared to the region below the $J/\psi$-mass. Estimates of the
branching ratios are of $O(10^{-7})$, which should be accessible at
the LHC. For further suggestions concerning polarization-dependent
asymmetries in \bxsll , we refer to \cite{Hewettpol,KS96}.

  While still on the subject of the inclusive decays \bxslll, we note
that both the dilepton mass distribution and the FB asymmetry are sensitive 
to non-SM effects. The case of SUSY was first studied in the classic paper
on this subject by Bertolini et al. \cite{Masieroetal}. Since then, more 
detailed studies have been reported in
the literature \cite{AGM94,CMW96}. For a recent update of the 
SUSY effects in \bxslll,
we refer to \cite{Gotoetal96} in which it is shown  that the distributions in
this decay may be distorted significantly above the SM-related uncertainties,
even if the inclusive decay rates are not significantly
effected. Based on this and earlier studies along the same lines,
it is conceivable that the FCNC decay \bxsll may open a window on new 
physics. This scenario is still a possibility even after the LEP anomaly in 
$Z^0 \to b\bar{b}$ decay has largely disappeared. The point is that flavour
conserving and flavour changing neutral currents have very different
underlying contributions both in the SM and extensions of it. Hence, 
our experimental colleagues are well advised to measure
the FCNC decay \bxsll as precisely as possible. This, and the related
exclusive decays, may turn out to be the first glance on beyond-the-SM
landscape!

\subsection{Summary and overview of rare $B$ decays in the SM}

\par
 The rare $B$ decay mode $B \to X_s \nu \bar{\nu}$, and some of the
exclusive channels associated with it,
 have comparatively larger branching ratios. The estimated inclusive
branching ratio in the SM is \cite{AGM92,Grossman}:
\begin{equation}\label{bxsnunu}
 {\cal B}(B \to X_s \nu \bar{\nu}) = (4.0 \pm 1.0) \times 10^{-5}~,
\end{equation}
where the main uncertainty in the rates is due
to the top quark mass. The scale-dependence,
which enters indirectly through the top quark mass, has
been brought under control through the NLL corrections, calculated in
\cite{BuBu93}. The corresponding CKM-suppressed decay $B \to X_d \nu
\bar{\nu}$ is related by the ratio of the CKM matrix element
squared \cite{AGM92}:
\begin{equation}\label{bxsdnunu}
 \frac{{\cal B}(B \to X_d \nu \bar{\nu})}
  {{\cal B}(B \to X_s \nu \bar{\nu})} = \left[
\frac{\Vtdabs}{\Vtsabs}\right]^2 ~.
 \end{equation}
Similar relations hold for the ratios of the exclusive decay rates which
 depend additionally on the ratios of the form factors squared,
which deviate  from unity through $SU(3)$-breaking terms, in close  
analogy with the exclusive radiative decays discussed earlier.
 These decays are particularly attractive  probes of
the short-distance physics, as  the long-distance
contributions are practically absent in such decays. Hence, relations
such as the one in (\ref{bxsdnunu}) provide, in principle, one of
the best methods for the
 determination of the CKM matrix element ratio $\Vtdabs/\Vtsabs$
\cite{AGM92}. From the practical point of view, however, these decay
modes are rather difficult to measure, in particular
at the hadron colliders and probably also at the $B$ factories. The
best chances are in the $Z^0$-decays at LEP,
from where the present
best upper limit stems \cite{ALEPHwarsaw}:
\begin{equation}\label{bsnunulim}
 {\cal B}(B \to X \nu \bar{\nu}) < 7.7 \times 10^{-4}.
\end{equation}
 
Some other rare $B$ decays such as $(B^0,B_s^0) \to \ell^+ \ell^-$ 
and $(B^0,B_s^0) \to \gamma \gamma$
have been recently updated in \cite{ALI96} and \cite{Buraswarsaw}, to
which we refer for details and references to the original literature.
%
%
\section{An Update of the CKM Matrix}
 The present knowledge of the magnitude of all nine CKM matrix elements is 
summarized in  Table \ref{tabckm}. We have discussed 
the experiments and theory which underlie five of the nine matrix elements 
listed there; these are the ones in which $b$ quarks are involved.
The remaining four are taken from the PDG review\cite{PDG96} from where 
references to the original literature can also be got. Note that the 
value given in this table for $\Vtbabs$ is from the direct CDF measurements
\cite{CDFvtb} and not the one following from the CKM unitarity which
gives $\Vtbabs = 0.9991 \pm 0.0004$ \cite{PDG96}.
\begin{table}
\hfil
\vbox{\offinterlineskip
\halign{&\vrule#&
   \strut\quad#\hfil\quad\cr
\noalign{\hrule}
height2pt&\omit&&\omit&\cr
& $\vert V_{ij} \vert$ && Present Value & & \cr
height2pt&\omit&&\omit&\cr
\noalign{\hrule}
height2pt&\omit&&\omit&\cr
& $\Vudabs$ && $ 0.9744 \pm 0.0010 ~\cite{PDG96}$ & \cr
& $\Vusabs$ && $ 0.2205 \pm 0.0011 ~\cite{PDG96}$ & \cr
& $\Vubabs$ && $ (3.1 \pm 0.8) \times 10^{-3} ~\cite{Gibbons96}$ & \cr
& $\Vcdabs$ && $ 0.204 \pm 0.017 ~\cite{PDG96}$ & \cr
& $\Vcsabs$ && $ 1.01\pm 0.18 ~\cite{PDG96}$ & \cr   
& $\Vcbabs$ && $ 0.0393 \pm 0.0028 ~\cite{Gibbons96}$ & \cr
& $\Vtdabs$ && $ (9.2 \pm 3.0) \times 10^{-3} \cite{AL96}$ & \cr
& $\Vtsabs$ && $ 0.033 \pm 0.009 $ \cite{ALI96}& \cr
& $\Vtbabs$ && $ 0.97 \pm 0.22  ~\cite{CDFvtb}$ & \cr
height2pt&\omit&&\omit&\cr
\noalign{\hrule}}}
\caption{Present values of the CKM matrix elements $\vert V_{ij} \vert$
 discussed in the text and in the PDG review
\protect\cite{PDG96}.} \label{tabckm}
\end{table}
Having updated the CKM matrix elements, we now discuss the present 
profile of the CKM unitarity triangle
which is obtained by constraining the apex of this triangle
given by the co-ordinates $(\rho,\eta)$ (see Fig.~\ref{triangle}) 
in the Wolfenstein parametrization \cite{Wolfenstein}.

Of the four parameters, $\lambda$, $A$, $\rho$ and $\eta$, the first two are:
\bea
\lambda &=& 0.2205\pm 0.0018~, \nonumber\\
A &=& 0.80 \pm 0.075~.
\label{Avalue}
\eea
The other two CKM parameters $\rho$ and $\eta$ are constrained by the
measurements of $\vert V_{ub}/V_{cb}\vert$, 
$\abseps$ (the CP-violating
parameter in the kaon system), $\xd$ (\bdbdbar\ mixing) and (in principle)
$\epsilon^\prime/\epsilon$ ($\Delta S=1$ CP-violation in the kaon system).
The constraints from $\epsilon^\prime/\epsilon$ are not included due
to the various experimental and theoretical uncertainties surrounding it at
present \cite{Buraswarsaw}.

 The experimental value of $\abseps$ is \cite{PDG96}:
\beq
\abseps = (2.280\pm 0.013)\times 10^{-3}~.
\eeq
Theoretically, $\abseps$ is essentially proportional to the imaginary part
of the box diagram for \kkbar\ mixing and is given by \cite{Burasetal}
\begin{eqnarray}
\abseps &=& C_{\abseps}
\hat{B}_K\left(A^2\lambda^6\eta\right)
\bigl(y_c\left\{\hat{\eta}_{ct}f_3(y_c,y_t)-\hat{\eta}_{cc}\right\}
 \nonumber \\
&~& ~~~~~~~~~+ 
~\hat{\eta}_{tt}y_tf_2(y_t)A^2\lambda^4(1-\rho)\bigr), 
\label{eps}
\end{eqnarray}
where $C_{\abseps} = G_F^2f_K^2M_KM_W^2/(6\sqrt{2}\pi^2\Delta M_K)$, 
 $y_i\equiv m_i^2/M_W^2$, and the functions $f_2$ and $f_3$ can be
found in Ref.~\cite{AL94}. Here, the $\hat{\eta}_i$ are QCD correction
factors, calculated at next-to-leading order in \cite{HN94}
($\hat{\eta}_{cc}$), \cite{etaB} ($\hat{\eta}_{tt}$) and \cite{HN95}
($\hat{\eta}_{ct}$). The theoretical uncertainty in the expression for 
$\abseps$ is in the renormalization-scale independent parameter
$\hat{B}_K$, which represents our ignorance of the hadronic matrix
element $\langle K^0 \vert {({\overline{d}}\gamma^\mu (1-\gamma_5)s)}^2
\vert {\overline{K^0}}\rangle$. Some recent calculations of $\hat{B}_K$
using lattice QCD methods \cite{Soni95} and the $1/N_c$ approach
\cite{BP95} are: $\hat{B}_K=0.83 \pm 0.03$ (Sharpe \cite{Sharpe94}),
$\hat{B}_K=0.86 \pm 0.15$ (APE Collaboration \cite{Crisafulli95}),
$\hat{B}_K=0.67 \pm 0.07$ (JLQCD Collaboration \cite{JLQCD}), 
$\hat{B}_K=0.78 \pm 0.11$ (Bernard and Soni \cite{JLQCD}), and
$\hat{B}_K=0.70 \pm 0.10$ (Bijnens and Prades \cite{BP95}).
The calculations given above are compatible with the range
\begin{equation}
 \hat{B}_K=0.75 \pm 0.10 ,
\label{BKrange}
\end{equation}
which has been used in the CKM analysis in \cite{AL96}.
 The present world average for $\Delta M_d$ is \cite{Gibbons96}
\beq
\Delta M_d = 0.464 \pm 0.018~(ps)^{-1} ~.
\label{deltamd}
\eeq
The mass difference $\Delta M_d$ is calculated from the \bdbdbar\ box
diagrams. Unlike the kaon system, where the contributions of both the $c$-
and the $t$-quarks in the loop are important, both $\delmd$ and
$\delms$ are dominated by $t$-quark exchange:
\begin{equation}
\Delta M_d = C_d\hat{\eta}_B y_t
f_2(y_t) \vert V_{td}^*V_{tb}\vert^2~,
\label{xd}
\end{equation}
where $C_d=G_F^2/(6\pi^2)M_W^2M_B(\fbb)$, 
 $\vert V_{td}^*V_{tb}\vert^2= A^2\lambda^{6}
[\left(1-\rho\right)^2+\eta^2]$. Here, $\hat{\eta}_B$ is the QCD
correction. In Ref.~\cite{etaB}, this correction was analyzed including the
effects of a heavy $t$-quark. It was found that $\hat{\eta}_B$ depends
sensitively on the definition of the $t$-quark mass, and that, strictly
speaking, only the product $\hat{\eta}_B(y_t)f_2(y_t)$ is free of this
dependence. In the fits presented here we use the value
$\hat{\eta}_B=0.55$, calculated in the $\overline{MS}$ scheme, following
Ref.~\cite{etaB}. Consistency requires that the top quark mass be rescaled
from its pole (mass) value of $\mt =175 \pm 9$ GeV to the value
$\overline{\mt}(\mt(pole))$ in the $\overline{MS}$ scheme, given above.

For the $B$ system, the hadronic uncertainty is given by $\fbb$, analogous
to $\hat{B}_K$ in the kaon system, except that in this case $\fbd$ has not
been measured. The present status of the lattice-QCD estimates for $\fbd$,
$\hat{B}_{B_d}$ and related quantities for the $B_s$ meson, obtained in the
quenched (now usually termed as the valence) approximation was summarized
 in \cite{Wittig96}, giving
\begin{eqnarray}
 \fbd &=& 170 ^{+55}_{-50} ~\mbox{MeV}, \nonumber\\
 \hat{B}_{B_d} &=& 1.02 ^{+0.05 ~~+0.03}_{-0.06 ~~-0.02},
\end{eqnarray}
where the first error on $\hat{B}_{B_d}$ is statistical and the second 
systematic, estimated by the UKQCD collaboration \cite{UKQCDBB}.
This compares well with a recent calculation by Gim\'enez and Martinelli,
obtaining $\hat{B}_{B_d} = 1.08 \pm 0.06 \pm 0.08$ \cite{Gimenez96}. 
 A modern estimate of $\fbb$ in the
QCD sum rule approach is that given in \cite{narison95}, which is
stated in terms of $f_\pi$, and on using $f_\pi =132$ MeV translates into
\begin{equation}
\fbd \sqrt{\hat{B}_{B_d}} = 197 \pm 18 ~\mbox{MeV}~.
\label{FBrangesr}
\end{equation} 
The CKM fits being presented use 
\begin{equation}
\fbd \sqrt{\hat{B}_{B_d}} = 200 \pm 40 ~\mbox{MeV}~,
\label{FBrange}
\end{equation}
which is compatible with the results from both lattice-QCD and QCD sum
rules for this quantity. The present experimental input
can be summarized as \cite{AL96}:
\begin{eqnarray}
\sqrt{\rho^2+\eta^2} &=& 0.363 \pm 0.073 ~~\mbox{(from
$|V_{ub}/V_{cb}|=0.08 \pm 20\%$),} \nonumber \\
(\fbd \sqrt{\hat{B}_{B_d}}/\mbox{1 GeV}) \sqrt{(1-\rho)^2 + \eta^2} &=&
0.202 \pm 0.017 ~~\mbox{(from $\delmd=0.464 \pm 0.018 ~(ps)^{-1}$),}
\nonumber \\
\hat{B}_K \eta [ 0.93 + (2.08 \pm 0.34) (1-\rho)] &=& 0.79 \pm 0.11
~\mbox{(from $\abseps=(2.280 \pm 0.013)\times 10^{-3}$).}
\label{ckmfiteqns}
\end{eqnarray}
The errors of the last two lines include the small experimental errors on
$\delmd$ (3.9\%) and $\abseps$ (0.6\%), as well as the larger errors on
$m_t^2$ (11\%) and $A^2$ (14\%).

In order to find the allowed unitarity triangles, the computer program
MINUIT is used to fit the CKM parameters $A$, $\rho$ and $\eta$ to the
experimental values of $\Vcbabs$, $\vert V_{ub}/V_{cb}\vert$, $\abseps$ and
$\xd$. Since $\lambda$ is very well measured, we have fixed it to its
central value given above. As discussed in \cite{AL96}, one can
perform two types of fits:
\begin{itemize}
\item
Fit 1: the ``experimental fit.'' Here, only the experimentally measured
numbers are used as inputs to the fit with Gaussian errors; the coupling
constants $f_{B_d} \sqrt{\hat{B}_{B_d}}$ and $\hat{B}_K$ are given fixed
values.
\item
Fit 2: the ``combined fit.'' Here, both the experimental and theoretical
numbers are used as inputs assuming Gaussian errors for the theoretical
quantities.
\end{itemize}

Since, there are still large theoretical uncertainties, we show here the
results only for Fit 2. The results corresponding to Fit 1 can be seen
in \cite{AL96}.
\begin{figure} 
\vskip -0.9truein
\centerline{\epsfxsize 2.5 truein \epsfbox {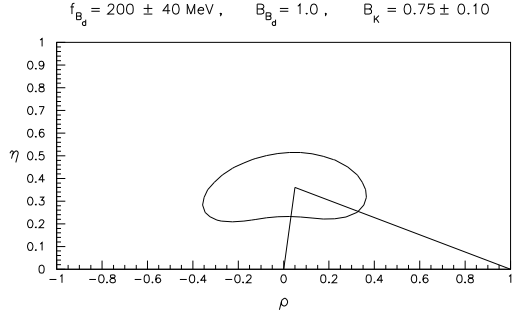}}
\vskip -1.2truein
\caption{Allowed region in $\rho$-$\eta$ space, from a simultaneous fit to
both the experimental and theoretical quantities given in the text.
 The theoretical errors are treated as Gaussian for
this fit. The solid line represents the region with $\chi^2=\chi_{min}^2+6$
corresponding to the 95\% C.L.\ region. The triangle shows the best 
fit. (Figure taken from \protect\cite{AL96}.)} 
\label{rhoeta2}
\end{figure}
 The resulting CKM triangle region is shown
in Fig.~\ref{rhoeta2}. As is clear from this figure, the allowed region is
still rather large at present. However, present data and theory do
restrict the parameters $\rho$ and $\eta$ to lie in the 
range which we have given earlier in eq.~(\ref{rhoetarange}).
The preferred values obtained from the ``combined fit" are
\beq
(\rho,\eta) = (0.05,0.36) ~,
\eeq
which gives rise to an almost right-angled unitarity triangle, with the
angle $\gamma$ being close to $90$ degrees. However, as we quantify below,
the allowed ranges of the CP violating angles $\alpha$, $\beta$, and
$\gamma$ estimated at the $95\%$ C.L. are still quite large, though
correlated.
\subsection{$\delms$ (and $\xs$) and the Unitarity Triangle}

Mixing in the \bsbsbar\ system is quite similar to that in the \bdbdbar\
system,
and the mass difference between the mass eigenstates $\delms$ is given by a
formula analogous to that of Eq.~(\ref{xd}):
\beq
\delms = C_s
\hat{\eta}_{B_s} y_t f_2(y_t) \vert V_{ts}^*V_{tb}\vert^2~,
\label{xs}
\eeq
where $C_s=G_F^2/(6\pi^2)M_W^2M_{B_s}(\fbbs)$.
To our accuracy $\vert V_{cb}\vert=\vert V_{ts}\vert$,
hence one of the sides of the unitarity triangle, $\vert
V_{td}/\lambda V_{cb}\vert$, can be obtained from the ratio of $\delmd$ and
$\delms$,
\beq
\frac{\delms}{\delmd} =
 \frac{\hat{\eta}_{B_s}M_{B_s}\left(\fbbs\right)}
{\hat{\eta}_{B_d}M_{B_d}\left(\fbb\right)}
\left\vert \frac{V_{ts}}{V_{td}} \right\vert^2.
\label{xratio}
\eeq
All dependence on the $t$-quark mass drops out, leaving the square of the
ratio of CKM matrix elements, multiplied by a factor which reflects
$SU(3)_{\rm flavour}$ breaking effects. The only real uncertainty in this
factor is the ratio of hadronic matrix elements. Whether or not $\xs$ can
be used to help constrain the unitarity triangle will depend crucially on
the theoretical status of the ratio $\fbbs/\fbb$. In \cite{AL96}, 
a range $\xi_s \equiv (f_{B_s} \sqrt{\hat{B}_{B_s}}) / (f_{B_d}
\sqrt{\hat{B}_{B_d}}) = (1.15 \pm 0.05)$ has been used, consistent with both 
earlier lattice-QCD \cite{Wittig96} and QCD sum rules \cite{Narison}. Recent 
lattice-QCD calculations reported in \cite{Gimenez96} yield 
$\xi_s^2=1.37 \pm 0.07$, in good agreement with these values. 
 (The SU(3)-breaking
factor in $\delms/\delmd$ is $\xi_s^2$.)

The mass and lifetime of the $B_s$ meson have now been measured at LEP and
Tevatron and their present values are $M_{B_s}=5369.3 \pm 2.0$ MeV and
$\tau(B_s)= 1.52 \pm 0.07 ~ps$ \cite{Richman96}. The QCD correction factor
$\hat{\eta}_{B_s}$ is equal to its $B_d$ counterpart, i.e.\
$\hat{\eta}_{B_s} =0.55$. The main uncertainty in $\delms$ (or,
equivalently, $\xs$) is now $\fbbs$. Using the determination of $A$ given
previously, and $\overline{\mt}=165 \pm 9$ GeV, one obtains \cite{AL96}:
\begin{eqnarray}
\delms &=& \left(12.8 \pm 2.1\right)\frac{\fbbs}{(230~\mbox{MeV})^2} 
~(ps)^{-1}~, \nonumber \\
\xs &=& \left(19.5 \pm 3.3\right)\frac{\fbbs}{(230~\mbox{MeV})^2}~.
\end{eqnarray}
The choice $f_{B_s}\sqrt{\hat{B}_{B_s}}= 230$ MeV corresponds to the
central value given by the lattice-QCD estimates, and with this our fits
give $\xs \simeq 20$ as the preferred value in the SM. Allowing the
coefficient to vary by $\pm 2\sigma$, and taking the central value for
$f_{B_s}\sqrt{\hat{B}_{B_s}}$, this gives \cite{AL96}
\begin{eqnarray} 
12.9 &\leq & \xs \leq 26.1~, \nonumber\\
8.6 ~(ps)^{-1} &\leq & \delms \leq 17.0 ~(ps)^{-1}~.
 \label{bestxs}
\end{eqnarray}
It is difficult to ascribe a confidence level to this range due to the
dependence on the unknown coupling constant factor. All one can say is that
the standard model predicts large values for $\delms$ (and hence $\xs$).

An alternative estimate of $\delms$ (or $\xs$) can also be obtained by
using the relation in Eq.~(\ref{xratio}). Two quantities are required.
First, we need the CKM ratio $\vert V_{ts}/V_{td} \vert$. In
\cite{AL96},  the allowed values  for the inverse
of this ratio as a function of $\fbd\sqrt{\hat{B}_{B_d}}$ was worked out.
From this one gets (at 95\% C.L.)
\beq
2.94 \leq \left\vert {V_{ts} \over V_{td}} \right\vert \leq 6.80~.
\eeq

The second ingredient is the SU(3)-breaking factor which we take to be
$\xi_s = 1.15 \pm 0.05$, or $1.21 \le \xi_s^2 \le 1.44$. The result of the
CKM fit can therefore be expressed as a $95\%$ C.L. range:
\beq
 11.4 \left(\frac{\xi_s}{1.15}\right)^2
 ~\leq ~\frac{\delms}{\delmd} ~\leq ~
 61.2 \left(\frac{\xi_s}{1.15}\right)^2 ~.
\eeq
Again, it is difficult to assign a true confidence level to $\delms/\delmd$
due to the dependence on $\xi_s$. However, the uncertainty due to the CKM
matrix element ratio has now been reduced to a factor 5.3 due to the
 constraints on the unitarity triangle.
 The allowed range for the ratio
$\delms/\delmd$ shows that this method is still poorer at present for the
determination of the range for $\delms$, as compared to the absolute value
for $\delms$ discussed above, which in comparison is uncertain by a factor
of 2. Both suffer from additional dependences on
$f_{B_s}\sqrt{\hat{B}_{B_s}}$ or $\xi_s$.

\begin{figure}
\vskip -1.0truein
\centerline{\epsfxsize 3.0 truein \epsfbox {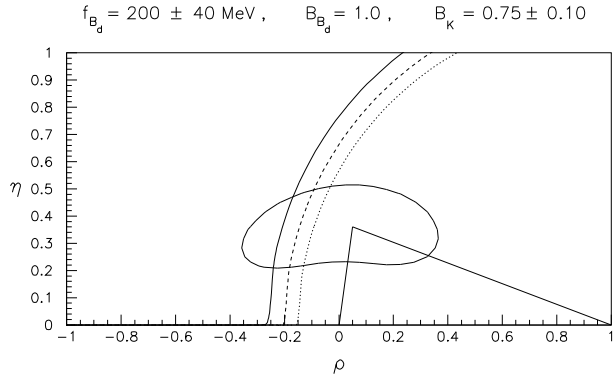}}
\vskip -1.4truein
\caption{Further constraints in $\rho$-$\eta$ space from the LEP bound
 $\delms/\delmd > 19.0$. The bounds are presented for 3 choices of 
the SU(3)-breaking
parameter: $\xi_s^2 = 1.21$ (dotted line), $1.32$ (dashed line) and $1.44$
(solid line). In all cases, the region to the left of the curve is ruled
out. (Figure taken from \protect\cite{AL96}.)}
\label{xslimit}
\end{figure}

The present lower bound from LEP $\delms > 9.2~(ps)^{-1}$ (95\% C.L.)
\cite{Gibbons96} and the present world average $\delmd = (0.464 \pm
0.018)~(ps)^{-1}$ can be used to put a bound on the ratio $\delms/\delmd$,
yielding $\delms/\delmd > 19.0$. This  is significantly  better than the
lower bound on this quantity from the CKM fits, using the central value for
$\xi_s$. The 95\% confidence limit on $\delms/\delmd$ can be turned into a
bound on the CKM parameter space $(\rho,\eta)$ by choosing a value for the
SU(3)-breaking parameter $\xi_s^2$. We assume three representative values:
$\xi_s^2 = 1.21$, $1.32$ and $1.44$, and display the resulting constraints
in Fig.~\ref{xslimit}. This graph shows that the LEP bound now restricts
the allowed $\rho$-$\eta$ region for all three values of $\xi_s^2$, though
this restriction is weakest for the largest value of $\xi_s^2$ assumed.
Thus the LEP bound on $\delms$ provides  more stringent lower bounds on the
parameters $\rho$ and $\eta$ than those obtained from the CKM fits
without this constraint:
\begin{eqnarray}
 0.25 &\leq & \eta \leq 0.52 , \nonumber \\
 -0.25 &\leq & \rho \leq 0.35 ~.
\label{rhoetarangexs}
\end{eqnarray}
%

Summarizing the discussion on $\xs$, we note that the lattice-QCD-inspired
estimate $f_{B_s} \sqrt{\hat{B}_{B_s}} \simeq 230$ MeV and the CKM fit
predict that $\xs$ lies between 13 and 26, with a central value around 20.
All of these values scale as $(f_{B_s}\sqrt{\hat{B}_{B_s}}/230
~\mbox{MeV})^2$. The present constraints  on the CKM parameters from the
bound on $\delms$ are now competitive with those from fits to other data,
and this will become even more pronounced with improved data. In
particular, one expects to reach a sensitivity of $\xs \simeq 15$ (or
$\delms \simeq 10~ps^{-1})$ at LEP combining all data and tagging
techniques, and similarly at the SLC, CDF and HERA-B. Of course, an actual
measurement of $\delms$ (equivalently $\xs$) would be very helpful in
further constraining the CKM parameter space. Note that the entire range
for $\xs$ worked out here is accessible at the LHC experiments.


\section{CP Violation in the $B$ System}

It is expected that the $B$ system will exhibit large CP-violating effects,
characterized by nonzero values of the angles $\alpha$, $\beta$ and
$\gamma$ in the unitarity triangle (Fig.~\ref{triangle}) \cite{BCPasym}.
The most promising method to measure CP violation is to look for an
asymmetry between $\Gamma(B^0\to f)$ and $\Gamma({\overline{B^0}}\to f)$,
where $f$ is a CP eigenstate. If only one weak amplitude contributes to the
decay, the CKM phases can be extracted cleanly (i.e.\ with no hadronic
uncertainties). Thus, $\sin 2\alpha$, $\sin 2\beta$ and $\sin 2\gamma$ can
in principle be measured in $\bdbarp \to \pi^+ \pi^-$, $\bdbarp\to J/\psi
K_S$ and $\bsbarp\to\rho K_S$, respectively.

Penguin diagrams \cite{penguins} will, in general, introduce some hadronic
uncertainty into an otherwise clean measurement of the CKM phases. In the
case of $\bdbarp\to J/\psi K_S$, the penguins do not cause any problems,
since the weak phase of the penguin is the same as that of the tree
contribution. Thus, the CP asymmetry in this decay still measures $\sin
2\beta$. For $\bdbarp \to \pi^+ \pi^-$, however, although the penguin is
expected to be small with respect to the tree diagram, it will still
introduce a theoretical uncertainty into the extraction of $\alpha$. This
uncertainty can, in principle, be removed by the use of an isospin analysis
\cite{isospin}, which requires the measurement of the rates for
$B^+\to\pi^+\pi^0$, $B^0\to\pi^+\pi^-$ and $B^0\to\pi^0\pi^0$, as well as
their CP-conjugate counterparts. Thus, even in the presence of penguin
diagrams, $\sin 2\alpha$ can in principle be extracted from the decays
$B\to\pi\pi$. Still, this isospin program is ambitious experimentally. If
it cannot be carried out, the error induced on $\sin 2\alpha$ is of order
$|P/T|$, where $P$ ($T$) represents the penguin (tree) diagram. The ratio
$|P/T|$ is difficult to estimate since it is dominated by hadronic physics.
However, one ingredient is the ratio of the CKM elements of the two
contributions: $|V_{tb}^* V_{td} / V_{ub}^* V_{ud} | \simeq
|V_{td}/V_{ub}|$. From the fits in \cite{AL96},
 the allowed range for the ratio of these CKM matrix elements is
\beq
1.4 \leq \left\vert {V_{td}\over V_{ub}} \right\vert \leq 4.6 ~,
\eeq
with a central value of about 3.

It is $\bsbarp\to\rho K_S$ which is most affected by penguins. In fact,
the penguin contribution is probably larger in this process than the tree
contribution. This decay is clearly not dominated by one weak (tree)
amplitude, and thus cannot be used as a clean probe of the angle $\gamma$.
Instead, two other methods have been devised, not involving CP-eigenstate
final states. The CP asymmetry in the decay $\bsbarp\to D_s^\pm K^\mp$ can
be used to extract $\sin^2 \gamma$ \cite{ADK}. Similarly, the CP asymmetry
in $B^\pm\to\dcp K^\pm$ also measures $\sin^2 \gamma$ \cite{growyler}.
Here, $\dcp$ is a $D^0$ or $\dbar$ which is identified in a CP-eigenstate
mode (e.g.\ $\pi^+\pi^-$, $K^+K^-$, ...).
Further discussion on CP violation is given in
 \cite{Buraswarsaw,EWPsize,DH2,GR96,Gronau96}.

\begin{figure}
\vskip -1.0truein
\centerline{\epsfxsize 3.0 truein \epsfbox {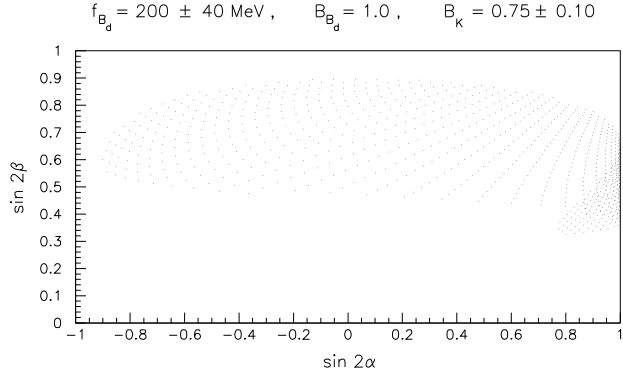}}
\vskip -1.4truein
\caption{Allowed region of the CP-violating quantities $\sin 2\alpha$ and 
$\sin 2\beta$ resulting from the ``combined fit" of the data for the ranges
for $\fbd\protect\sqrt{\hat{B}_{B_d}} $ and $\hat{B}_K$ given in the text.
(Figure taken from \protect\cite{AL96}.)}
\label{alphabeta2}
\end{figure}

\begin{figure}
\vskip -1.0truein
\centerline{\epsfxsize 3.0 truein \epsfbox {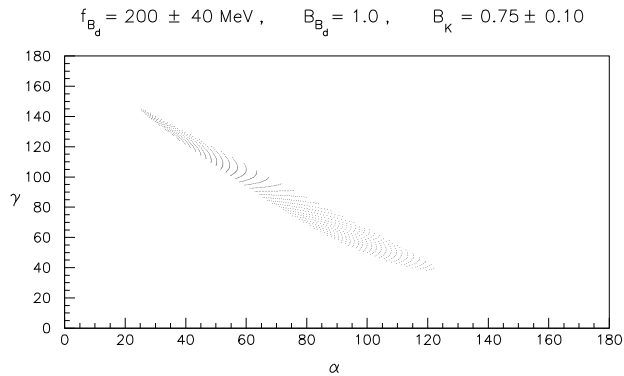}}
\vskip -1.4truein
\caption{Allowed values (in degrees) of the angles $\alpha$ and $\gamma$ 
resulting from the ``combined fit" of the data for the ranges for 
$\fbd\protect\sqrt{\hat{B}_{B_d}} $ and $\hat{B}_K$ given in the text.
(Figure taken from \protect\cite{AL96}.)} 
\label{alphagam}
\end{figure}

The CP-violating asymmetries can be expressed straightforwardly in terms
of the CKM parameters $\rho$ and $\eta$. The 95\% C.L.\ constraints on
$\rho$ and $\eta$ found previously can be used to predict the ranges of
$\sin 2\alpha$, $\sin 2\beta$ and $\sin^2 \gamma$ allowed in the standard
model.
Since the CP asymmetries all depend on $\rho$ and $\eta$, the ranges for
$\sin 2\alpha$, $\sin 2\beta$ and $\sin^2 \gamma$ 
 are correlated. That is, not all values in the ranges are
allowed simultaneously. 
Given a value for $\fbd\sqrt{\hat{B}_{B_d}}$, the CP asymmetries are fairly
constrained. However, since there is still considerable uncertainty in the
values of the coupling constants, a more reliable profile of the CP
asymmetries at present is given by the ``combined fit" (Fit 2) \cite{AL96}. 
The resulting correlations are shown in Figs.~\ref{alphabeta2}
and \ref{alphagam}. From
Fig.~/ref{alphabeta2} one sees that the smallest value of $\sin 2\beta$ 
occurs 
in a small region of parameter space around $\sin 2\alpha\simeq 0.8$-0.9.
Excluding this small tail, one expects the CP-asymmetry in $\bdbarp\to
J/\Psi K_S$ to be at least 20\% (i.e., $\sin 2 \beta > 0.4)$. Note that the
LEP bound $\delms/\delmd > 19.0$ removes a  part of the  small $\sin
2\beta$ region in this tail. This is easy to understand if one recalls the
relation $\sin 2\beta = 2 \eta (1 - \rho) /((1-\rho)^2  + \eta^2)$. As seen
from Fig.~\ref{xslimit}, the LEP bound removes the large negative-$\rho$
values, which amounts to removing small $\sin 2  \beta$ values. The allowed
region in $\sin 2 \alpha$ is not affected significantly from the LEP-bound. 
Hence the following ranges for the CP-violating rate asymmetries 
parametrized by $\sin
2\alpha$, $\sin 2\beta$ and and $\sin^2 \gamma$ are determined at 95\% C.L.
to be
\begin{eqnarray}
&~& -0.90 \leq \sin 2\alpha \le 1.0~, \nonumber \\
&~& 0.40 \leq \sin 2\beta \le 0.94~, \\
&~& 0.34 \leq \sin^2 \gamma \le 1.0~. \nonumber
\end{eqnarray}

%

%
%

\section{Summary and Outlook}

  We have discussed  some aspects of $B$ decays in 
the context of SM. Flavour physics, in particular $B$ physics, provides an 
excellent laboratory in testing calculational techniques in QCD, 
involving both perturbative and non-perturbative aspects. 
The applications presented here are by no means exhaustive but are fairly 
representative of the kind of problems being studied in $B$ decays and the 
techniques being used to tackle them. Not all experimental observations are
calculable from first principles in QCD - this remains an ambitious
and long-term goal. Nevertheless, the present quantitative rapport between
experiment and theory (SM) in $B$ decays is impressive.
New experimental facilities will churn out a wealth of data encouraging us
to ask increasingly sophisticated questions and seek their answers. 

 A good part of $B$
decays is accountable in QCD by virtue of the fact that the
mass of the $b$ quark is large enough to warrant perturbative calculations
and the expansion parameter $\alpha_s(m_b)/\pi) \leq 0.1$ is small, so that
leading and next-to-leading corrections should be sufficient.
This, coupled with the working hypothesis that $b$ quark can be treated as
heavy, enables one to do a systematic expansion of the Green's functions 
in the parameter $\bar{\Lambda}/m_b = O(0.1)$. The resulting
framework has found many applications. Illustrative of these
are the semileptonic branching ratio ${\cal B}_{SL}(B)$, the electromagnetic
penguin decay rate $\BBGAMAXS$ and
the average charmed hadron multiplicity in $B$ decays $\langle n_c 
\rangle$, which are all in fair agreement with data. Some of the present 
theoretical dispersion in these quantities is expected to be 
considerably reduced as and when the complete NLO QCD corrections are
available. 

The only visible question mark in inclusive $B$ decays is the considerably 
shorter observed lifetime of the $\Lambda_b$ baryon, which is theoretically 
neither anticipated nor easy to accommodate. To firm up present 
estimates, one has to reliably calculate the mesonic and 
baryonic matrix elements of the local four-quark operators present in the 
effective Hamiltonian based on the SM.
This is an ambitious calculation for lattice QCD and one which will 
probably not
be carried through in this century. We have little choice but to sharpen 
other tools such as the QCD sum rules to draw definite conclusions. 
The apparent mismatch in lifetimes may owe itself to our imprecise
understanding of the non-leptonic decays, but one can 
not exclude the possibility that it may after all have an experimental 
origin, like the once omnipresent (and now defunct) $Z^0 \to b\bar{b}$ 
anomaly. This remains to be settled in future experiments. In particular,
at HERA-B and in experiments at Tevatron and the LHC, the $\Lambda_b$-lifetime
will be measured very precisely using fully reconstructed $\Lambda_b$'s.

   As emphasized in the introduction, $B$ decays enter
in five of the nine CKM matrix elements. The best measured of
these is the matrix element $\Vcbabs$ (see Table \ref{tabckm}), which is 
determined with
$\pm 7\%$ accuracy, with remarkably consistent results from the exclusive
and inclusive decays. This can be taken as an excellent test of the 
parton-hadron duality in semileptonic decays. 
 In exclusive decays, this precision has been made
possible due to the theoretical developments in the context of HQET of
which the decay $B \to D^* \ell \nu_\ell$ remains the show-piece case.
 More work is needed to reach similar precision in
other matrix elements of which two, $\Vtdabs$ and $\Vubabs$, are crucial
in testing the CKM unitarity (see Fig.~\ref{triangle}).
The former, together with $\Vtsabs$, will be measured in a variety of ways
involving $B^0$ - $\overline{B^0}$ mixings and rare $B$ decays. Present 
determination and theoretical
proposals have been discussed here. Once again, the matrix elements of the
four-quark operators play a crucial role and they have to be determined
as accurately as possible. Fortunately,  experiments will be able to
put direct and model-independent bounds on some of these matrix elements.
The case in point is the radiative decays $B \to \rho \gamma$ and $B \to 
\omega \gamma$, where data on charged $B^\pm$ and neutral $B^0$ decays can
be used to disentangle the contributions of the four-quark operators and the
electromagnetic penguin operator with the help of isospin symmetry.
Apart from testing the CKM unitarity, rare $B$ decays are sensitive to 
new physics. The case in point here is the decays \bxsll  and the 
related exclusive modes. Invariant dilepton mass and FB asymmetry in 
these decays, measured precisely, may reveal deviations from the SM. 
Such deviations, for example, are anticipated in SUSY models. 

   Finally, the overriding interest in $B$ decays is that they will test the
CKM paradigm for CP violation. Present estimates of the CP-violating 
asymmetries predict a large value for $\sin 2 \beta$. Since this asymmetry
is measurable in a large number of experimental facilities being built,
and there are no theoretical uncertainties in the interpretation of data,
there is good reason to be optimistic that soon one would have first
observations of CP violation in $B$ decays which one can also transcribe in 
terms of the underlying CKM parameters, in particular $\eta$ and $\rho$.
 However, to quantitatively test the CKM paradigm
one needs the measurement of at least one more CP asymmetries, 
related to the angles $\alpha$ and/or $\gamma$. Some estimates of these
asymmetries, related problems and possible resolutions are discussed in 
these lectures and elsewhere. The different ways of testing the CKM unitarity
through CP asymmetries, rare decays and mixing 
will surely lead to an overdetermination of the CKM parameters, which is the
goal of $B$ physics.
  

\smallskip
\noindent
{\bf Acknowledgements}:
I would like to thank Hrachia Asatrian, Vladimir Braun, Christoph Greub, Tak 
Morozumi, 
and Matthias Neubert for helpful discussions. Matthias Neubert kindly
provided Figure 2 based on the work in ref. \cite{NS96}. The warm 
hospitality of Marek Jezabek 
and the organizers of the Zakopane school is gratefully acknowledged.\\

{\it Post scriptum:}
 
As this manuscript was being completed, I heard the sad news of the passing 
away of Professor Abdus Salam, one of the principal architects of the 
standard model and uncontestedly the staunchest supporter of the third 
world science. The scientific world is poorer without him. For me 
personally he was a role model - an ideal teacher, a great scientific leader
and a compassionate human being - bubbling with ideas, always 
enthusiastic, full of passion and free of prejudices. Alas, he is no more!
These lectures are dedicated in gratitude to him.


\newpage
\vspace*{2mm}
\begin{thebibliography}{99}
%
%
\bibitem{GSW}
    S.L. Glashow, Nucl. Phys. {\bf 22} (1961) 579;
    S. Weinberg, Phys. Rev. Lett. {\bf 19} (1967) 1264;
    A. Salam, in {\it Elementary Particle Theory}, ed. N. Svartholm
    (Almqvist and Wiksell, Stockholm) (1968).

\bibitem{Cabibbo} N. Cabibbo, Phys. Rev. Lett. {\bf 10} (1963) 531.

\bibitem{KM}
    M. Kobayashi and K. Maskawa, Prog. Theor. Phys. {\bf 49} (1973) 652.

\bibitem{GIM} S.L. Glashow, J. Iliopoulos and L. Maiani,
            Phys. Rev.  {\bf D2} (1970) 1285.
\bibitem{top95}
F. Abe et al. (CDF Collaboration), Phys. Rev. Lett. {\bf 74} (1995) 2626;\\
S. Abachi et al. (D0 Collaboration), Phys. Rev. Lett. {\bf 74} (1995) 2632.

\bibitem{CDFvtb}
F. Abe et al. (CDF Collaboration), FERMILAB-CONF-95-237-E (1995);\\
K. Kondo, invited talk at the 4th KEK Topical Conference, Tsukuba, Japan,
October 29 -31, 1996.

\bibitem{Fuji96}
K. Fujii, invited talk at the 4th KEK Topical Conference, Tsukuba, Japan,
October 29 -31, 1996.

\bibitem{Wolfenstein} L. Wolfenstein, Phys. Rev. Lett. {\bf 51} (1983)
1845.

\bibitem{AL96} A. Ali and D. London, preprint DESY 96-140,
UdeM-GPP-TH-96-45, [hep-ph/9607392], to appear in the {\it Proc.\ of QCD
Euroconference 96}, Montpellier, July 4-12, 1996.

\bibitem{ALI96} A. Ali, preprint DESY 96-106 [hep-ph/9606324]; to appear in
the Proceedings of the XX International Nathiagali Conference on
Physics and Contemporary Needs, Bhurban, Pakistan, June 24-July 13, 1995 
(Nova Science Publishers, New York, 1996).

\bibitem{GW77} S.L. Glashow and S. Weinberg, Phys. Rev. {\bf D15} (1977) 
1958.

\bibitem{PDG96} R.M. Barnett et al.\ (Particle Data Group), Phys.\ Rev.\
 D54 (1996) 1.

\bibitem{BuBu94}
      A.J. Buras and G. Buchalla, Phys. Lett. {\bf B336} (1994) 263.

\bibitem{AKL94}
    R. Aleksan, B. Kayser and D. London, Phys. Rev. Lett. {\bf 73} (1994) 18.

\bibitem{Jarlskog} C. Jarlskog, Phys. Rev. Lett. {\bf 55} (1985) 1039;
                   Z. Phys. {\bf C29} (1985) 491;

and in {\it CP Violation},
 ed. C. Jarlskog (World Scientific, Singapore) (1989) 3.

\bibitem{Jansen96} For a review and references , see K. Jansen, Nucl. 
Phys. B (Proc. Suppl.) {\bf 47} (1996) 196.

%
%
 \bibitem{CM}
 N. Cabibbo and L. Maiani, Phys. Lett {\bf B79} (1978) 109.

\bibitem{Suzuki}
 M. Suzuki, Nucl. Phys. {\bf B145} (1978) 420.

\bibitem{Alipiet}
     A. Ali and E. Pietarinen,
      Nucl. Phys. {\bf B154} (1979) 519.

\bibitem{ACCMM}
     G. Altarelli et al.,
     Nucl. Phys. {\bf B208} (1982) 365.
       
\bibitem{PHam83}
Q. Hokim and X.Y. Pham, Phys. Lett. {\bf B122} (1983) 297.

\bibitem{FCz95}
A. Falk et al., Phys. Lett. {\bf B 326} (1994) 145;
A. Czarnecki, M. Jezabek, and J.H. K\"uhn, Phys. Lett. {\bf B 346} (1995) 
335.

\bibitem{JK89}
M. Jezabek and J.H. K\"uhn, Nucl. Phys. {\bf B320} (1989) 20.
\bibitem{Behrends}
 R.E. Behrends, R.J. Finkelstein and A. Sirlin, Phys. Rev. {\bf 101}
 (1956) 866;
 S.M. Berman, Phys. Rev. {\bf 112} (1958) 267;
 T. Kinoshita and A. Sirlin, Phys. Rev. {\bf 113} (1959) 1652.
%
%
\bibitem{AM74}
G. Altarelli and L. Maiani, Phys. Lett. {\bf B52} (1974) 351;\\
M.K. Gaillard and B.W. Lee, Phys. Rev. Lett. {\bf 33} (1974) 108.

\bibitem{VSZ}
A. I. Vainshtein, V.I. Zakharov and M.A. Shifman, JETP {\bf 45} (1977)
   670.
%
\bibitem{LLAP} G. Altarelli et al., Phys. Lett. {\bf B99} (1981) 141;
       Nucl. Phys. {\bf B187} (1981) 461.

\bibitem{BW90}
       A.J. Buras and P.H. Weisz, Nucl. Phys. {\bf B333} (1990) 66.

\bibitem{BBL95}
G. Buchalla, A.J. Buras, and M.E. Lautenbacher,
MPI-Ph/95-104; TUM-T31-100/95; FERMILAB-PUB-95/305-T; SLAC-PUB 7009;
[hep-ph/9512380].
%
%
%
\bibitem{Chayetal}
J. Chay, H. Georgi and B. Grinstein, Phys. Lett. {\bf B247} (1990) 399.

\bibitem{Bigietal}
I. Bigi, N. Uraltsev and A. Vainshtein, Phys. Lett. {\bf B293} (1992)
 430; [E: {\bf B297} (1993) 477];
B. Blok and M. Shifman, Nucl. Phys. {\bf B399} (1993) 441, {\it ibid.}
459;\\
I. Bigi et al., Phys. Rev. Lett. {\bf 71} (1993) 496 and
in {\it Proc. of the Annual Meeting of the Division of
Particles and Fields of the APS}, Batavia, Illinois, 1992, edited by C.
Albright et al. (World Scientific, Singapore), 610.

\bibitem{MW94}
A.V. Manohar and M.B. Wise, Phys. Rev. {\bf D49} (1994) 1310.

\bibitem{falketalbsll}
        A. F. Falk, M. Luke and M. J. Savage, Phys. Rev.
        {\bf D49} (1994) 3367.

\bibitem{BKSV94}
B. Blok et al., Phys. Rev. {\bf D49} (1994) 3356 [E:{\bf D50} (1994), 3572].

\bibitem{Bigi}
I. Bigi et al., in: B Decays, edited by S. Stone, Second Edition (World
Scientific, Singapore, 1994) 132;\\
I.Bigi, preprint UND-HEP-95-BIG02 (1995) [hep-ph/9508408].

\bibitem{AHHM96}
A.Ali, G. Hiller, L.T. Handoko, and T. Morozumi, Preprint DESY 96-206,
Hiroshima Univ. report HUPD-9615 [hep-ph-9609449]; to appear in Phys.
Rev. D.
%
\bibitem{BKbook}
For a discussion of relativistic kinematics, see the classic
book by E. Byckling and K. Kajantie: Particle Kinematics (John Wiley \&
Sons, New York) (1972).

\bibitem{Bagan94}
E. Bagan, P. Ball, V.M. Braun, and P. Gosdzinsky, Nucl. Phys. {\bf B432}
(1994) 3.
\bibitem{Bagan95a}
E. Bagan, P. Ball, V.M. Braun, and P. Gosdzinsky, Phys. Lett. {\bf B342}
(1995) 362 [E: {\it ibid} {\bf B374} (1996) 363].
\bibitem{Bagan95b}
E. Bagan, P. Ball, B. Fiol, and P. Gosdzinsky, Phys. Lett. {\bf B351}
(1995) 546.

\bibitem{kinetic}
        P. Ball and V.M. Braun, Phys. Rev. {\bf D49} (1994) 2472;
        V. Eletsky and E. Shuryak, Phys. Lett. {\bf B276} (1992) 191;
        M. Neubert, Phys. Lett. {\bf B322} (1994) 419.
\bibitem{Neubert96} M. Neubert, preprint CERN-TH/96-208 (1996)
[hep-ph-9608211].

\bibitem{Martinelli96} G. Martinelli, preprint ROME 1155/96 
(1996) [hep-ph-9610455].

\bibitem{GKLW96} M. Gremm, A. Kapustin, Z. Ligeti, and M.B. Wise,
       Phys.\ Rev.\ Lett.\ {\bf 77} (1996) 20.

\bibitem{mtmsbar} N. Gray, D.J. Broadhurst, W. Grafe,  and
K. Schilcher, Z. Phys. {\bf C48} (1990) 673.

\bibitem{Neubert95}
M. Neubert,  in {\it Proc. of the 17th Int.   
Symp. on Lepton and Photon Interactions}, Beijing, P.R. China, 10 -15 August
1995 (World Scientific 1996; Editors: Zheng Zhi-Peng and Chen He-Seng).
%
%
\bibitem{Tomasz95}
T. Skwarnicki,  in {\it Proc. of the 17th Int.
Symp. on Lepton and Photon Interactions}, Beijing, P.R. China, 10 -15 August
1995 (World Scientific 1996; Editors: Zheng Zhi-Peng and Chen He-Seng).

\bibitem{Perret95} P. Perret, in {\it Proc. of the Int. Europhys. Conf.
on High Energy Physics}, Brussels, Belgium, 27 July - August 1995
(World Scientific 1996; Editors: J. Lemonne, C. Vander Velde and F.
Verbeure).

\bibitem{Calderini96}
G. Calderini, presented at the 31. Rencontres de Moriond: QCD and High
Energy Hadronic Interactions, Les Arcs, France, March 1996;//
D. Buskulic et al. (ALEPH Collaboration), preprint CERN-PPE/96-117 (1996).

\bibitem{NS96}
M. Neubert and C.T. Sachrajda, preprint CERN-TH/96-19; SHEP 96-03
[hep-ph/9603202] (to appear in Nucl. Phys. B).

\bibitem{Richman96} J. Richman, plenary talk at the International Conference
on High Energy Physics, Warsaw, ICHEP96 (1996).

\bibitem{Uraltsev96}
N.G. Uraltsev, Phys. Lett. {\bf B376} (1996) 303.

\bibitem{Guberinaetal79}
B. Guberina et al., Phys. Lett. {\bf B89} (1979) 811.

\bibitem{Rosner96}
J. Rosner, Phys. Lett. {\bf B379} (1996) 267.

\bibitem{DELPHISIGMA}
P. Abreu et al. (DELPHI Collaboration),
 in {\it Proc. of the Int. Europhys. Conf. on High
Energy Physics}, Brussels, Belgium, 27 July - August 1995
(World Scientific 1996; Editors: J. Lemonne, C. Vander Velde and F.
Verbeure).

\bibitem{CD96}
P. Colangelo and F. De Fazio, Phys. Lett. {\bf B387} (1996) 371.

\bibitem{AMPR96}
G. Altarelli, G. Martinelli, S. Petrarca, and F. Rapuano,
Phys. Lett. {\bf B382} (1996) 409.

\bibitem{BBD96}
M. Beneke, G. Buchalla, and I. Dunietz, Phys. Rev. {\bf D54} (1996) 4419.
 
\bibitem{HQET}  H.D. Politzer and   M. Wise,
               Phys. Lett. {\bf B206} (1988) 681;
               {\it ibid.} {\bf B208} (1988) 504;\\
                 M. Voloshin and   M. Shifman,
               Sov. J. Nucl. Phys. {\bf 45} (1987) 292;
               {\it ibid.} {\bf 47} (1988) 511;\\
                 E. Eichten and   B. Hill,
               Phys. Lett. {\bf B234} (1990) 511;\\
                 H. Georgi,
               Phys. Lett. {\bf B240} (1990) 447;\\
                 B. Grinstein,
               Nucl. Phys. {\bf B339} (1990) 253.
%
\bibitem{IW}   N. Isgur and   M. Wise,
               Phys. Lett. {\bf B232} (1989) 113;
               {\it ibid.} {\bf B237} (1990) 527.
%

\bibitem{Luke} M.E. Luke, \plb{252}{90}{447}.

\bibitem{Boyd} C.G. Boyd and D.E. Brahm, Phys. Lett. {\bf B257} (1991) 393.

\bibitem{Neubert} M. Neubert and V. Rieckert, Nucl. Phys.
{\bf B382} (1992) 97;\\  
M. Neubert, Phys. Lett. {\bf B264} (1991) 455, Phys. Rev. {\bf D46}
(1992) 2212.

\bibitem{hybrid} M.B. Voloshin and M.A. Shifman, Sov. J. Nucl. Phys. {\bf 
45} (1987) 292.

\bibitem{Vainshtein95}
A. Vainshtein, in {\it Proc. of the Int. Europhys. Conf. on High
Energy Physics}, Brussels, Belgium, 27 July - August 1995
(World Scientific 1996; Editors: J. Lemonne, C. Vander Velde and F.
Verbeure).

\bibitem{Cz96} A. Czarnecki, Phys.\ Rev.\ Lett.\  {\bf 76} (1996) 4124.

\bibitem{Gibbons96} L. Gibbons (CLEO Collaboration),
 invited talk at the
International Conference on High Energy Physics, Warsaw, Poland
July 25-31, 1996 (to appear in the proceedings.)
\bibitem{Bartelt93}
J. Bartelt et al. (CLEO Collaboration), Phys. Rev. Lett. {\bf 64} (1990) 16.

%
\bibitem{CLEOrare2}
M.S. Alam et al. (CLEO Collaboration), Phys. Rev. Lett. {\bf 74} (1995) 2885.

\bibitem{CLEOrare1}
 R. Ammar et al. (CLEO Collaboration), Phys. Rev. Lett. {\bf 71} (1993) 674.

\bibitem{InamiLim}
        T. Inami and C.S. Lim,
        Prog. Theor. Phys. {\bf 65} (1981) 297.

\bibitem{BSGAM}
     S. Bertolini, F. Borzumati and A. Masiero, Phys. Rev. Lett.
     {\bf 59} (1987) 180;\\
     R. Grigjanis et al., Phys. Lett. {\bf B213} (1988) 355;\\
     B. Grinstein, R. Springer, and M.B. Wise, Phys. Lett. {\bf 202}
                  (1988) 138; Nucl Phys. {\bf B339} (1990) 269;\\
     G. Cella et al., Phys. Lett. {\bf B248} (1990) 181.

\bibitem{Ciuchini}
     M. Ciuchini et al.,
     Phys. Lett. {\bf B316} (1993) 127; Nucl. Phys. {\bf B415} (1994) 403;\\
     G. Cella et al., Phys. Lett. {\bf B325} (1994) 227; \\
     M. Misiak, Nucl. Phys. {\bf B393} (1993) 23;
     [E. {\bf B439} (1995) 461].

\bibitem{Misiak96}
     M. Misiak, contribution to the International Conference on High
Energy Physics, Warsaw, 25 - 31 July 1996.

\bibitem{ag1}  A. Ali and C. Greub,
              Z. Phys. {\bf C49} (1991) 431;
              Phys. Lett. {\bf B259} (1991) 182.

\bibitem{ag2}  A. Ali and C. Greub ,
               Phys. Lett. {\bf B287} (1992) 191.

\bibitem{ag3}  A. Ali and C. Greub,
              Z. Phys. {\bf C60} (1993) 433.
\bibitem{ag95} A. Ali and C.
Greub, Phys. Lett. {\bf B361} (1995) 146.

\bibitem{Pott95}
N. Pott, Phys. Rev. D54 (1996) 938.

\bibitem{Yao94}
K. Adel and Y.-p. Yao, Phys. Rev. {\bf D49} (1994) 4945.

\bibitem{GHW96}
C. Greub, T. Hurth and D. Wyler, Phys. Lett. {\bf B380} (1996) 385;
            Phys. Rev. {\bf D54} (1996) 3350.

\bibitem{Buras94}
       A.J. Buras, M. Misiak, M. M\"unz, and S. Pokorski,
       Nucl. Phys. {\bf B424} (1994) 374.
\bibitem{AGM92}
A. Ali, C. Greub and T. Mannel, DESY Report 93-016 (1993), and in
{\it $B$-Physics Working Group Report,
ECFA Workshop on a European $B$-Meson Factory},
ECFA 93/151, DESY 93-053 (1993), edited by R. Aleksan and A. Ali.

\bibitem{Ciuchini94}
       M. Ciuchini et al., Phys. Lett. {\bf B334} (1994) 137.
\bibitem{GH96}
C. Greub and T. Hurth, preprint SLAC-PUB-7267, ITP-SB-96-46 (1996)
[hep-ph-9608449].

\bibitem{bsgamld}
D. Atwood, B. Blok, and A. Soni, Int. J. Mod. Phys. {\bf A11} (1996) 3743;\\
H.-Y. Cheng, Phys. Rev. {\bf D51} (1995) 6228;\\
J.M. Soares, Phys. Rev. {\bf D53} (1996) 241;\\
J. Milana, Phys. Rev. {\bf D53} (1996) 1403;\\
G. Eilam, A. Ioannissian, and R.R. Mendel, Z. Phys. {\bf C71} (1996) 95.

\bibitem{DHT95}
N.G. Deshpande, X.-G. He, and J. Trampetic, Phys. Lett. {\bf B367} (1996) 
362.

\bibitem{GP95}
     E. Golowich and S. Pakvasa, Phys. Rev. {\bf D51}, 1215 (1995).

\bibitem{Ricciardi}
G. Ricciardi, Phys. Lett. {\bf B355} (1995) 313.

\bibitem{BH95}
        T.E. Browder and K. Honscheid,
        Prog. Part. Nucl. Phys. {\bf 35} (1995) 81.
\bibitem{BSW} 
M. Bauer, B. Stech, and M. Wirbel, Z. Phys. {\bf C34} (1987) 103. 

\bibitem{Hewett96}
J.L. Hewett and J.D. Wells, preprint SLAC-PUB-7290 (1996) [hep-ph-9610323].

%
\bibitem{abs93}
A. Ali, V.M. Braun and H. Simma, Z. Phys. {\bf C63} (1994) 437.
%
\bibitem{bksnsr}
   P. Ball, TU-M\"unchen Report  TUM-T31-43/93 (1993);\\
   P. Colangelo et al., Phys. Lett. {\bf B317} (1993) 183;\\
   S. Narison, Phys.\ Lett.\ {\bf B327} (1994) 354;\\
   J. M. Soares, Phys.\ Rev.\ {\bf D49} (1994) 283.

\bibitem{KS94}
G. Korchemsky and G. Sterman, Phys. Lett. {\bf B340} (1994) 96.

\bibitem{Shifmangamma}
R.D. Dikeman, M. Shifman, and R.G. Uraltsev,
Int. J. Mod. Phys. {\bf A11} (1996) 571.

\bibitem{Sudakov} V. Sudakov, Sov. Phys. JETP {\bf 3} (1956) 65 ;\\ G.
Altarelli, Phys. Rep. {\bf 81} (1982) 1.

\bibitem{klp95}
A. Kapustin, Z. Ligeti and H.D. Politzer, Phys. Lett. {\bf B357} (1995) 
653.

\bibitem{JR94}
R. Jaffe and L. Randall, Nucl. Phys. {\bf B412} (1994) 79.

\bibitem{Bigietal2}
I. Bigi et al., Phys. Rev. Lett. {\bf 71}, (1993) 496; Int. J. Mod.
Phys. {\bf A9}, 2467 (1994).

\bibitem{Neubert94}
M. Neubert, Phys. Rep. {\bf 245} (1994) 259.

\bibitem{bqmass}
E. Bagan, P. Ball, V.M. Braun, and H.G. Dosch, Phys. Lett. {\bf B278},
  457 (1992). 

\bibitem{neubertbsg}
M. Neubert, Phys. Rev. {\bf D49} (1994) 4623.

%
\bibitem{aag96}
A. Ali, H.M. Asatrian, and C. Greub (to be published).

\bibitem{CLEOwarsaw}
 R. Ammar et al. (CLEO Collaboration), contributed paper to the International
Conference on High Energy Physics, Warsaw, 25 - 31 July 1996, CLEO CONF
96-05.

\bibitem{wyler95}
A. Khodzhamirian, G. Stoll, and D. Wyler, Phys. Lett. {\bf B358} (1995) 129.

\bibitem{ab95}
A. Ali and V.M. Braun, Phys. Lett. {\bf B359} (1995) 223.

\bibitem{BBK89}
I.I. Balitsky, V.M. Braun, and A.V. Kolesnichenko, Nucl. Phys. {\bf 312}
(1989) 509.
\bibitem{DGP96}
J.F. Donoghue, E. Golowich, and A.A. Petrov, preprint UMHEP-433 (1996)
[hep-ph-9609530].
\bibitem{Alihera}
A. Ali, in {\it Future Physics at HERA}, Proceedings of the Workshop,
DESY, Hamburg 1995/96 (editors: G. Ingelman, A. De Roeck, R. Klanner),
Vol.~1 (1996) 446.
\bibitem{AGM94}
 A. Ali, G. F. Giudice and T. Mannel, Z. Phys. {\bf C67} (1995) 417.
 \bibitem{LRsymmetry}
        K. Fujikawa and A. Yamada, Phys. Rev. {\bf D49} (1994) 5890;\\
        P. Cho and M. Misiak, Phys. Rev. {\bf D49} (1994) 5894.
\bibitem{MisiakBM94}
        M. Misiak in ref. \protect\cite{Ciuchini};
A.J. Buras and M. M\"unz, Phys. Rev. {\bf D52} (1995) 186.
\bibitem{long}
        C.~S.~Lim, T.~Morozumi and A.~I.~Sanda,
                 Phys. Lett. {\bf 218} (1989) 343;\\
        N.~G.~ Deshpande, J.~Trampetic and K.~Panose,
                 Phys. Rev. {\bf D39} (1989) 1461;\\
        P.~J.~O'Donnell and H.~K.~K.~Tung, Phys. Rev. {\bf D43} (1991)
R2067.\\
        N. Paver and Riazuddin, Phys. Rev. {\bf D45} (1992) 978.

\bibitem{amm91}
        A. Ali, T. Mannel and T. Morozumi, Phys. Lett. {\bf B273} (1991) 505.
\bibitem{CMW96}
P. Cho, M. Misiak, and D. Wyler, Phys. Rev. {\bf D54} (1996) 1944.

\bibitem{HWS87}
W.S. Hou, R.S. Willey and A. Soni, Phys. Rev. Lett. {\bf 58} (1987) 1608
[E. {\bf 60} (1988) 2337].

 \bibitem{UA1R}
C. Albajar et al. (UA1), Phys. Lett. {\bf B262} (1991) 163.

\bibitem{Hewettpol}
J. Hewett, Phys. Rev. {\bf D53} (1996) 4964.

\bibitem{KS96}
F. Kr\"uger and L.M. Sehgal, Phys. Lett. {\bf B380} (1996) 199.

\bibitem{Masieroetal}
S. Bertolini, F. Borzumati, A. Masiero, and G. Ridolfi,
Nucl. Phys. {\bf B353} (1991) 591.

\bibitem{Gotoetal96}
T. Goto, Y. Okada, Y. Shimizu, and M. Tanaka, preprint KEK-TH-483;
OU-HET 247; TU-504 (1996) [hep-ph-9609512].

%
\bibitem{Grossman}
Y. Grossman, Z. Ligeti, and E. Nardi, Nucl. Phys. {\bf B465} (1996) 369.
%
\bibitem{BuBu93}
G. Buchalla, and A.J. Buras, Nucl. Phys. {\bf B400} (1993) 225.

\bibitem{ALEPHwarsaw}
Contributed paper by the ALEPH collaboration to the International
Conference on High Energy Physics, Warsaw, ICHEP96 PA10-019 (1996).

%
 \bibitem{Buraswarsaw}
A.J. Buras, preprint TUM-HEP-259/96, MPI-PhT/96-111 (1996) [hep-ph/9610461],
 invited talk at the
International Conference on High Energy Physics, Warsaw, Poland,
July 25-31, 1996 (to appear in the proceedings.)

%
%
\bibitem{AL94} A. Ali and D. London, \zpc{65}{95}{431}.

\bibitem{Burasetal} A.J. Buras, W. Slominski, and H. Steger, Nucl.\ Phys.\
 B238 (1984) 529; {\it ibid.}  {\bf B245} (1984) 369.

\bibitem{HN94} S. Herrlich and U. Nierste, Nucl.\ Phys.\ {\bf B419} (1994)
292.

\bibitem{etaB} A.J. Buras, M. Jamin and P.H. Weisz, Nucl.\ Phys.\
{\bf B347} (1990) 491.

\bibitem{HN95} S. Herrlich and U. Nierste, Phys.\ Rev.\  {\bf D52} (1995)
6505.

\bibitem{Soni95}
A. Soni, preprint [hep-lat/9510036] (1995).

\bibitem{BP95} J. Bijnens and J. Prades, Nucl.\ Phys.\  {\bf B444} (1995)
523.

\bibitem{Sharpe94} S. Sharpe, Nucl.\ Phys.\ B (Proc.\ Suppl.)  {\bf 34}
(1994) 403.

\bibitem{Crisafulli95} M. Crisafulli et al.\ (APE Collaboration), Phys.\
Lett.\  {\bf B369} (1996) 325.

\bibitem{JLQCD} S. Aoki et al.\ (JLQCD Collaboration), in {\it Lattice 1995}.
 The numbers cited for $B_K$ from the JLQCD
collaboration as well as from the work of Soni and Bernard are quoted by
Soni in his review \protect\cite{Soni95}.

\bibitem{Wittig96} H. Wittig, preprint DESY 96-110 (1996) [hep-ph/9606371].

\bibitem{UKQCDBB} A.K. Ewing et al.\ (UKQCD Collaboration), preprint,
[hep-lat-9508030] (1995).

\bibitem{Gimenez96} V. Gim\'enez and G. Martinelli, preprint ROME 96/1153,
FTUV 96/25- IFIC 96/30 (1996) [hep-ph-9610024].

\bibitem{narison95} S. Narison, Phys.\ Lett.\  {\bf B351} (1995) 369.

\bibitem{Narison} S. Narison, Phys.\ Lett.\  {\bf B322} (1994) 247; S.
Narison and A. Pivovarov, {\it ibid}  {\bf B327} (1994) 341.  
%
\bibitem{BCPasym} For reviews, see, for example, Y. Nir and
H.R. Quinn, in {\it $B$ Decays}, edited by S. Stone (World 
Scientific, Singapore, 1992) 362; I. Dunietz, {\it ibid} 393.

\bibitem{penguins}  D. London  and R. Peccei,
Phys. Lett. {\bf B223} (1989) 257;\\
 B. Grinstein, Phys. Lett. {\bf B229} (1989) 280;\\
M. Gronau, Phys. Rev. Lett. {\bf 63} (1989) 1451,
Phys. Lett. {\bf B300} (1993) 163.

\bibitem{isospin}  M. Gronau  and D. London,
Phys. Rev. Lett. {\bf 65} (1990) 3381.

\bibitem{ADK}
 R. Aleksan, I. Dunietz, and B. Kayser,
Z. Phys. {\bf C54} (1992) 653.

\bibitem{growyler} M. Gronau and D. Wyler, 
Phys. Lett. {\bf B265} (1991) 172.
 See also M. Gronau and D. London,
Phys. Lett. {\bf B253} (1991) 483;
I. Dunietz, Phys. Lett. {\bf B270} (1991) 75.

\bibitem{EWPsize}
 R. Fleischer, Phys. Lett. {\bf B332} (1994) 419;\\
 N.G. Deshpande and  X.-G. He, Phys. Lett. {\bf B345} (1995) 547;\\
 G. Kramer and W.F. Palmer, Phys. Rev. {\bf D52} (1995) 6411;
 A.J. Buras and R. Fleischer, {\bf B365} (1996) 390.

\bibitem{DH2} N.G. Deshpande and  X.-G. He, Phys. Rev. Lett. {\bf 75}
(1995) 3064.

\bibitem{GR96}
M. Gronau and J.L. Rosner, Phys. Rev. {\bf D53} (1996) 2516.

\bibitem{Gronau96}
M. Gronau, preprint TECHNION-PH-96-39 (1996) [hep-ph/9609430].

\end{thebibliography}
\end{document}